\documentclass[11pt,a4paper]{article}

\usepackage[utf8]{inputenc}
\usepackage[T1]{fontenc}
\usepackage{lmodern}
\usepackage[a4paper,margin=1in]{geometry}
\usepackage[protrusion=true,expansion=false]{microtype}
\usepackage{parskip}
\usepackage{xcolor}
\PassOptionsToPackage{hyphens}{url}
\usepackage{url}

\usepackage{amsmath,amssymb}

\usepackage{graphicx}
\usepackage{subcaption}
\usepackage{multirow}
\usepackage{booktabs}
\usepackage{array}
\usepackage{tabularx}
\usepackage{longtable}
\usepackage{threeparttable}
\usepackage{algorithm}
\usepackage{algpseudocode}

\usepackage{enumitem}
\usepackage{eurosym}
\usepackage[numbers,sort&compress]{natbib}

\usepackage{tikz}
\usetikzlibrary{arrows.meta, positioning, fit, backgrounds}
\usetikzlibrary{calc}

\usepackage[hidelinks]{hyperref}

\setlist[itemize]{leftmargin=1.2em,itemsep=0.2em,topsep=0.2em}
\setlist[enumerate]{leftmargin=1.6em,itemsep=0.2em,topsep=0.2em}

\newcolumntype{L}{>{\raggedright\arraybackslash}X}

\begin{document}

\title{\textbf{SCARFACE: a harmonized spatio-temporal dataset integrating socio-economic, environmental, and agricultural indicators for the Po Valley (Italy), 2011--2024}}
\author{
Paolo Maranzano$^{1,2}$, Pietro Colombo$^{3}$, Felicetta Carillo$^{4}$, \\
Riccardo Borgoni$^{1}$, Riccardo Pajno$^{1}$, Matteo Borrotti$^{1}$,\\
Luca Ferrero$^{5}$ and Ezio Bolzacchini$^{5}$\\
\vspace{0.2em}
\small
$^{1}$University of Milano-Bicocca, Department of Economics, Management and Statistics (DEMS),\\
\small
Piazza dell'Ateneo Nuovo n.1, 20100, Milano, Italy\\
\small
$^{2}$Fondazione Eni Enrico Mattei, Corso Magenta n.63, 20123, Milano, Italy\\
\small
$^{3}$School of Mathematics and Statistics, University of Glasgow, UK\\
\small
$^{4}$Italian Council for Agricultural Research and Economics -- \\
\small
Research Centre for Agricultural Policies and Bioeconomy (CREA-PB),\\
\small
$^{5}$University of Milano-Bicocca, Department of Earth and Environmental Sciences (DISAT)\\
\small
Corresponding author: \texttt{paolo.maranzano@unimib.it}
}
\date{}
\maketitle

\begin{abstract}
We present \textit{Sequestering CARbon through Forests, AgriCulture, and land usE (SCARFACE)}, a harmonized spatio-temporal dataset that integrates climate, air quality, airborne pollutant emissions, land cover, soil properties, agro-industry dynamics and socio-economic indicators, to jointly investigate interconnected processes linking agricultural systems, atmospheric dynamics, emissions, and socioeconomic conditions in the Po Valley, Northern Italy. The spatial reference unit is the Agrarian Sub Region (ASR), that is, groups of contiguous municipalities that are considered homogeneous with respect to physical geography, agronomic characteristics, and prevailing agricultural production systems. The dataset adopts an annual panel structure from 2011 to 2024 defined over the 256 ASRs partitioning the Po Valley and comprises more than 2,700 indicators sourced from national and international public institutions. Heterogeneous data are harmonized within a processing workflow, tailored to the specific characteristics of each dataset, that guarantee spatial and temporal consistency of the output dataset. The resource supports reuse in applied econometrics, spatio-temporal modeling, clustering, and policy analysis focused on agriculture, air quality, and land use in a major European hotspot.
\end{abstract}

\vspace{0.5em}
\noindent\textbf{Keywords:} High-dimensional spatio-temporal dataset; Po Valley in Northern Italy; Agrarian regions; Agricultural activities; Air quality and pollution emissions; Meteorology; Spatial block kriging; Land cover; Livestock; Socio-economic indicators.

\section{Background \& Summary}\label{sec:intro}
Recent advances in environmental data science have led to the development of large-scale, multi-variable datasets based on high-resolution long-term observational records and model-based outputs. A common feature of these datasets is the integration of multiple data sources, such as satellite-derived information and ground monitoring networks, through geostatistical interpolation. These efforts enabled the generation of consistent spatio-temporal data capturing environmental variability across large domains, supporting applications in climate analysis, air quality monitoring, and ecosystem assessment.

Notable examples include global climate data repositories such as the HCLIM database \citep{LundstadEtAl2023}, which provide harmonized historical climate observations at the global scale. In parallel, high-resolution gridded datasets of climate extremes and related indicators have been developed to support the analysis of environmental variability and its impacts across sectors \citep{MistryEtAl2019}.

Substantial efforts have also been directed toward the production of harmonized, high-resolution geospatial datasets describing key components of the Earth system. These include global soil information products based on machine learning and environmental covariates \citep{HenglEtAl2017,GuptaEtAl2022,HenglEtAl2026}, global land cover datasets derived from satellite observations \citep{FriedlEtAl2010}, and large-scale maps of vegetation and land use dynamics \citep{ParenteEtAl2024}. In addition, global human settlement and population datasets derived from remote sensing and statistical modeling have enabled detailed spatial representations of human presence and sociodemographic patterns \citep{PesaresiEtAl2013,TatemEtAl2017}.

Other contributions have expanded the availability of agricultural and environmental datasets with strong relevance for cross-domain analysis. For example, \cite{LiangEtAl2026} presented a 30\,m multi-year dataset of major crop distributions in China for the period 2013--2024, integrating Sentinel-2 satellite data and phenological modeling to generate annual maps of cotton, maize, wheat, and rice with high validation accuracy and consistency with official statistics. Similarly, \cite{DiEtAl2026} developed annual 30\,m crop type maps for Northeast China spanning 2001--2022 by combining Landsat, MODIS, and temporal reflectance fusion, thereby extending fine-resolution crop mapping to earlier time periods and demonstrating strong agreement with both ground-truth samples and government statistics. At a finer observational scale, \cite{ChongEtAl2026} released a multi-sensor and multi-temporal dataset that integrates aerial and ground observations together with reference measurements of biomass and leaf area index, thus supporting data-driven crop monitoring methods across multiple crop species. Beyond crop mapping, \cite{CatarinoEtAl2025} provided a pan-European 50\,m map of natural pest control potential based on Copernicus high-resolution layers linking semi-natural habitat structure to regulating ecosystem services relevant for agricultural management. Complementing these gridded and remote-sensing products, \cite{ReesEtAl2025} built a rich farm-level dataset on Swiss arable farms covering soil management practices, farmer characteristics, production conditions, and linked secondary census information, thereby offering detailed micro-level evidence on management decisions and soil-health-related practices.

The Po Valley in Northern Italy (depicted in Figure~\ref{fig:PoValley_multipanel}) is characterized by strong interactions among agricultural systems, environmental processes, and human activities \citep{raffaelli2020improving,bigi2012analysis}. In fact, the Po valley is one of Europe’s air pollution hotspots, due to stable atmospheric conditions and low wind speeds lead to aerosol accumulation within the planetary boundary layer \citep{FerreroEtAl2011,DiemozEtAl2019,CrovaEtAl2021} a phenomenon evident even from satellite observations \citep{DiNicolantonioEtAl2007,DiNicolantonio2009,FerreroEtAl2019}. Most importantly, the Po Valley is proxy of a more common phenomenon holding for the entire Italian peninsula due to its complex orography which results in a multitude of basin valleys of various sizes, in which urban, industrial and agricultural activities lie \citep{FerreroEtAl2012,  FerreroEtAl2014} making it a proxy of similar geographical contexts. In such contexts, the availability of harmonized (i.e., spatially-consistent, and temporally-comparable), high-resolution spatio-temporal datasets integrating environmental, agricultural, and socio-economic dimensions is essential to support robust analysis, modeling, and evidence-based policy design. A noticeable example of this approach is the Agrimonia dataset \citep{Agrimonia2023} that integrates multiple data sources, such as satellite retrievals, in-situ monitoring networks, to simultaneously characterize livestock, air quality, airborne pollutant emissions and meteorological conditions for a specific part of the Po Valley, namely the Extended Lombardy.


The above contributions highlight the rapid growth of high-resolution agricultural datasets based on satellite imagery, gridded environmental layers, and farm-level observations, while also illustrating that most available resources remain focused on specific thematic domains rather than on cross-domain harmonization across environmental, agricultural, and socio-economic dimensions. In this spirit, \textit{Sequestering CARbon through Forests, AgriCulture, and land usE (SCARFACE)} is a high-dimensional harmonized spatio-temporal dataset that systematically integrates several domains, such as climate, extreme events, air quality, pollution emissions, land cover, soil properties, agro-industry dynamics and socio-economic indicators, to jointly investigate the complex interconnected processes linking agricultural systems, atmospheric dynamics, emissions, and socio-economic conditions in the whole Po Valley. For example, carbon sequestration in agriculture, forestry, and land use policies is a critical aspect of climate change mitigation \citep{Plieninger2011}. The potential for carbon sequestration in these sectors offers a powerful tool for controlling global climate change \citep{PapadopoulosEtAl2014,KauerEtAl2015,AbbasEtAl2020,BhattacharyyaEtAl2022} but requires a careful assessment with all the other atmospheric phenomena, pollutants (e.g. NH3 and PMx) and dynamics.

A distinctive feature of the SCARFACE dataset is the explicit inclusion of a large set of indicators derived aggregating farm-level data to areal values. In particular, we considered indicators describing multiple dimensions of agricultural activities, that is, the techno-economic specialization and production structure (e.g., input, output, and the relative importance of crop and livestock production) of agricultural holdings; the environmental impact of farming activities (measured through a dedicated agricultural carbon footprint indicator); key agronomic practices such as the use of fertilizers, phytosanitary products, and manure management. Further indicators are extracted from the balance sheet and income statement to enable the characterization of their economic performance, resource allocation, and economic sustainability of farms.

SCARFACE complements techno-productive information with livestock headcounts from the Italian national veterinary census, enabling a consistent and spatially explicit representation of intensive animal farming systems. This information is particularly critical in the Po Valley, where specific meteorological and topographical conditions favor pollutant accumulation and exacerbate air quality issues in livestock-intensive regions such as Lombardy represent major hotspots of agricultural emissions \citep{MarongiuEtAl2024,MarongiuEtAl2025}. Indeed, livestock density is strongly associated with emissions of methane and ammonia (NH$_3$), as well as primary particulate matter (PM$_{2.5}$, PM$_{10}$) and their secondary precursors \citep{Agrimonia2023}. Moreover, ammonia emissions originating from manure management and fertilizer application play a key role in the formation of secondary inorganic aerosols, significantly contributing to ambient particulate matter concentrations \citep{MarongiuEtAl2022,ColomboEtAl2023}. Therefore, integrating high-resolution livestock data is essential to capture the agricultural contribution to air pollution and to support robust assessments of emission mitigation strategies in one of Europe's most polluted regions.

Another valuable feature of SCARFACE, is the presence of atmospheric indicators, which represents a key added value for investigating the environmental impacts of agriculture and for supporting integrated assessments of emission mitigation and air quality management. First, the dataset enables the identification of the relative contribution of different agricultural processes to total emissions of key airborne pollutants by explicitly incorporating a detailed breakdown of sector-specific emissions by agricultural processes--including manure management, enteric fermentation, agricultural soils, and agricultural waste burning--collected from satellite-derived emissions inventories. Furthermore, the dataset integrates gridded air quality datasets describing ambient concentrations of pollutants that are strongly influenced by agricultural emissions, such as PM$_{2.5}$, PM$_{10}$ \citep{DasEtAl2021}, and nitrogen compounds \citep{WyerEtAl2022}. Second, the dataset incorporates multiple satellite-derived products describing atmospheric conditions, meteorology, extreme weather events, land cover, and physical geography. These variables capture the physical drivers of environmental variability and enable a consistent analysis of agricultural pressures and atmospheric patterns over time.

Finally, the dataset integrates socio-economic and administrative information to contextualize environmental and agricultural dynamics within human systems. Specifically, the dataset contains detailed information on demographic structure, economic conditions, and local development patterns. The combination of environmental, agricultural, and socio-economic domains enables a comprehensive representation of coupled human–environment systems across space and time.

The dataset is designed as a versatile resource supporting both methodological and applied developments, as well as policy-relevant analyses. Its structure enables panel data analyses at moderate spatial and temporal resolutions \citep{Elhorst2010,ParentLeSage2012}, advanced spatio-temporal modeling in the presence of heterogeneous covariates and high-dimensional settings \citep{Wikle2015,CressieWikle2015}. In addition, the dataset lends itself naturally to spatial and spatio-temporal clustering exercises \citep{CerquetiMattera2025,MorelliEtAl2026}, facilitating the identification of regional typologies and underlying patterns in agricultural and environmental systems. These features make SCARFACE a valuable tool for the validation of innovative statistical and econometric methods, while simultaneously enabling empirical, data-driven applications aimed at testing research hypotheses on complex socio-environmental processes. Beyond methodological applications, by integrating heterogeneous data sources into a unified and harmonized framework, the dataset provides a robust empirical basis for reproducible, cross-domain policy-oriented analyses, particularly in relation to agricultural transitions, air quality management, and climate variability in one of Europe’s most critical environmental hotspots.

The remainder of this paper is organized as follows. Section \ref{Sec:Methods} provides a comprehensive overview of the SCARFACE database. In particular, it describes its statistical structure, the geographic area of interest and temporal coverage (Subsection \ref{sec:StudyArea}), the processing workflow adopted to harmonize the multiple data sources involved (Subsection \ref{sec:OverviewHarmonization}), the key statistical methodologies used for temporal and spatial aggregation (Subsection \ref{sec:spatial_alignment}), and the original data sources and corresponding data records (Subsection \ref{sec:DataSources}). Section \ref{sec:datarecords} describes the content of the Zenodo public repository in which the users will find the dataset, the replication code and other companion documents. Section \ref{sec:Validation} presents quality checks on the harmonized data, with specific discussion by data type (e.g., gridded or pointwise) and processing methodology. Section \ref{sec:usagenotes} provides practical guidance for users of the SCARFACE database, including information on missing data, potential uses, and key caveats to consider when analyzing the data. Finally, Section \ref{sec:codedata} provides information on code and data availability. The main text is accompanied by a Appendix that includes additional notes on the statistical methodologies employed in the workflow, as well as numerous auxiliary tables that complement the description of the data presented herein.

\section{Methods}\label{Sec:Methods}

\subsection{Overview of the SCARFACE dataset}\label{sec:StudyArea}
The SCARFACE dataset focuses on the Po Valley area in Northern Italy, one of the most important agricultural and industrial regions in Europe \citep{Yearbook2024}. The study area covers the four main administrative regions of Piedmont, Lombardy, Veneto, and Emilia-Romagna, which together form a highly productive agricultural basin characterized by intensive livestock farming, extensive crop production, and a strong agro-food industry.

The spatial reference unit adopted in SCARFACE is the \emph{Agrarian Sub-Region} (ASR), the English rendering of the Italian \emph{Regione Agraria}, a territorial classification defined by the Italian National Statistics Office \citep{ASRdef}. ASRs represent groups of contiguous municipalities that are considered homogeneous with respect to physical geography, agronomic characteristics, and prevailing agricultural production systems. The Po Valley can be partitioned into $m=256$ time-invariant ASRs with different sizes and shapes.

\begin{figure}[htbp]
\centering
\begin{subfigure}[b]{0.48\textwidth}
    \centering
    \includegraphics[width=\textwidth]{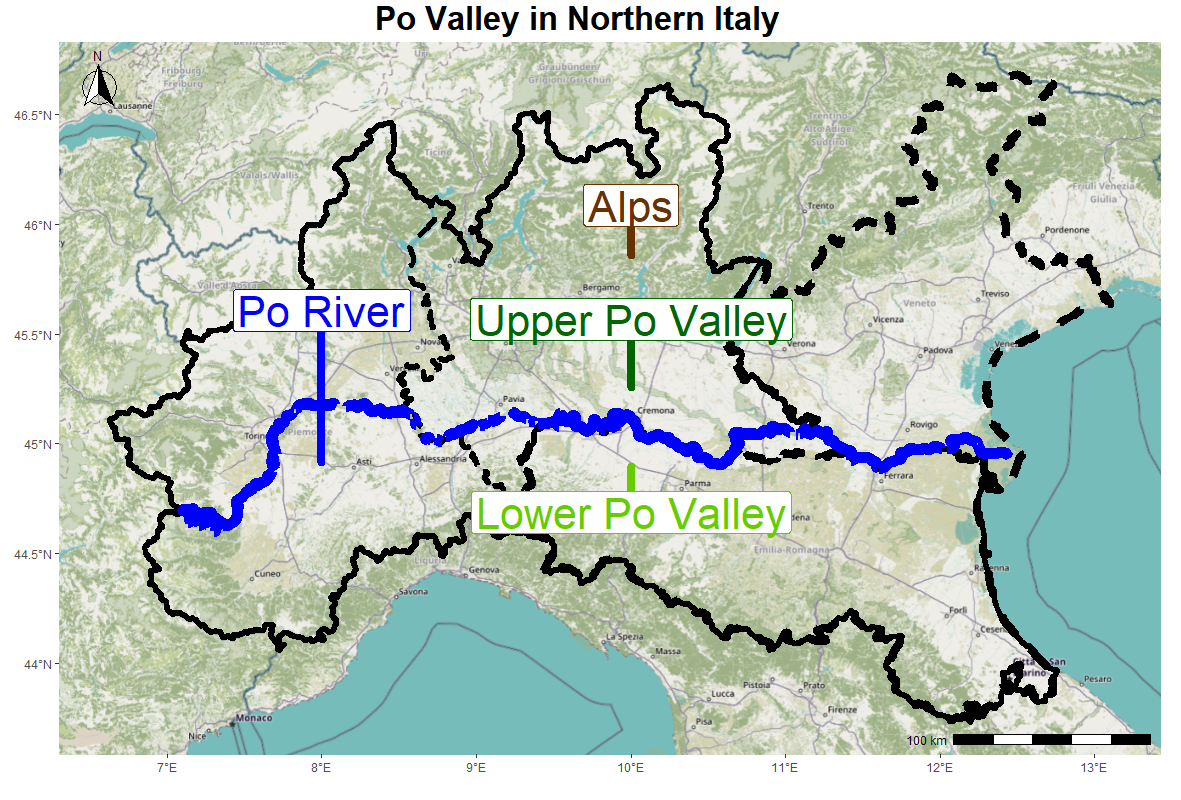}
    \caption{Po Valley study area}
    \label{fig:PoValley_area}
\end{subfigure}
\hfill
\begin{subfigure}[b]{0.48\textwidth}
    \centering
    \includegraphics[width=\textwidth,height=5cm]{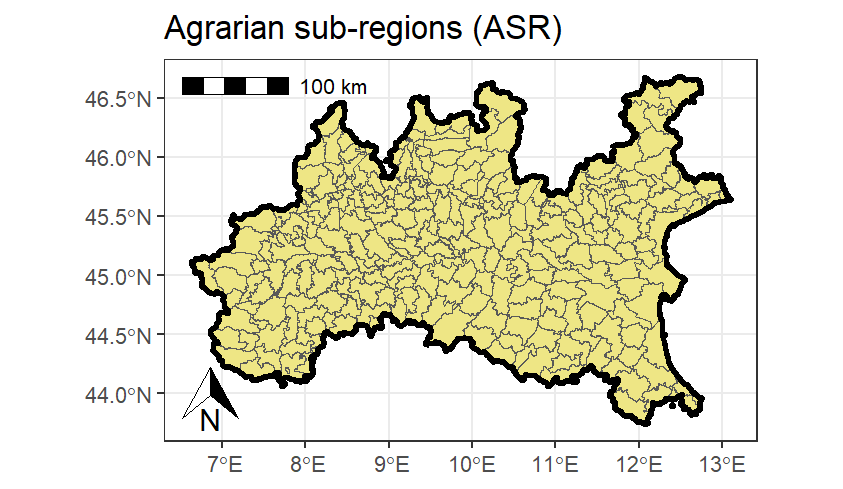}
    \caption{Agrarian Sub-Regions (ASRs)}
    \label{fig:PoValley_ASR}
\end{subfigure}
\caption{Spatial framework of the SCARFACE dataset. Panel (a) shows the geographical configuration of the Po Valley study area in Northern Italy. Panel (b) displays the $m=256$ Agrarian Sub-Regions (ASRs) used as the output reference unit for the Po Valley .}
\label{fig:PoValley_multipanel}
\end{figure}

The spatial framework of the dataset is illustrated in Figure~\ref{fig:PoValley_multipanel}. The left figure shows the geographical extent of the Po Valley study area, while the right figure displays the ASRs used as the reference spatial units for data harmonization and analysis.

The SCARFACE dataset integrates information for the period from 2011 to 2024 (i.e., $T=14$ time stamps), with the initial and final temporal coverage depending on the availability of the individual data sources. Therefore, the final database adopts an annual panel structure defined over the ASRs and composed of a total number of spatio-temporal observations equal to $N=m\times\,T=256\times\,14=3584$ for each variable.

Overall, SCARFACE comprises a set of $p=2748$ variables (plus three unique identifiers, that is, year, ASR and geometry) that include administrative records, gridded environmental products, satellite-derived land information, and survey-based socio-economic indicators sourced from national and international public institutions, covering a wide range of thematic domains. Table~\ref{tab:core_domains} summarizes the main thematic domains and data sources integrated in the dataset, together with their original spatial and temporal supports and the harmonization procedures used to align them to the ASR level. Farm activity and agro-economic indicators are derived from the Farm Accountancy Data Network (FADN) survey coordinated by the Italian Council for Agricultural Research and Economics (CREA). Emissions data are obtained from the EDGAR inventories developed by the European Commission, while air quality information is sourced from both the European Environment Agency (EEA) and the Copernicus Atmosphere Monitoring Service (CAMS). Meteorological variables are retrieved from the ERA5-Land reanalysis produced by the European Centre for Medium-Range Weather Forecasts (ECMWF), and extreme weather indicators are provided by the European Drought Observatory (EDO). Land cover information is based on the CORINE Land Cover dataset from Copernicus and the Global Dynamic Land Cover (GDLC) dataset. Livestock data are obtained from the Italian National Livestock Registry (BDN) managed by the Ministry of Health, while socio-economic indicators are produced by ISTAT. Finally, geographical features and administrative metadata are derived from a combination of Amazon Web Service (AWS), ISTAT and Eurostat.

\begin{table}[htbp]
\centering
\resizebox{\textwidth}{!}{
\begin{threeparttable}
\caption{Overview of core thematic domains integrated in the SCARFACE dataset}
\label{tab:core_domains}
\begin{tabular}{p{2.6cm} p{5.0cm} p{2.4cm} p{2.0cm} p{2.4cm} p{4.8cm} p{1.6cm}}
\toprule
\textbf{Domain} & \textbf{Data source} & \textbf{Spat. resolution} & \textbf{Temp. resolution} & \textbf{Period} & \textbf{ASR method} & \textbf{Num. indic.} \\
\midrule
Farm activities & FADN (CREA) & Farm-level microdata & Annual & 2011--2023 & Population-weighted direct aggregation (sum/mean) & 434 \\
Emissions & EDGAR GHG v2025; EDGAR AP v8.1 (JRC, European Commission) & $0.1^{\circ}$ grid & Annual & 2011--2024 (GHG); 2011--2022 (AP) & Spatial block kriging & 96 \\
Air quality (1) & European air quality interpolated data (EEA Datahub) & 1 km grid & Annual & 2011--2022 & Spatial block kriging & 8 \\
Air quality (2) & CAMS European air quality reanalyses (CAMS Datahub) & $0.1^{\circ}$ grid & Hourly & 2013--2024 & Temporal aggregation + spatial block kriging & 680 \\
Meteorology & ERA5-Land reanalysis (ECMWF / Copernicus CDS) & $0.1^{\circ}$ hourly grid & Hourly & 2011--2024 & Temporal aggregation + spatial block kriging &  1170 \\
Extreme weather & European Drought Observatory (EDO) & $0.1^{\circ}$ grid & Infra-annual & 2011--2024 & Temporal aggregation + spatial block kriging &  208 \\
Land cover (1) & CORINE Land Cover (Copernicus) & 100 m grid & Multi-year snapshots & 2012 and 2018 & Reclassification + piecewise constant + area shares & 9 \\
Land cover (2) & GDLC dataset (Global Dynamic Land Cover) & 100 m grid & Multi-year snapshots & 2015--2019 & Reclassification + piecewise constant + area shares &  6 \\
Livestock & BDN livestock registry (Italian Ministry of Health) & Municipality (LAU) & Annual & 2011--2024 & Direct aggregation (sum) &  16 \\
Socio-economic indicators & Sistema informativo ``A misura di comune'' (ISTAT) & Municipality (LAU) & Annual & 2014--2023 & Direct aggregation (sum/mean) & 95 \\
Geographical features & Terrain elevation and morphology indicators (AWS Terrain Tiles; ISTAT) & Raster / administrative polygons & Static & Time-invariant & Spatial extraction + aggregation & 10 \\
Administrative metadata & NUTS territorial attributes (Eurostat GISCO) & Administrative regions & Static & Time-invariant & Spatial join to ASR & 16 \\
\bottomrule
\end{tabular}
\begin{tablenotes}[flushleft]
\footnotesize
\item \textit{Note:} All variables are harmonized at the ASR level through temporal aggregation, spatial block kriging, or direct aggregation depending on the nature of the original data source. The number of indicators (Num. Indic.) is given by the sum of annual and seasonal indicators (for CAMS, EDO and ERA5) or the sum of mean and total indicators plus the corresponding variances (for FADN data).
\end{tablenotes}
\end{threeparttable}
}
\end{table}


It should be noted that, while all datasets included in SCARFACE are provided by national and international public institutions and are, in principle, publicly available, farm activity and agro-economic indicators derived from the FADN represent the only exception, as they are accessed under a research data agreement with the data owners, that is, CREA. Specifically, the analysis relies on farm-level microdata that cannot be publicly disclosed without appropriate anonymization and spatial aggregation. The methodology adopted to ensure confidentiality is described in detail in the following Section \ref{Sec:Data_FADN}.

\subsection{The harmonization workflow in brief}\label{sec:OverviewHarmonization}
The input data used in SCARFACE differ substantially in spatial support (i.e., municipal polygons, regular grids, raster cells, point observations), temporal resolution (e.g., hourly, seasonal, annual, quasi-static), coordinate reference systems, and measurement units. For instance, environmental variables such as meteorological indicators and air pollutant concentrations are typically provided on regular grids, whereas administrative and census statistics are available at municipal level. Integrating these heterogeneous sources therefore requires a combination of spatio-temporal aggregation and interpolation procedures tailored to the specific characteristics of each dataset.

The temporal coverage of the underlying data sources is also heterogeneous across datasets. For example, meteorological variables are available for the entire study period, whereas some socio-economic indicators are available only for a subset of years. Other information, including land cover and terrain descriptors, is essentially time-invariant or updated only at sparse time intervals. As a result, the harmonized panel may contain missing observations for certain variables and time periods. These gaps reflect differences in the availability of the original data sources and are therefore preserved in the released dataset.

More broadly, the SCARFACE database integrates several types of information, including census-based administrative statistics (e.g., livestock counts), quasi-static geographical descriptors (e.g., land cover and terrain morphology), spatio-temporal environmental variables (e.g., meteorological conditions and atmospheric pollutant concentrations), and survey-based farm-level information (e.g., techno-economic specialization on agricultural farms). Each of these data types requires dedicated pre-processing and harmonization procedures before being integrated into the final ASR-level panel.

\begin{figure}[htbp]
\centering
\resizebox{1\textwidth}{!}{
\begin{tikzpicture}[
    font=\small,
    >=Latex,
    node distance=8mm and 12mm,
    box/.style={
        rounded corners=3pt,
        draw=black,
        thick,
        align=center,
        minimum height=10mm,
        text width=#1,
        inner sep=4pt
    },
    input/.style={box=#1, fill=blue!8},
    process/.style={box=#1, fill=orange!10},
    output/.style={box=#1, fill=green!10},
    header/.style={
        font=\bfseries,
        align=center,
        text width=#1
    },
    arrow/.style={->, thick},
    group/.style={
        rounded corners=4pt,
        draw=gray!60,
        dashed,
        inner sep=6pt
    }
]
\node[header=3.6cm] (h1) at (0,0) {Input data types};
\node[header=5cm] (h2) at (6.5,0) {Main transformations};
\node[header=5cm] (h3) at (13.0,0) {Spatial harmonization};
\node[header=3.4cm] (h4) at (19.0,0) {Unified output};
\node[input=3.8cm, below=10mm of h1] (muni) {\textbf{Municipal administrative data}\\[1mm]
\footnotesize (e.g., livestock, socio-economic indicators)};
\node[input=3.8cm, below=8mm of muni] (grid) {\textbf{Gridded data}\\[1mm]
\footnotesize (e.g., meteorology, air quality, emissions, drought indicators)};
\node[input=3.8cm, below=8mm of grid] (raster) {\textbf{Static raster} \\[1mm]
\footnotesize (e.g., land cover, terrain and morphometric features)};
\node[process=4.0cm, right=22mm of muni] (agg1) {Direct aggregation\\[1mm]
\footnotesize Sum/mean by municipality};
\node[process=4.0cm, right=22mm of grid] (temp) {Temporal aggregation\\[1mm]
\footnotesize Hourly/daily/monthly $\rightarrow$ seasonal \& annual summaries};
\node[process=4.0cm, right=22mm of raster] (recl) {Reclassification \& extraction\\[1mm]
\footnotesize Land cover grouping, terrain summaries, area shares};
\node[process=4.0cm, right=22mm of agg1] (asr1) {ASR aggregation\\[1mm]
\footnotesize Municipality $\rightarrow$ ASR via fixed 1:N mapping};
\node[process=4.0cm, right=22mm of temp] (bk) {Local block kriging (BK)\\[1mm]
\footnotesize Cells $\rightarrow$ ASR mean \& variance};
\node[process=4.0cm, right=22mm of recl] (asr2) {ASR-level summaries\\[1mm]
\footnotesize Shares, averages, minima, maxima, standard deviations};
\node[output=3.8cm, right=22mm of bk] (final) {SCARFACE harmonized database\\[1mm]
\footnotesize 256 ASRs in the Po Valley\\
\footnotesize WGS84 (EPSG:4326)\\
\footnotesize Annual panel structure};
\draw[arrow] (muni) -- (agg1);
\draw[arrow] (grid) -- (temp);
\draw[arrow] (raster) -- (recl);
\draw[arrow] (agg1) -- (asr1);
\draw[arrow] (temp) -- (bk);
\draw[arrow] (recl) -- (asr2);
\draw[arrow] (asr1.east) -- ++(0.7,0) |- (final.west);
\draw[arrow] (bk.east) -- (final.west);
\draw[arrow] (asr2.east) -- ++(0.7,0) |- (final.west);
\node[inner sep=0pt, outer sep=0pt] (finalbottom) at ($(asr2.south)+(110pt,-6pt)$) {};
\begin{scope}[on background layer]
    \node[group, fit=(muni) (grid) (raster) (h1)] {};
    \node[group, fit=(agg1) (temp) (recl) (h2)] {};
    \node[group, fit=(asr1) (bk) (asr2) (h3)] {};
    \node[group, fit=(final) (h4) (finalbottom)] {};
\end{scope}
\end{tikzpicture}
}
\caption{Overview of the data harmonization workflow used to generate the SCARFACE dataset. Different types of input data (i.e., municipal administrative data, gridded environmental datasets, and static raster products) are processed through dataset-specific transformations, including direct aggregation, temporal aggregation, reclassification, and local spatial block kriging. All variables are ultimately aligned to a common spatial support defined by the $m=256$ ASRs of the Po Valley and to a unified coordinate reference system (WGS84, EPSG:4326).}
\label{fig:workflow}
\end{figure}
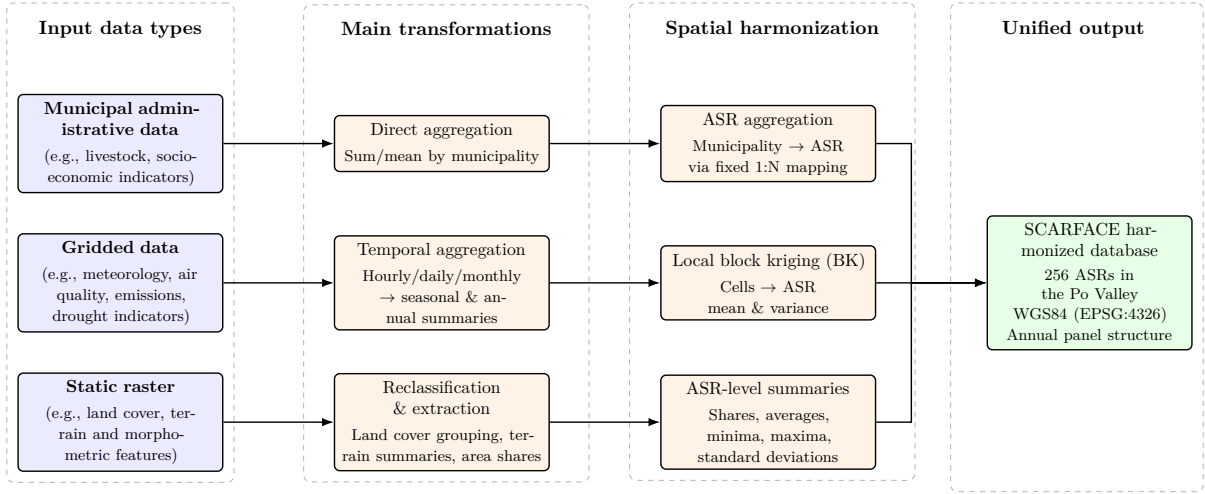

To ensure comparability across spatial and temporal domains, all variables were harmonized to the common ASR spatial support and organized at yearly frequency, allowing the construction of a consistent spatio-temporal database describing environmental conditions, agricultural activities, and socio-economic characteristics of the Po Valley. Specifically, municipal-level administrative data (e.g., livestock and socio-economic indicators) were aggregated to ASR level using direct aggregation (e.g., average or sum) due to the one-to-multiple time-fixed relationship existing between ASRs and municipalities; gridded environmental datasets (e.g., meteorology, air-quality indicators, and emissions inventories) were first temporally aggregated to annual and seasonal summaries where necessary and then spatially transformed to ASR-level estimates using a local spatial block kriging algorithm; static raster products (e.g., land cover, land use and morphometric indicators) were properly reclassified and converted into ASR-level shares or summary statistics. Figure~\ref{fig:workflow} illustrates the structured harmonization workflow implemented to transform the heterogeneous input sources into a consistent spatio-temporal database.

\subsection{Spatial alignment of gridded datasets via spatial block kriging}\label{sec:spatial_alignment}
Several environmental datasets included in SCARFACE, such as meteorological variables, atmospheric concentrations, and emission inventories are originally provided on regular spatial grids (see Table \ref{tab:core_domains}) and need to be realigned to the ASR spatial support to be consistent with the final structure of the dataset.

The need to integrate information collected on different spatial supports is known in spatial statistics as the \textit{Change of Support Problem}, extensively discussed in \cite{gotway2002combining}. In our case, the original data consist of grid-cell observations, whereas the target quantities correspond to averages over irregular polygonal domains (i.e., the ASRs). When the objective is to obtain spatial averages over areas, spatial block kriging (BK) represents a natural and statistically principled solution \citep{chiles2012geostatistics}.

The spatial prediction is implemented using a local ordinary BK algorithm \citep{pebesma2004multivariable}, such that, for each combination of variable, year and sector (when applicable), only the spatial field observed at grid-cell location and a subset of nearby observations is used to predict the average value over each ASR polygon. The kriging predictor is obtained as a weighted linear combination of nearby observations, where the set of contributing observations is restricted to the closest sampling locations around the target block. The locality of the prediction is controlled by the size of the kriging neighborhood (i.e., the number of neighboring observations included in the predictor), whereas the weights depend on a spatial covariance function and on the distance between the cells'centroids.

In practice, the up-scaling procedure consists of three main steps. First, a spatial covariance model is selected. Several isotropic covariance families are considered, including spherical, exponential, Gaussian, and Matérn models. For each candidate covariance model, prediction performance is evaluated using repeated out-of-sample predictions obtained through random $5$-fold cross-validation (CV) \citep[we refer the readers to the recent literature review by][for a detailed overview on cross-validation methods for spatial and spatio-temporal dataset]{OttoEtAl2024}. The covariance model producing the lowest root mean squared prediction error (RMSE) is selected. Second, for the selected covariance model, a grid of candidate neighborhood size is tuned through CV, and the value minimizing the RMSE is retained. Third, using the selected covariance structure and neighborhood configuration, ordinary BK is performed over the ASR polygons. The procedure yields both the predicted spatial mean for each ASR and the corresponding kriging prediction variance, which quantifies the interpolation uncertainty associated with the area-level spatial prediction. The spatial alignment procedure based on local BK is summarized in Algorithm~1 in the Appendix.

\subsection{Data sources and processing}\label{sec:DataSources}
This section summarizes sources, spatial and temporal resolutions, and the key transformations used to produce the harmonized yearly ASR-level variables.

\subsubsection{Administrative information and time-invariant geographical features}
Administrative boundaries and geographical features are used in SCARFACE to support spatial harmonization across datasets and to provide time-invariant contextual variables. Municipality polygons and identifiers constitute the primary spatial reference used for linking several data sources and for aggregating municipal indicators to the ASR level. The relationship between municipalities and ASRs is time-invariant and follows a one-to-many structure: each municipality is uniquely associated with a single ASR, while each ASR may include multiple municipalities\footnote{In the data repository, we provide a matching table reporting the complete municipality--ASR correspondence to facilitate reproducibility of the spatial aggregation procedures.}.

Elevation data were retrieved from the AWS Terrain Tiles digital elevation model (DEM) service with approximate ground resolution of about 30\,m \citep{TilezenTerrain}. Elevation values were extracted for each ASR polygon and used to compute summary statistics including minimum, maximum, mean, and standard deviation of altitude.\footnote{Further technical details on the computation of terrain tiles and ground resolution are available at \href{https://github.com/tilezen/joerd/blob/master/docs/data-sources.md\#what-is-the-ground-resolution}{https://github.com/tilezen/joerd/blob/master/docs/data-sources.md\#what-is-the-ground-resolution} (accessed on March 14th, 2026).} In addition, the total land area of each ASR was also calculated in square meters based on the corresponding polygon geometries. The internal composition of the territory according to elevation bands, namely the shares of ASR area classified as plains (elevation $\leq 200$ m), hills (200 m $<$ elevation $\leq 600$ m), and mountains (elevation $>$ 600 m), were obtained by intersecting the elevation raster with ASR polygons and computing the proportion of cells belonging to three altitude classes.

Finally, additional administrative and territorial attributes were obtained from the Eurostat GISCO database for the nomenclature of territorial units for statistics (NUTS) regions \citep{NUTS}. These include hierarchical identifiers and descriptive variables for the NUTS0 (country), NUTS1 (macro-regions), NUTS2 (regions), and NUTS3 (provinces) levels, together with territorial classifications such as urban–rural typology, remoteness, metropolitan status, coastal and mountain indicators, border status, and land surface measures.

The main time-invariant geographical descriptors included in SCARFACE are summarized in Table~12 of the Appendix.

\subsubsection{EDGAR grid emissions by sector and pollutant}
Sector-specific air pollution emissions data for the period 2011--2024 were obtained from the Emissions Database for Global Atmospheric Research (EDGAR)\footnote{\href{https://edgar.jrc.ec.europa.eu/}{https://edgar.jrc.ec.europa.eu/} (accessed on April 2nd, 2026)}, a comprehensive global dataset developed by the \cite{EDGAR}. Greenhouse gas (GHG) emissions were retrieved from the EDGAR v2025 global greenhouse gas inventory\footnote{\href{https://edgar.jrc.ec.europa.eu/dataset_ghg2025}{https://edgar.jrc.ec.europa.eu/dataset\_ghg2025} (accessed on April 11th, 2026)}, while air pollutant emissions were obtained from the EDGAR v8.1\footnote{\href{https://edgar.jrc.ec.europa.eu/dataset_ap81}{https://edgar.jrc.ec.europa.eu/dataset\_ap81} (accessed on April 11th, 2026)} \citep{CrippaEtAl2024_ESSD}. EDGAR provides globally consistent, gridded annual emission estimates by sector and pollutant at a spatial resolution of 0.1$^\circ$.

The analysis focuses on agricultural emission sources, including agricultural soils (AGS), agricultural waste burning (AWB), enteric fermentation (ENF), manure management (MNM), and indirect N$_2$O emissions from agriculture (N2O), corresponding to the Intergovernmental Panel on Climate Change (IPCC) sector classifications reported in Table~\ref{tab:edgar_sectors}. GHGs include CH$_4$, N$_2$O, CO$_2$, biogenic CO$_2$ (CO$_2$bio), and total greenhouse gases expressed as GWP$_{100}$ according to IPCC AR5. Air pollutant emissions include PM$_{10}$, PM$_{2.5}$, NO$_x$, NMVOC, NH$_3$, black carbon (BC), sulfur dioxide (SO$_2$), organic carbon (OC), and carbon monoxide (CO).

\begin{table}[htbp]
\centering
\caption{Agricultural emission sectors included in the SCARFACE dataset based on the EDGAR inventories. The table reports the corresponding IPCC classification codes and the GHGs and air pollutants considered for each sector.}
\label{tab:edgar_sectors}
\begin{tabular}{llllp{5.5cm}}
\toprule
Sector & Description & IPCC 1996 & IPCC 2006 & Emitted species \\
\midrule
AGS & Agricultural soils & 4C+4D1  & 3C2+3C3 & N$_2$O, NH$_3$, NO$_x$ \\
    &                    & 4D2+4D4 & 3C4+3C7 & NMVOC, PM$_{10}$, \\
AWB & Agricultural waste & 4F & 3C1b & PM$_{2.5}$, CO$_2$, CH$_4$, N$_2$O, \\
    & burning            & & & CO, NO$_x$, NMVOC, PM$_{10}$, PM$_{2.5}$, BC, OC \\
ENF & Enteric fermentation & 4A & 3A1 & CH$_4$, NMVOC, NH$_3$ \\
MNM & Manure management & 4B & 3A2 & CH$_4$, N$_2$O, NH$_3$, \\
 & & & & NMVOC, NO$_x$, PM$_{10}$, PM$_{2.5}$ \\
N2O & Indirect N$_2$O emiss. & 4D3 & 3C5+3C6 & N$_2$O\\
& from agriculture & & & \\
\bottomrule
\end{tabular}
\end{table}

\begin{figure}
\centering
\includegraphics[height=8cm]{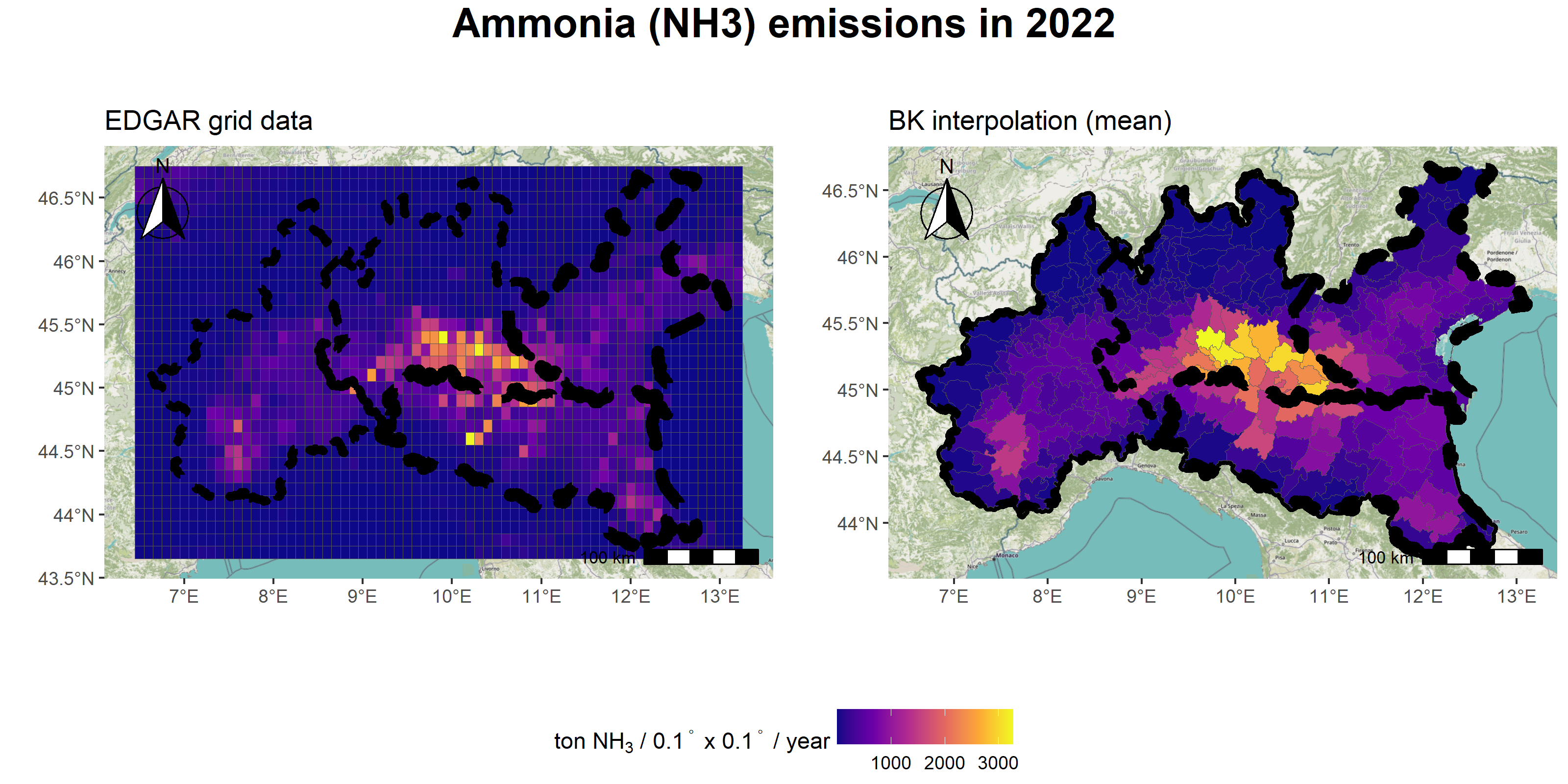}
\caption{Example of the spatial alignment procedure applied to EDGAR emission data. The left panel shows the original EDGAR grid for NH$_3$ emissions in 2022 ($0.1^{\circ}\times0.1^{\circ}$ spatial resolution), while the right panel displays the corresponding ASR-level estimates obtained through local block kriging.}
\label{fig:EDGARasr_NH3_2022}
\end{figure}

Since EDGAR emissions are provided as annual gridded estimates, no temporal aggregation was required. Emission data were spatially aggregated from the original grid to ASRs using the local BK procedure described in detail in Section \ref{sec:spatial_alignment}. An example of the spatial alignment procedure is shown in Figure~\ref{fig:EDGARasr_NH3_2022}, which illustrates the conversion of EDGAR grid emissions for NH$_3$ in 2022 into ASR-level averages using local BK.

\subsubsection{Air quality}
Air pollutant concentration data were obtained from two complementary sources, namely the European air quality interpolated dataset provided by the \cite{EEA2023} and the \cite{CAMS} European air quality reanalyses\footnote{\href{https://ads.atmosphere.copernicus.eu/datasets/cams-europe-air-quality-reanalyses?tab=overview}{https://ads.atmosphere.copernicus.eu/datasets/cams-europe-air-quality-reanalyses?tab=overview} (accessed on April 2nd, 2026).}. These datasets differ in terms of spatial resolution, temporal resolution, and pollutant coverage, and were jointly used to provide a comprehensive characterization of air quality conditions over the study area. While the EEA dataset provides a consistent set of regulatory pollutants over a long time horizon, the CAMS dataset extends the coverage to a broader set of atmospheric components and chemical species, albeit with heterogeneous temporal availability across pollutants. This extended coverage is particularly relevant in agricultural contexts, where emissions of ammonia (i.e., NH$_3$) from livestock production and fertilizer application play a key role in the formation of secondary particulate matter \citep{granella2024formation}, including ammonium nitrate (i.e., NH$_4$NO$_3$) and ammonium sulfate (i.e., (NH$_4$)$_2$SO$_4$) in both PM$_{10}$ and PM$_{2.5}$ \citep{SuttonEtAl2013, BeheraEtAl2013, Agrimonia2023}. 


The EEA dataset provides annually interpolated concentrations derived from monitoring station observations reported under the European air quality regulatory framework and spatially interpolated using geostatistical methods. The data are distributed on a regular grid with a spatial resolution of 1\,km $\times$ 1\,km across Europe. Among others, we considered annual concentration for particulate matter (PM$_{10}$ and PM$_{2.5}$), nitrogen dioxide (NO$_2$), and nitrogen oxides (NO$_x$). Since these data are already available at annual frequency, no temporal aggregation was required. All pollutant concentrations are expressed in $\mu g\,m^{-3}$.

The CAMS European air quality reanalyses provide hourly concentration fields generated through a combination of chemical transport models and data assimilation techniques (ensemble reanalysis). Data are available on a coarser regular grid with a spatial resolution of 0.1$^{\circ}$ (approximately $\sim$10\,km) and surface-level coverage. Hourly CAMS variables were first temporally aggregated at the grid-cell level to derive summary indicators: for each grid cell, annual statistics including cumulated total, mean, minimum, maximum, and standard deviation were computed. In addition, seasonal indicators were derived by computing seasonal means, minima, maxima and standard deviations for the four climatological seasons, that is, Winter (December–February), Spring (March–May), Summer (June–August), and Fall (September–November). 

\begin{figure}
\centering
\includegraphics[height=10cm]{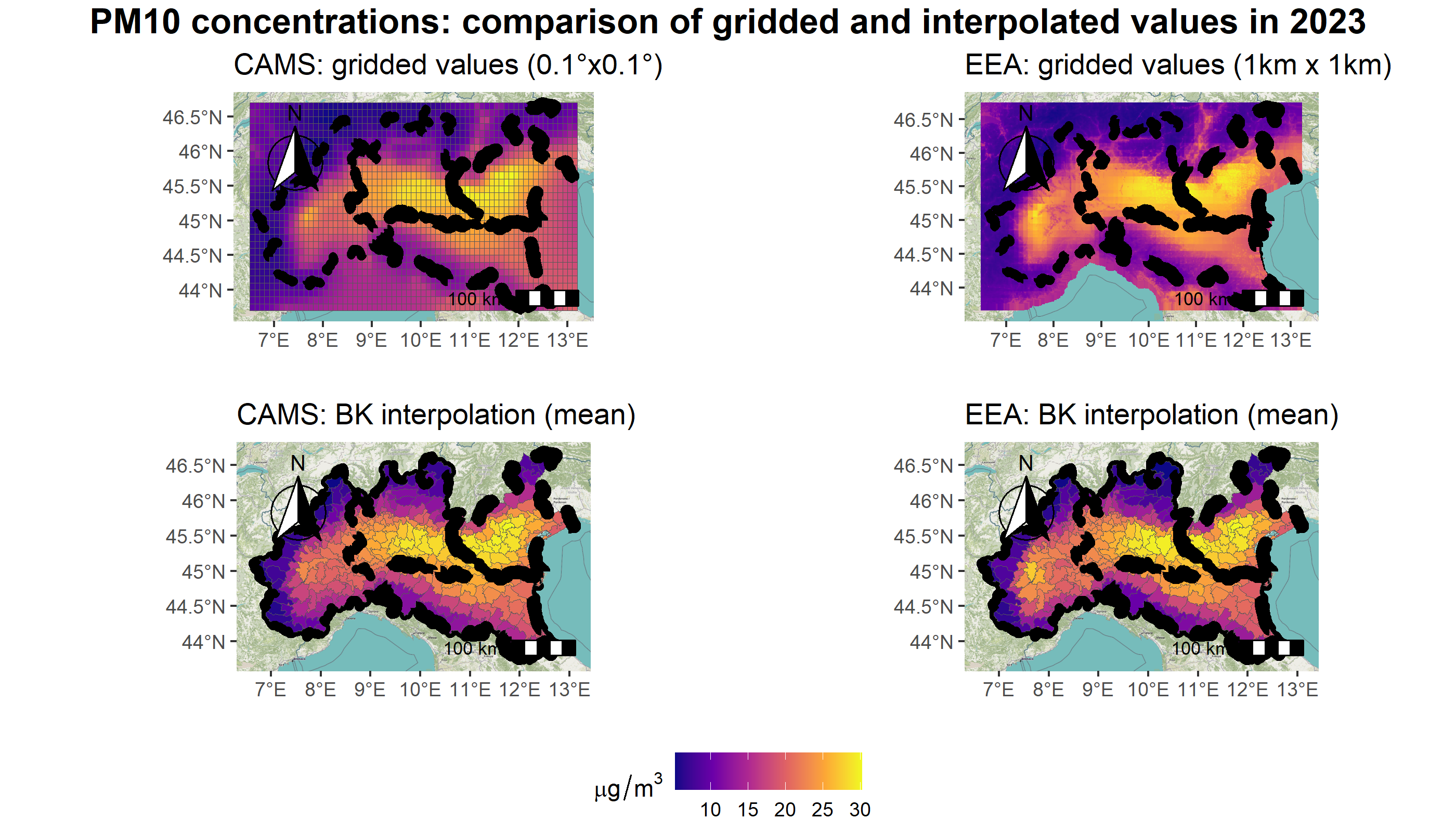}
\caption{Comparison of original gridded data with area-level interpolation of PM$_{10}$ concentrations ($\mu g/m^3$) in 2023. Original data from EEA are regular cells of 1km$\times$1km, whereas data from CAMS are regular cells of 0.1$^\circ\times$0.1$^\circ$.}
\label{fig:EEA_CAMS_comparison_2023}
\end{figure}

For both datasets, spatial harmonization was performed by transferring grid-level values to the ASRs using the local BK procedure. In Figure ~\ref{fig:EEA_CAMS_comparison_2023} we provide an exemplificative plot that compares the original gridded data with the interpolated outcomes for both data sources. The plot clearly shows the consistency among the two sources and among gridded and interpolated values.

Furthermore, a summary of the air quality datasets, including their spatial and temporal characteristics and pollutant coverage, is provided in Table~\ref{tab:air_quality_datasets}, while a detailed description of all pollutant-specific variables and temporal coverage is reported in Table~4 of the Appendix.

\begin{table}[htbp]
\centering
\caption{Air quality datasets integrated in the SCARFACE database.}
\label{tab:air_quality_datasets}
\begin{tabular}{p{4cm} p{3.3cm} p{2.5cm} p{5.0cm}}
\toprule
\textbf{Dataset and source} & \textbf{Spatial resolution} (grid) & \textbf{Temporal resolution} & \textbf{Pollutants (unit: $\mu g\,/m^{3}$)} \\
\midrule
EEA interpolated air quality (European Environment Agency) 
& 1\,km $\times$ 1\,km 
& Annual 
& PM$_{10}$, PM$_{2.5}$, NO$_2$, NO$_x$ \\[0.4em]
\midrule
CAMS air quality reanalysis (Copernicus Atmosphere Monitoring Service) 
& 0.1$^{\circ}$$\times$0.1$^{\circ}$ 
& Hourly $\rightarrow$ annual \& seasonal 
& NH$_3$, PM$_{2.5}$, PM$_{10}$, NO$_2$, NO, SO$_2$, O$_3$, CO; secondary PM$_{2.5}$ components (ammonium, nitrate, sulfate, organic matter, elemental carbon); dust and sea salt \\
\bottomrule
\end{tabular}
\end{table}

\subsubsection{Meteorology}
Meteorological variables were obtained from the ERA5-Land global reanalysis dataset \citep{MunozSabater2021} produced by the European Centre for Medium-Range Weather Forecasts \citep[ECMWF, ][]{Hersbach2020}. The dataset is generated by forcing the land surface model with atmospheric fields from the ERA5 reanalysis, ensuring physically consistent estimates of meteorological, soil, hydrological, and vegetation variables \citep{MunozSabater2021,Hersbach2020}. The dataset offers hourly estimates of key land and near-surface meteorological, hydrological, soil, and vegetation variables on a regular grid with resolution of $0.1^{\circ}\times0.1^{\circ}$. Wind speed ($ws$), wind direction ($wd$), and relative humidity ($rh$) were derived from ERA5-Land variables using standard meteorological formulations. Wind speed was derived from the horizontal wind components ($u$ and $v$) \citep{Stull1988} as the Euclidean norm of the wind vector. For example, wind speed at 10\,m was calculated as $ws = \sqrt{u10^{2} + v10^{2}}$, while wind direction was calculated as $wd = 180 - \left(\mathrm{atan2}(u10/ws,v10/ws)\cdot180/\pi\right)$ \cite{Stull1988}. Relative humidity was estimated from 2 m air temperature ($t2m$) and dew point temperature ($d2m$) using a Magnus-type approximation of the saturation vapor pressure relationship \citep{Lawrence2005,Thompson2023}. Table~5 of the Appendix provides the complete list of meteorological and land-surface variables included in the dataset, together with their descriptions, units of measurement, and classification.

\begin{figure}[htbp]
\centering
\includegraphics[width=1\textwidth]{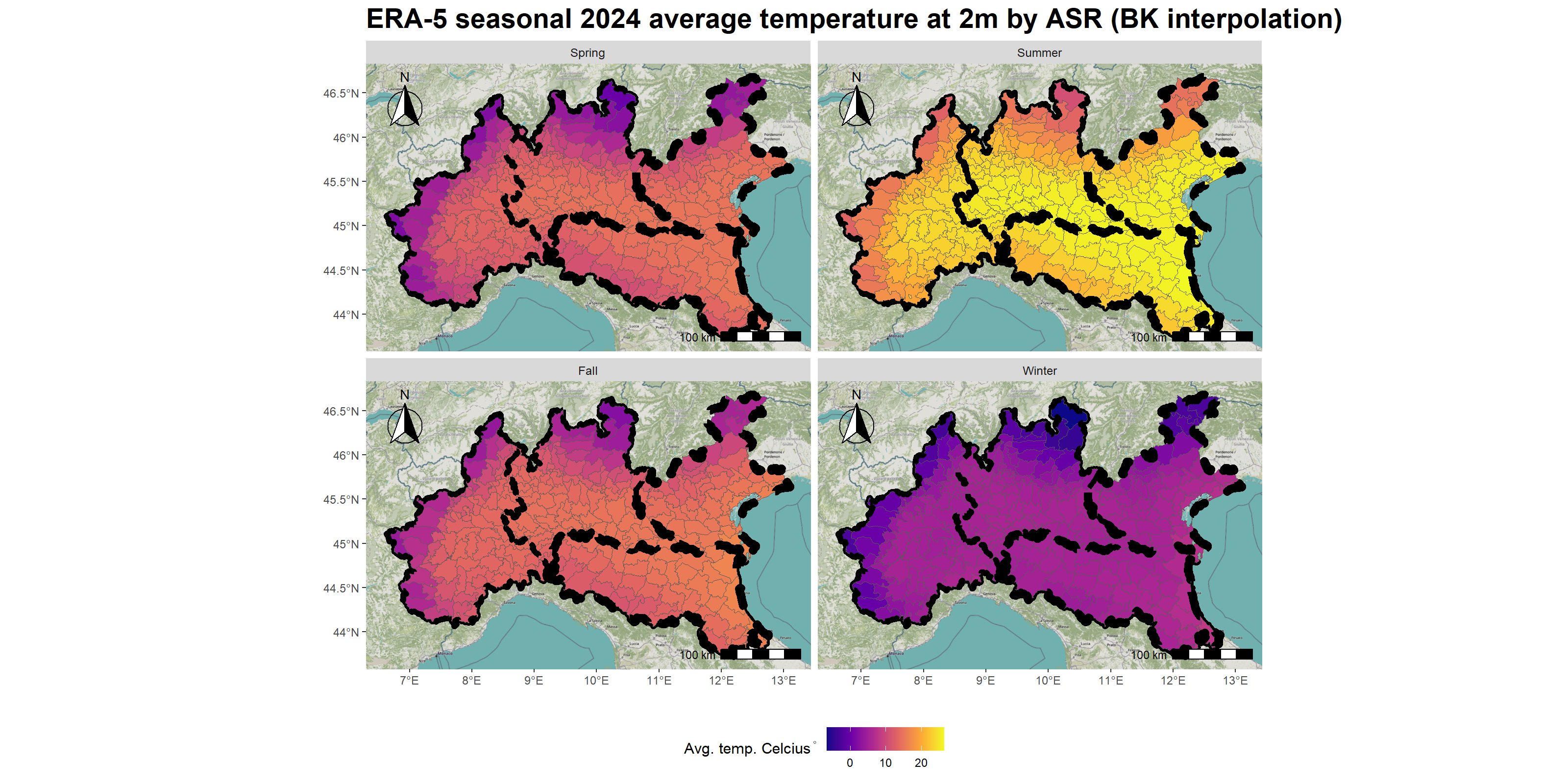}
\caption{Seasonal average air temperature at 2\,m for the Po Valley in 2024 derived from ERA5-Land data using local block kriging. Panels correspond to the four climatological seasons (spring, summer, fall, and winter), illustrating the spatial and seasonal variability of near-surface temperature across the study area.}
\label{fig:ERA5_temp_ASR}
\end{figure}

Hourly ERA5-Land variables covering Northern Italy were extracted for the period 2011–2024 and aggregated across both space and time. Prior to aggregation, variables were harmonized to consistent physical units (e.g., temperatures converted from Kelvin to degrees Celsius and water fluxes from meters to millimeters). In the same spirit of CAMS hourly concentrations, also ERA5 data were first temporally aggregated to derive annual and seasonal indicators (mean, minimum, maximum, and standard deviation) and the annualized grid-cell indicators were spatially aggregated to the ASR level through the BK algorithm. An example of the meteorological indicators included in SCARFACE is shown in Figure~\ref{fig:ERA5_temp_ASR}, which reports the seasonal average air temperature at 2\,m for 2024 after BK spatial harmonization of ERA5-Land data to the ASR level.

\subsubsection{Drought and weather extremes indicators}
Drought-related and weather extremes indicators were obtained from the European Drought Observatory (EDO)\footnote{\href{https://drought.emergency.copernicus.eu/tumbo/edo/download/}{https://drought.emergency.copernicus.eu/tumbo/edo/download/} (accessed on April 2nd, 2026)}, part of the Copernicus Emergency Management Service. From an agricultural perspective, these indicators complement the ERA5-Land meteorological variables by providing information on water availability, soil moisture dynamics, and extreme temperature events, which are key drivers of agricultural vulnerability under climate change  \citep{willetts2022review,lesk2016influence,Zscheischler2020Record}. These factors critically influence crop productivity, irrigation demand, and livestock stress, particularly under compound hot and dry conditions that can severely impact agricultural systems \citep{lesk2016influence, Zscheischler2020Record}.

Soil moisture conditions are described through the Soil Moisture Index (SMI) and its anomalies, which provide normalized measures of deviations from long-term climatological conditions. Meteorological drought is captured through the Standardized Precipitation-Evapotranspiration Index (SPEI), a widely used multiscalar indicator that accounts for both precipitation and atmospheric evaporative demand \citep{Vicente2010New}. The SPEI framework has been extensively applied for drought monitoring across multiple temporal scales and climatic contexts, providing a flexible tool for assessing water balance anomalies \citep{Begueria2014Standardized}; computed at multiple accumulation periods (1, 6, and 12 months), allowing the identification of short-, medium-, and long-term water deficits. Soil moisture anomalies and SPEI indices capture drought conditions affecting plant growth and yield variability, while heat-wave indicators are closely related to crop heat stress and animal welfare. In addition, the Combined Drought Indicator (CDI) integrates information on precipitation, soil moisture, and vegetation response to provide a synthetic classification of drought conditions. Temperature extremes are represented through heat-wave and cold-wave indicators, including both duration and intensity metrics.

The original EDO products are characterized by heterogeneous temporal resolutions, including 10-days, monthly, and daily indicators, depending on the variable. To ensure consistency with the SCARFACE framework, all variables were aggregated at the grid-cell level to derive annual and seasonal summary statistics computed by aggregating observations within the four climatological seasons defined above. After temporal aggregation, the gridded EDO indicators were spatially aligned to the ASR level using the usual local BK procedure.

Table~6 of the Appendix provides an overview of the drought and temperature extremes indicators derived from the European Drought Observatory and included in the SCARFACE dataset.

\subsubsection{Livestock consistency data from the National Livestock Registry}
Information on livestock consistency was obtained from the Italian Veterinary Information System (\textit{Sistema Informativo Veterinario}), which provides official statistics from the National Livestock Registry (\textit{Banca Dati Nazionale dell'Anagrafe Zootecnica}, BDN). The BDN system collects administrative information on livestock farms and animal stocks in Italy and provides annual statistics at the municipal level by livestock species and farming mode. Yearly municipal data for bovine and buffalo farms and heads, as well as swine farms and heads, were collected for the period 2011--2024.


For swine production, annual records were harmonized into three broad categories of livestock mode, that is, \textit{Stabled}, corresponding to intensive or fully housed systems; \textit{SemiWild}, corresponding to semi-extensive systems; and \textit{Unknown}, which includes observations with unspecified or mixed farming modes. A similar procedure was applied to bovine and buffalo data. In particular, we harmonized the farming modes into \textit{Stabled}, that includes observations classified as stable or intensive systems, whereas \textit{SemiWild} includes transhumant, outdoor, or extensive farming systems. Records with unspecified or mixed farming modes were classified as \textit{Unknown}. Municipal annual totals of farms and heads were subsequently computed for each municipality--mode combination.

Eventually, the harmonized municipal livestock indicators were then aggregated to the ASR level by summing the number of farms and the number of heads for all municipalities belonging to the same ASR. This procedure was applied separately by species, year, and livestock mode, producing ASR-level annual indicators of livestock presence and farming structure. The resulting output is an annual panel of livestock headcounts at the ASR level for each species, directly comparable across regions and years and not subject to spatial interpolation procedures.

\subsubsection{Socio-economic variables}
Socio-economic indicators were obtained from the experimental statistical dataset \textit{Sistema informativo a misura di comune}\footnote{\href{https://www.istat.it/statistica-sperimentale/aggiornamento-degli-indicatori-del-sistema-informativo-a-misura-di-comune/}{https://www.istat.it/statistica-sperimentale/aggiornamento-degli-indicatori-del-sistema-informativo-a-misura-di-comune/} (accessed on April 3rd, 2026).} produced by the Italian National Institute of Statistics (ISTAT). The dataset provides harmonized socio-economic indicators at the municipal level for the entire Italian territory. The indicators cover multiple thematic domains, including demography, families, labor market conditions, entrepreneurship, public finance, social services, political representation, cultural infrastructure, environmental conditions, and transport for the period 2014--2023. The full set of socio-economic indicators included in SCARFACE is listed in Table~8 of the Appendix. The harmonized municipal indicators were subsequently aggregated to the ASR level for each year. Aggregation was performed separately for each variable depending on its statistical interpretation. Indicators expressed as rates, shares, averages, or indices (e.g., employment rates, dependency indices, educational attainment shares, or voter turnout) were aggregated using the arithmetic mean across municipalities belonging to the same ASR. In contrast, indicators representing counts or totals (e.g., population, number of families, number of vehicles, or number of childcare beneficiaries) were aggregated using sums.

\begin{table}[htbp]
\centering
\caption{Summary of socio-economic indicators included in the SCARFACE dataset.}
\label{tab:socioeconomic_summary}
\begin{tabular}{ll}
\hline
Characteristic & Description \\
\hline
Source & ISTAT \textit{a misura di comune} \\
Temporal coverage & 2014--2023 \\
Number of indicators & 95 socio-economic indicators \\
Thematic domains & Demography; Families; Labor market; Economy and \\ 
& entrepreneurship; Public finance; Governance and political \\
& participation; Culture; Environment; Transport; Education;\\
& Social services \\
Original data format & Annual municipal indicators \\
Aggregation to ASR & Arithmetic mean for rates, shares, averages and indices; \\
& sum for count variables (e.g., population, families, vehicles) \\
\hline
\end{tabular}
\end{table}

\subsubsection{Land cover}
Land use and land cover (LULC) information in SCARFACE was derived from two complementary global land cover products: the CORINE Land Cover (CLC) inventory\footnote{\href{https://land.copernicus.eu/en/products/corine-land-cover}{https://land.copernicus.eu/en/products/corine-land-cover} (accessed on April 11th, 2026)} from the Copernicus Land Monitoring Service and the Global Dynamic Land Cover (GDLC) dataset\footnote{\href{https://land.copernicus.eu/en/products/global-dynamic-land-cover}{https://land.copernicus.eu/en/products/global-dynamic-land-cover} (accessed on April 11th, 2026)}.

\begin{figure}
\centering
\includegraphics[height=8cm]{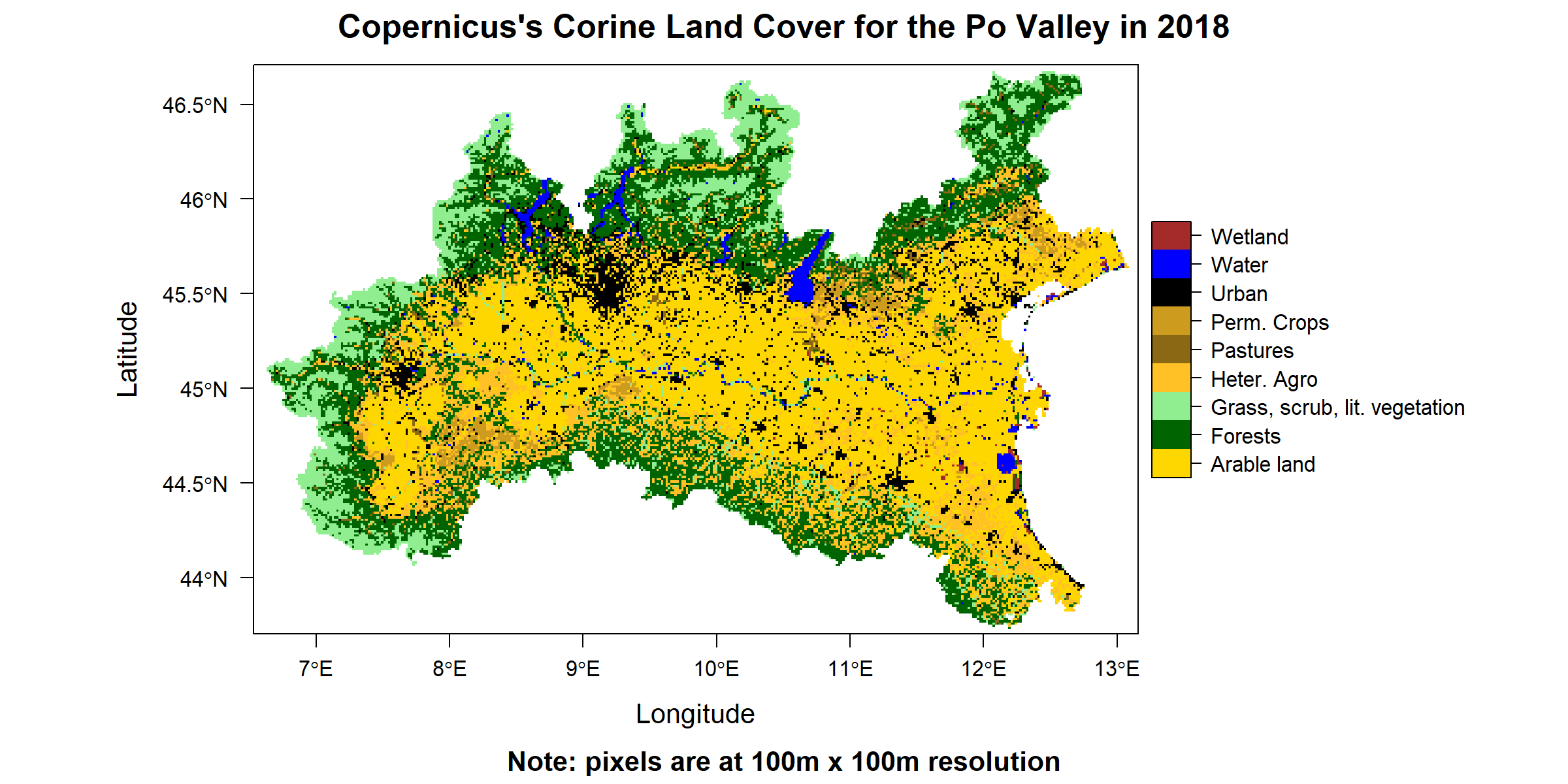}
\caption{CORINE Land Cover classification for the Po Valley in 2018. The map shows the reclassified CLC raster at 100\,m spatial resolution. The Po Valley is largely dominated by agricultural land uses, particularly arable land and heterogeneous agricultural areas, with forested regions concentrated along the surrounding Alpine and Apennine mountain ranges.}
\label{fig:CLC_example}
\end{figure}

The CLC dataset provides high-resolution land cover maps for Europe with a spatial resolution of 100\,m and 44 thematic classes describing different land cover types (e.g., arable land, forests, vineyards, and urban areas) derived from satellite image interpretation. In this study, the CLC 2012 and CLC 2018 releases were used. The GDLC dataset provides annual global land cover maps at comparable spatial detail and allows a more temporally resolved description of land cover dynamics. GDLC maps were available for the years 2015–2019. An example of the CORINE land cover data used in the SCARFACE dataset is shown in Figure~\ref{fig:CLC_example}, illustrating the original 100\,m resolution raster for the Po Valley in 2018 prior to the reclassification step.

Prior to spatial aggregation, the original thematic classes of each dataset were reclassified into a common set of nine land cover categories designed to capture the main agricultural, natural, and artificial land uses relevant for the analysis, while keeping a simpler yet effective interpretation. The correspondence between the original classification systems and the harmonized categories is reported in Table~10 and Table~11 of the Appendix.

Land cover indicators were then computed by intersecting the raster land cover layers with the ASRs and summing the area belonging to each land cover category within each ASR. The resulting areas were subsequently normalized by the total ASR area to obtain land cover shares. Since the land cover datasets are not available for every year of the study period, a piecewise-constant assumption was adopted when constructing the panel dataset. Specifically, CLC 2012 was used to represent land cover conditions for the period 2011–2017 and CLC 2018 for the period 2018–2023. For the GDLC dataset, the earliest available map (2015) was used for the period 2011–2014, while the latest map (2019) was used for the period 2020–2024. Table~\ref{tab:lulc_datasets} summarizes the main characteristics of the land cover datasets integrated in SCARFACE.

\begin{table}[htbp]
\centering
\caption{Land cover datasets integrated in the SCARFACE database.}
\label{tab:lulc_datasets}
\begin{tabular}{p{4.2cm} p{3.6cm} p{2.5cm} p{6.0cm}}
\toprule
\textbf{Dataset} & \textbf{Source} & \textbf{Classes / Resolution} & \textbf{Note} \\
\midrule
CORINE Land Cover (CLC) 
& Copernicus Land Monitoring Service 
& 44 classes; 100m
& Data available for 2012 (CLC12) and 2018 (CLC18). Piecewise-constant assumption: CLC12 used for 2011–2017 and CLC18 for 2018–2024 under a piecewise-constant assumption \\[0.4em]
Global Dynamic Land Cover (GDLC) 
& Global land cover dataset 
& GDLC classes; 100m 
& Data available for 2015-2019. Piecewise-constant assumption: the 2015 map was applied to 2011–2014 and the 2019 map to 2020–2024. \\
\bottomrule
\end{tabular}
\end{table}

\subsubsection{Farm activity and agro-economic indicators from FADN} \label{Sec:Data_FADN}
Farm-level information on agricultural activities and their environmental impacts was obtained from the Italian Farm Accountancy Data Network (FADN) survey\footnote{\href{https://www.crea.gov.it/en/web/politiche-e-bioeconomia/-/fadn-farm-accountancy-data-network}{https://www.crea.gov.it/en/web/politiche-e-bioeconomia/-/fadn-farm-accountancy-data-network} (accessed on April 4, 2026)}, coordinated by CREA\footnote{\href{https://www.crea.gov.it/en/home}{https://www.crea.gov.it/en/home} (accessed on April 4, 2026)}, which serves as the Italian liaison body of the European FADN system. FADN is an annual stratified survey designed to collect harmonized structural, technical, and economic data on agricultural holdings across the European Union. The survey covers farms above a minimum economic size threshold (8{,}000~\euro\ of Standard Output in the Italian implementation) and adopts a stratified sampling design based on administrative regions (NUTS~2 level), technical--economic specialization, and economic size classes. 

The available microdata from the FADN cover the period 2011--2023 and include georeferenced farms participating in repeated annual survey waves. The database contains a large set of farm-level variables describing agricultural production systems and management practices, including livestock endowment, agricultural land use, machinery availability, labor input, agronomic practices (e.g., fertilizers, manure, phytopharmaceuticals, and energy use), techno-economic specialization, and economic balance of sheet indicators. Moreover, FADN provides an estimate of the environmental impact associated with agricultural activities, referred to as the \textit{agricultural carbon footprint} indicator, which is expressed in CO$_2$-equivalent emissions. This carbon footprint indicator is not a measure directly obtained through interviews; rather, it is an ex post estimate derived from collected data by applying conversion coefficients to the activities carried out on the farm. Overall, we considered a set of 109 indicators from FADN, classified according to six thematic domains. the complete list of variable is provided in Table~13 of the Appendix.

To obtain population-representative estimates, survey observations are projected to represent broader populations within each domain by using calibrated expansion factors \citep[see Section 11 of][for a broader presentation]{Kim2026} that incorporate the stratified sampling design along spatial and temporal dimensions, enabling consistent inference from the sample to the population of farms operating in the Po Valley. However, the upscaling procedure inherently introduces an additional layer of uncertainty, as the resulting area-level estimates depend on both sampling variability and the properties of the weighting scheme. To explicitly account for this aspect, we systematically quantifies the uncertainty associated with the aggregation process by providing estimated sampling variances for both mean and total domain-level of each variable. 

\paragraph{Census-based spatio-temporal weighting system}
A \textit{census-based post-stratification procedure} was developed to reconstruct a spatio-temporal weighting system linking the FADN sample to the underlying farm population at the ASR level. The procedure relies on the Italian Agricultural Censuses of 2010 and 2020, which provide population counts of farms by municipality, economic size class, and broad technical specialization. To preserve consistency with the FADN survey, we retained only farms with at least 8,000\euro of Standard Output for every municipality. The construction of the weighting system follows five main steps.

\begin{enumerate}
    \item The ASR-level composition of the farm population was reconstructed for census years by aggregating municipal census counts according to two economic size classes (small farms with Standard Output below 100\,000\euro and large farms above this threshold) and three broad technical specialization classes (i.e., crop-oriented, livestock-oriented, and mixed farming systems);
    \item The total number of farms in each ASR was linearly interpolated for intercensal years (2011--2019) using the trend observed between the two census benchmarks (2010 and 2020). For the most recent years (2021--2023), the number of farms was held constant at the 2020 level. This choice was adopted because extrapolating the strong structural decline observed between the two census years would have generated implausible projections for several ASRs, including negative or near-zero farm counts\footnote{Fixing the totals at the most recent census level therefore provides a conservative and numerically stable approximation of the underlying population size in the absence of more recent census benchmarks};
    \item Assuming that the internal distribution of farms across economic size and technical specialization remains stable over time, annual stratum-specific population counts were reconstructed for each ASR by applying the census-based composition to the interpolated (or fixed) totals;
    \item These reconstructed population totals were then linked to the observed FADN sample counts to construct the spatio-temporal weighting framework;
    \item Because the survey sample does not always cover all combinations of economic size and specialization within each ASR-year domain, a calibration procedure based on raking was applied to ensure consistency between the weighted survey counts and the known population margins.
\end{enumerate}

Regarding the latter point, due to a lack of representativeness of the surveys at spatial resolution finer than the NUTS2 regional one, the sample does not always include observations for all combinations of economic size and technical specialization within each ASR--year domain. As a result, direct cell post-stratification weights cannot always be computed for every stratum. To address this limitation, we apply a calibration procedure based on \textit{iterative proportional fitting}, also referred to as \textit{raking} \citep{Kolenikov2014,LomaxNorman2016}. This procedure adjusts the survey weights so that the weighted sample reproduces the reconstructed population margins by economic size and technical specialization within each ASR and year, while remaining robust to sparse or incomplete sample coverage across strata.

The calibration procedure is applied sequentially at the ASR--year level. When the sample fully covers the cross-classification of size and specialization, direct cell weights are used (i.e., no calibration is applied). When some strata are not represented in the sample, raking is used to adjust the weights so that the weighted survey counts match the known marginal totals. If calibration with both margins fails or the support of the sample is insufficient, a hierarchical fallback strategy is implemented, including calibration on a single margin or the use of temporal donor information from adjacent years within the same ASR. This strategy ensures that the weighting system remains stable even in the presence of sparse sample coverage while preserving the consistency of the weighted totals with the reconstructed population margins.

\begin{figure}
\centering
\includegraphics[height=5cm]{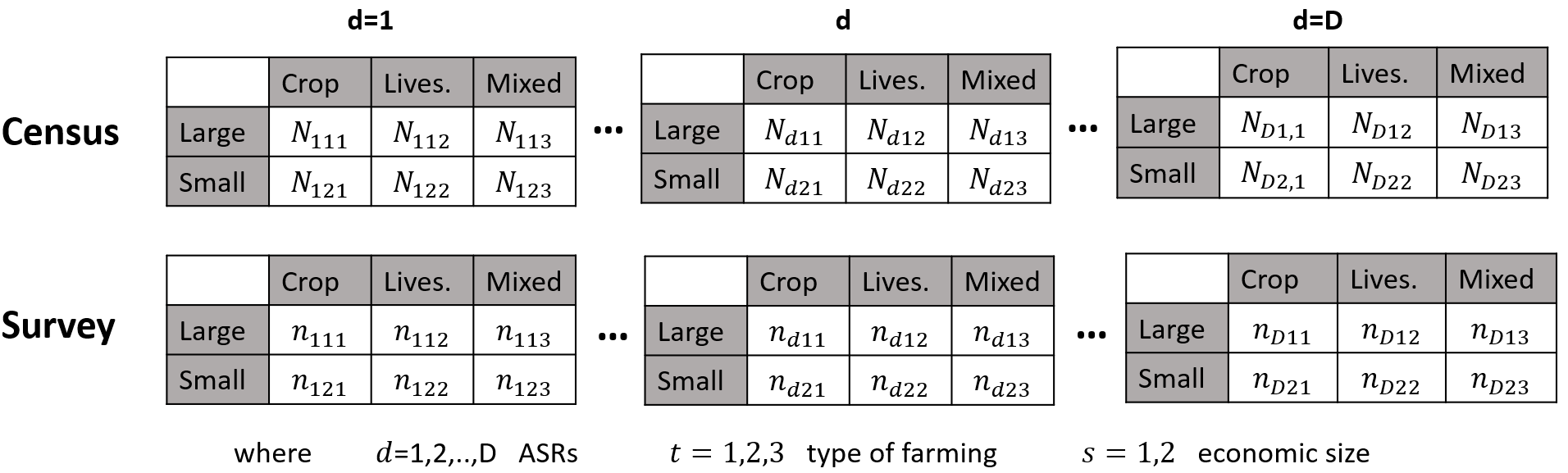}
\caption{Census-based post-stratification quantities used to construct the spatio-temporal weighting system for the FADN sample. Annual population counts of farms by economic size and technical specialization ($N_{dsty}$) are reconstructed at the ASR level using the 2010 and 2020 Italian Agricultural Censuses and interpolated for intermediate years. These reconstructed population totals are matched with the FADN sample counts ($n_{dsty}$) to derive post-stratification weights through the iterative proportional fitting procedure.}
\label{fig:new_poststratification}
\end{figure}

Let $d=1,\ldots,m$ denote the ASRs (with $m=256$), $s\in\{1,2\}$ the economic size classes (small, large), $t\in\{1,2,3\}$ the technical specialization classes (crop, livestock, mixed), and $y$ the year. Let $N_{dsty}$ denote the reconstructed number of farms in the population and $n_{dsty}$ the number of sampled farms in the FADN survey for the corresponding stratum. The final calibration weights $w_{dsty}$ are obtained through raking so that the weighted sample reproduces the population margins, that is,
\begin{equation}
\sum_{t} w_{dsty} n_{dsty} = N_{dsy} \qquad \sum_{s} w_{dsty} n_{dsty} = N_{dty},
\end{equation}
where $N_{dsy}$ and $N_{dty}$ denote the population totals by economic size and technical specialization per each combination of ASR and year, respectively. A schematic representation of the quantities involved in the post-stratification procedure is reported in Figure~\ref{fig:new_poststratification}.

\paragraph{Area-level estimation of the total and the mean}
Using the calibrated weights, farm-level indicators were aggregated to the ASR--year level through design-based \cite{HorvitzThompson1952} estimators (HT). For a generic farm-level variable $X_{dstyk}$ observed for farm $k$ belonging to stratum $(d,s,t,y)$, the HT estimator of the ASR total for year $y$ and area $d$ is
\begin{equation}
\hat{T}_{dy}(X)=\sum_{s=1}^{2}\sum_{t=1}^{3}\sum_{k=1}^{n_{dsty}} w_{dsty}X_{dstyk}.
\end{equation}
The corresponding estimator of the ASR mean for area $d$ in year $y$ is obtained as
\begin{equation}
\hat{\mu}_{dy}(X)=\frac{\hat{T}_{dy}(X)}{N_{dy}}, \qquad N_{dy}=\sum_{s=1}^{2}\sum_{t=1}^{3} N_{dsty}.
\end{equation}

\begin{figure}
\centering
\includegraphics[height=10cm]{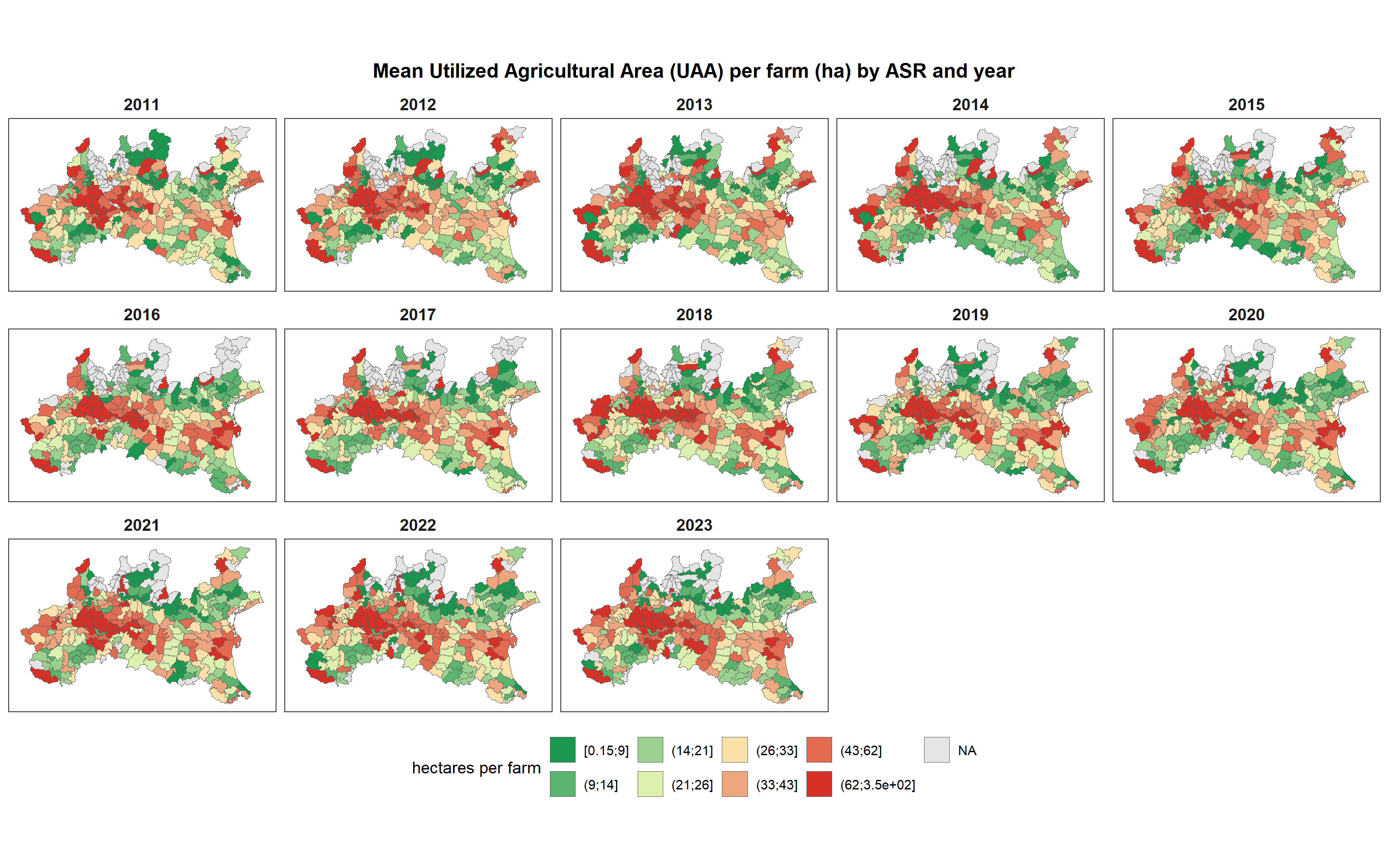}
\caption{Example of Horvitz--Thompson direct estimates derived from the FADN sample for the average utilized agricultural area (UAA) obtained through the post-stratified weighting system.}
\label{fig:HTmap}
\end{figure}

Figure~\ref{fig:HTmap} provides an example of the resulting ASR-level estimates for the utilized agricultural area (UAA), illustrating both the spatial distribution of the HT mean estimator and the associated sampling variability. It should be noted that the black areas in the figure correspond to ASRs for which no farms were sampled in a given survey year. In these cases, the corresponding estimates are recorded as missing, even though the agricultural census indicates that farms are present in those areas.

\paragraph{Variance regularization through generalized variance functions}
Sampling uncertainty of every FADN variable was initally quantified through the design-based variance of the HT estimator under stratified sampling. Using the previously introduced quantities, the variance of the estimated total for area $d$ in year $y$ can be expressed as
\begin{equation}
\widehat{\mathrm{Var}}(\hat{T}_{dy}(X))= \sum_{s=1}^{2}\sum_{t=1}^{3}N_{dsty}^{2}\left(1-\frac{n_{dsty}}{N_{dsty}}\right)\frac{S_{dsty}^{2}}{n_{dsty}},
\end{equation}
where $S_{dsty}^{2}$ denotes the sample variance of the target variable within stratum $(d,s,t,y)$. The variance of the estimated mean for area $d$ in year $y$ is then
\begin{equation}
\widehat{\mathrm{Var}}(\hat{\mu}_{dy}(X))= \frac{\widehat{\mathrm{Var}}(\hat{T}_{dy}(X))}{N_{dy}^{2}}.
\end{equation}

Direct HT variance estimates may be highly unstable for domains with very small sample sizes. This is a well-known issue in domain estimation from complex surveys, where direct design-based variance estimators can exhibit large variability due to sparse observations within certain domain--year combinations. To stabilize the variance structure of the resulting ASR-level estimates, a variance regularization procedure based on \textit{generalized variance functions} (GVF) was adopted.

The GVF approach models the systematic relationship between sampling variability and key design quantities describing domain precision. In particular, the logarithm of either the variance of the HT estimator or its squared coefficient of variation is modeled as a function of domain-level precision indicators, including the domain sample size (i.e., $n$), the domain population size (i.e., $N$), and the sampling fraction (i.e., $n/N$). Spatial and temporal heterogeneity are accounted for through random effects at the ASR and year levels. The resulting model provides predicted variance values that reflect the empirical relationship between estimator precision and sampling characteristics observed in the survey data.

To ensure robustness across heterogeneous variables and domains, several alternative GVF specifications are estimated and compared systematically. Candidate models differ both in the response formulation (variance or squared coefficient of variation) and in the set of precision indicators included in the regression. The final specification is selected according to diagnostic criteria that balance fit to the original design-based variances against the ability to shrink implausibly large estimates without distorting well-supported ones.

To avoid distorting reliable variance estimates, GVF predictions are not used directly. Instead, they are combined with the original design-based variances through a controlled blending rule that depends on the effective sample size within each domain. When the domain sample size is extremely small (i.e., $n=\{1,2\}$), the GVF prediction dominates the final variance estimate, whereas for domains with sufficiently large samples the direct variance is retained. For intermediate sample sizes, the final variance is obtained as a weighted average of the direct variance and the GVF prediction.

This strategy preserves the design-based interpretation of the HT estimator while reducing the influence of extremely large or unstable variance estimates that arise from small samples or irregular domain composition. In practical terms, the procedure regularizes the variance structure across ASR--year domains while leaving well-supported estimates essentially unchanged.

The complete description of the GVF model specification, the automatic model selection procedure, and the variance blending mechanism is provided in Section S3 of the Appendix.

\section{Data records}\label{sec:datarecords}
The SCARFACE dataset is made available to the public through the project's Zenodo repository (\href{https://doi.org/10.5281/zenodo.19600020}{https://doi.org/10.5281/zenodo.19600020}). Among others, the repository contains the following files:
\begin{itemize}
    \item \texttt{SCARFACE\_DatasetSingleObject} (either \texttt{.xlsx} or \texttt{RData} file) containing the source-specific dataset that constitute the SCARFACE framework. Data are organized into source-specific sheets/\texttt{data.frame} that can be matched using the primary keys \texttt{ASR} (unique geographical/spatial ID) and \texttt{Year} (unique temporal ID). Geometries can be added matching the shapefile contained in \texttt{ASRs\_Geometries.zip} (with primary key \texttt{ASR});
    \item \texttt{SCARFACE\_DatasetExtended} (either \texttt{.xlsx} or \texttt{RData} file) file containing the whole SCARFACE dataset in a single sheet/\texttt{data.frame}. Individual dataset were matched using a full join approach using the primary keys \texttt{ASR} (unique geographical/spatial ID) and \texttt{Year} (unique temporal ID). Geometries can be added matching the shapefile contained in \texttt{ASRs\_Geometries.zip} (with primary key \texttt{ASR});
    \item \texttt{ASRs\_Geometries.zip}, that contains the shapefile with the geometries of the m=256 ASRs polygon;
\end{itemize}
In addition to the dataset, the Zenodo repository contains metadata, geometries, supplementary materials and the code developed to replicate the complete harmonization workflow.

\section{Technical Validation}\label{sec:Validation}
The technical validation strategy adopted in SCARFACE was designed to address four main quality dimensions: 
\begin{enumerate}
    \item plausibility of the original and harmonized values;
    \item internal consistency of the transformations applied during harmonization
    \item uncertainty introduced by interpolation and estimation procedures; and 
    \item agreement with external benchmark sources whenever available.
\end{enumerate}

\subsection{Validation of administrative information and geographical features}
Administrative boundaries, spatial identifiers, time-invariant geographical features, and land cover were validated through a set of structural and spatial consistency checks.

First, the municipality--ASR crosswalk was verified to ensure that each municipality was uniquely assigned to one and only one ASR. Second, the completeness of the ASR partition was checked to confirm that all municipalities belonging to the study area were represented in the released correspondence table.

Administrative attributes linked from the NUTS territorial classification were also checked for consistency across hierarchical levels. In particular, the coherence between NUTS0, NUTS1, NUTS2, and NUTS3 identifiers associated with each ASR was verified through relational checks against the original GISCO reference files.

For the geographical descriptors derived from spatial layers, geometry validity was verified for all ASR polygons, and centroid coordinates were checked to ensure that they fell within the expected geographical extent of the Po Valley study area. Elevation-derived indicators, such as minimum, maximum, mean, and standard deviation of altitude, were inspected for implausible values and compared with the known topographic structure of the study region. In addition, the shares of ASR area classified as plains, hills, and mountains were verified to sum to hundred up to numerical tolerance.

Land cover indicators derived from CORINE Land Cover and the Global Dynamic Land Cover dataset were validated through cross-classification consistency checks and spatial accounting checks. First, the reclassification of the original land cover legends into the common nine-class scheme adopted in SCARFACE was verified to ensure that all original classes were mapped uniquely and exhaustively into one harmonized category. Second, we verified the mutual consistency among GDLC and CLC by comparing the shares of land cover attributed to each class and across data source. Third, the raster-to-polygon aggregation procedure was validated by checking that the class-specific areas computed within each ASR summed to the total ASR area up to numerical tolerance. The corresponding land cover shares were also checked to ensure that they summed to hundred.




\subsection{Validation of gridded environmental variables}
The validation procedures implemented on gridded environmental datasets, that is meteorological variables, emission inventories, and air-quality indicators, assessed the temporal and spatial aggregation consistency through cross-validation (CV) and the inspection of the resulting spatio-temporal patterns. 


First, for gridded datasets with high-frequency observations (e.g., ERA5-Land meteorological variables and EDO drought indicators), we validated the plausibility of the derived annual and seasonal summary statistics. Variable-specific checks were applied to verify expected physical constraints and temporal patterns, including non-negativity of precipitation and runoff variables, boundedness of relative humidity, and consistency of seasonal temperature ordering. These checks ensure that temporal aggregation from hourly or sub-annual observations produces physically meaningful indicators.

Second, the spatial alignment of gridded data to the ASR support was validated through the CV strategy described in Section~\ref{sec:spatial_alignment}. We recall that, for each variable, year and sector (whenever applied), candidate covariance families and size of the local kriging neighborhood were compared using repeated 5-fold CV, and the model with the lowest RMSE was selected. The results reported in Table~\ref{tab:avgcorr_interp_obs} indicate consistently high correlations between interpolated and observed values across pollutants and sectors (either considering emissions or concentrations), confirming the strong performance of the interpolation procedure. Correlations are generally higher for sector-specific components, while aggregated totals appear comparatively more challenging to reproduce, likely reflecting the compounded variability arising from multiple emission sources. Similarly, total concentrations show higher correlation than total emissions. It should be noted that the table is presented for illustrative purposes, while analogous results for all interpolated datasets are provided in the Appendix.

\begin{table}[!h]
\centering
\caption{Average Pearson's linear correlations between interpolated and observed values at the grid locations. Correlations are averaged over 2011--2022 for EDGAR emissions of air pollutants, over 2011--2024 for EDGAR emissions of GHGs and for EEA concentrations, and over 2013--2024 for CAMS concentrations.}
\label{tab:avgcorr_interp_obs}
\centering
\resizebox{\ifdim\width>\linewidth\linewidth\else\width\fi}{!}{
\fontsize{9}{11}\selectfont
\begin{tabular}[t]{cccccccc}
\toprule
Type & Pollutant & AWB & AGS & ENF & MNM & Indirect N2O & TOTALS\\
\midrule
GHG (EDGAR) & CH4 & 0.934 & 0.934 & 0.778 & 0.777 & / & 0.781\\
GHG (EDGAR) & CO2 & / & 0.853 & / & / & / & 0.438\\
GHG (EDGAR) & CO2bio & 0.933 & / & / & / & / & 0.254\\
GHG (EDGAR) & GWP\_100\_AR5\_GHG & 0.934 & 0.927 & 0.784 & 0.787 & 0.921 & 0.525\\
GHG (EDGAR) & N2O & 0.934 & 0.923 & / & 0.775 & 0.922 & 0.721\\
\addlinespace
AP (EDGAR) & BC & 0.935 & / & / & / & / & 0.683\\
AP (EDGAR) & CO & 0.935 & / & / & / & / & 0.739\\
AP (EDGAR) & NH3 & 0.936 & 0.927 & / & 0.785 & / & 0.853\\
AP (EDGAR) & NMVOC & 0.938 & 0.908 & / & 0.803 & / & 0.778\\
AP (EDGAR) & NOx & 0.936 & 0.924 & / & 0.786 & / & 0.586\\
AP (EDGAR) & OC & 0.935 & / & / & / & / & 0.747\\
AP (EDGAR) & PM10 & 0.935 & 0.909 & / & 0.849 & / & 0.615\\
AP (EDGAR) & PM2.5 & 0.935 & 0.926 & / & 0.840 & / & 0.690\\
AP (EDGAR) & SO2 & 0.934 & / & / & / & / & 0.253\\
\addlinespace
AQ (CAMS) & CO & / & / & / & / & / & 0.998\\
AQ (CAMS) & NH3 & / & / & / & / & / & 0.993\\
AQ (CAMS) & NO & / & / & / & / & / & 0.979\\
AQ (CAMS) & NO2 & / & / & / & / & / & 0.994\\
AQ (CAMS) & O3 & / & / & / & / & / & 0.995\\
AQ (CAMS) & PM10 & / & / & / & / & / & 0.999\\
AQ (CAMS) & PM2.5 & / & / & / & / & / & 0.999\\
AQ (CAMS) & SO2 & / & / & / & / & / & 0.994\\
\addlinespace
AQ (EEA) & NO2 & / & / & / & / & / & 0.948\\
AQ (EEA) & NOx & / & / & / & / & / & 0.990\\
AQ (EEA) & PM10 & / & / & / & / & / & 0.988\\
AQ (EEA) & PM25 & / & / & / & / & / & 0.990\\
\bottomrule
\multicolumn{8}{l}{\textsuperscript{} AWB = Agricultural Waste Burning; AGS = Agricultural Soil Use; ENF = Enteric Fermentation;}\\
\multicolumn{8}{l}{MNM = Manure Management; Indirect N2O = Indirect N2O emissions.}\\
\end{tabular}}
\end{table}

Then, the resulting ASR-level maps were inspected to verify that the major spatial patterns of the Po Valley were preserved. In particular, higher pollutant concentrations and emissions were expected in the central lowland and in urban-industrial areas, while lower values were observed in peripheral and mountainous regions. Similarly, meteorological indicators were checked to ensure consistency with known seasonal contrasts and spatial gradients.

\begin{figure}
\centering
\includegraphics[height=10cm]{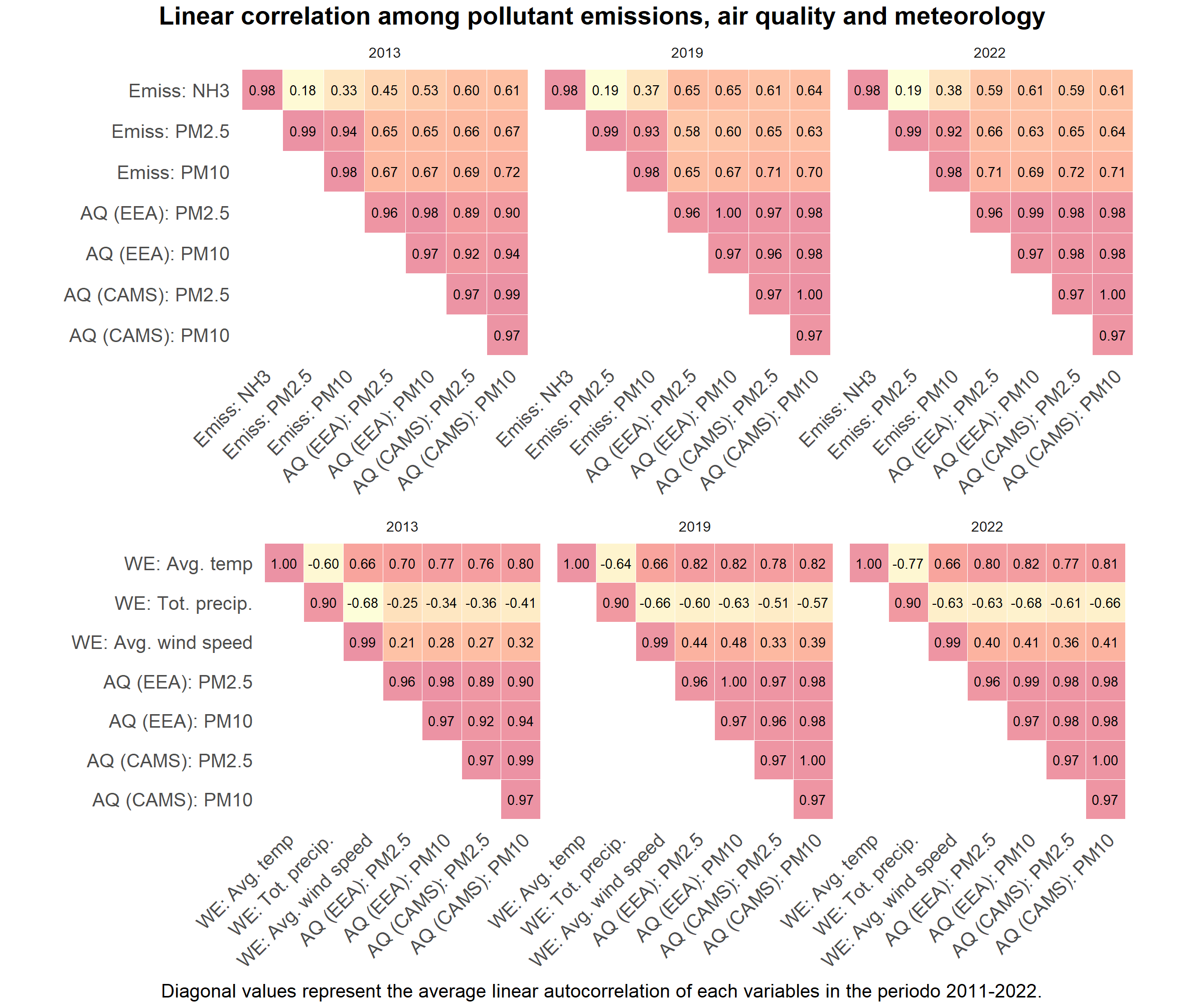}
\caption{Linear correlation among pollutant emissions, air quality indicators, and meteorological variables for selected years (2013, 2019, and 2022). Upper panels report correlations between emissions (from EDGAR) and air quality (from EEA and CAMS) indicators, while lower panels include meteorology and air quality. Diagonal elements represent the average temporal autocorrelation of available years between 2011 and 2024.}
\label{fig:CorrGrid}
\end{figure}

As shown in Figure~\ref{fig:CorrGrid}, the main relationships among emissions, air quality indicators, and meteorological variables can be summarized through a compact correlation framework. The figure is presented for illustrative purposes only, providing a synthetic overview of the large amount of environmental information included in the dataset. All environmental variables exhibit a high degree of temporal consistency, as indicated by the strong autocorrelation values observed along the diagonal \citep{BigiGhermandiHarrison2012,BigiGhermandi2014}. Moreover, the correlation patterns between emissions, air quality indicators, and meteorological variables are coherent with well-established atmospheric dynamics characterizing the Po Valley, that is, pollutant concentrations tend to be positively associated with emissions and temperature, while showing negative correlations with precipitation, reflecting the role of wet deposition and atmospheric dispersion processes \citep{PernigottiEtAl2012a,PernigottiEtAl2012b}.

\subsection{Validation of livestock and socio-economic indicators}
Since the original socio-economic indicators from ISTAT and the livestock information from the BDN are already annual and municipality-based, the primary aim of their validation is to ensure that the harmonization and aggregation steps preserve the statistical meaning of each variable while producing spatially consistent ASR-level indicators. Variable-specific range checks were applied to identify implausible values, particularly for rates, shares and totals, which are expected to fall within bounded intervals. For instance, yearly totals were inspected for negative or discontinuous values and for unexpected year-to-year changes; the sum of municipal values within each ASR was verified against the stored ASR-level aggregate; and for variables aggregated by arithmetic mean, the resulting ASR-level values were checked to ensure consistency with the underlying municipal values.

\subsection{Validation of the FADN survey indicators}
Farm-level indicators derived from the CREA--FADN microdata required a tailored validation approach because their harmonization involves post-stratification, calibration weighting, domain estimation, and variance regularization. The validation here is intended to show that the released area-level indicators are coherent with census information, statistically interpretable, and accompanied by uncertainty measures that remain stable even under sparse local sample coverage.

First, we validated the stability and plausibility of the GVF procedure used to regularize the variances associated with the HT estimators by evaluating its ability to reduce excessively large and implausible variances while preserving the overall pattern of design-based uncertainty across ASR--year domains.

\begin{figure}
\centering
\includegraphics[height=8cm]{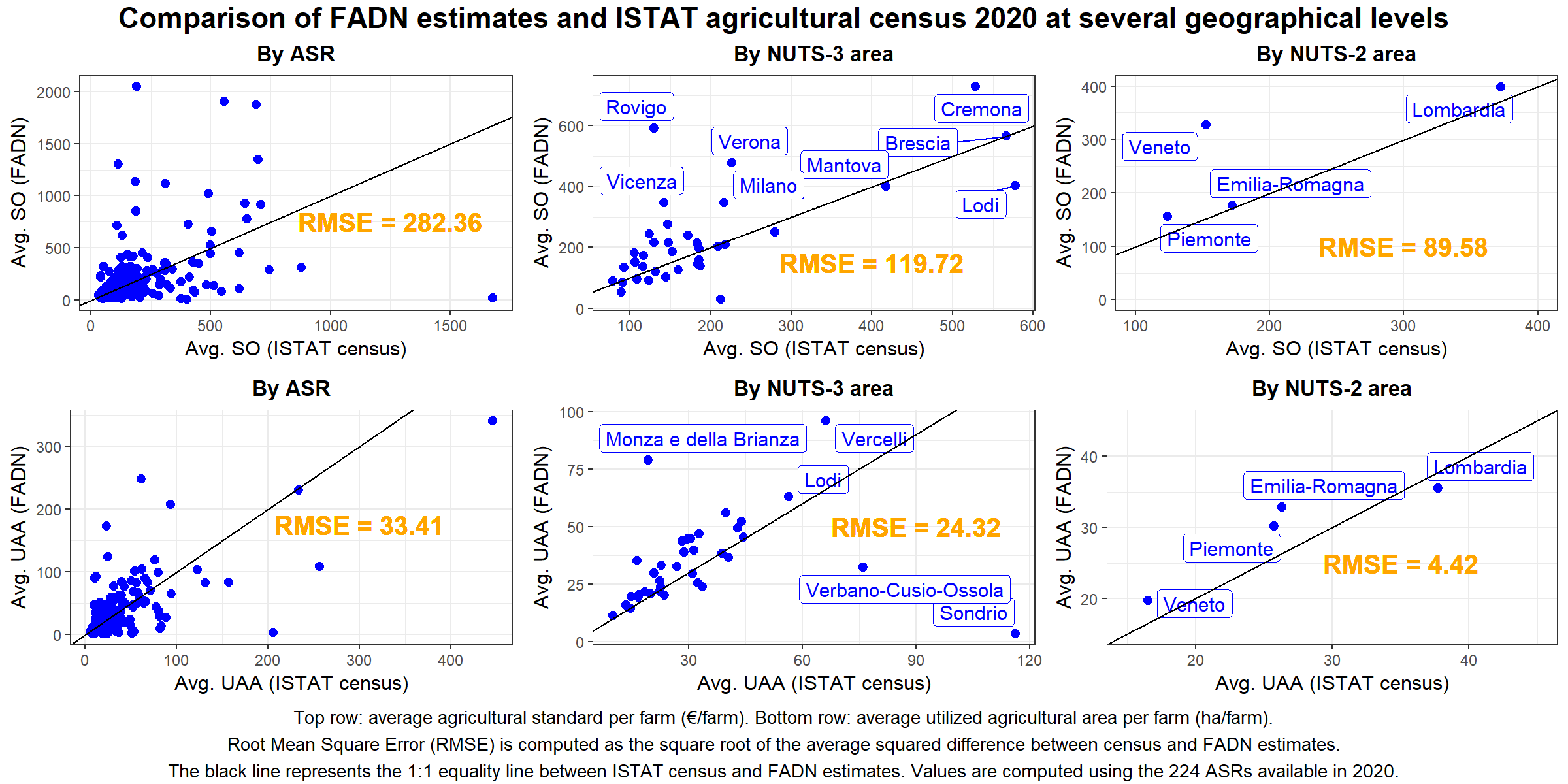}
\caption{Comparison of SCARFACE estimates and ISTAT agricultural census 2020 by ASRs (left panel), NUTS-3 areas (central panel) and NUTS2 areas (right panel). Top row: average agricultural standard per farm (\euro/farm); bottom row: average utilized agricultural area per farm (ha/farm). The black line represents the 1:1 equality line between ISTAT census and SCARFACE estimates, while the letter 'R' indicates the Pearson's linear correlation index. RMSEs are computed using the 224 ASRs available in 2020.}
\label{fig:AvgSO2020}
\end{figure}

Second, the plausibility of the resulting ASR-level direct estimates was evaluated by comparing selected indicators with municipality-level data sourced from the Seventh ISTAT agricultural census 2020\footnote{\href{https://www.istat.it/statistiche-per-temi/censimenti/agricoltura/7-censimento-generale/}{https://www.istat.it/statistiche-per-temi/censimenti/agricoltura/7-censimento-generale/} (accessed on April 15th, 2026).}. In particular, area-level estimates of variables such as Utilized Agricultural Area (UAA) or Standard Output (SO) were aggregated and compared with corresponding census-based values at broader territorial scales (e.g., NUTS-3 or NUTS2) to assess the ability of the weighting system to recover known structural patterns in the agricultural population. Figure~\ref{fig:AvgSO2020} illustrates the agreement between FADN estimates and the 2020 ISTAT agricultural census across three spatial scales: ASRs, administrative provinces (NUTS-3), and administrative regions (NUTS2). We recall that FADN data are representative of the Italian agricultural sector at the regional NUTS2 level. A clear and substantial reduction in RMSE is observed when moving from the ASR level to NUTS-3 and further to NUTS2 aggregations, indicating that discrepancies between census and estimated values decrease with increasing spatial aggregation. At the same time, even at the ASR level, the correlation between direct estimates and census data remains strong, suggesting that the estimates are overall consistent and provide realistic representations of the underlying agricultural variables.

\section{Usage Notes}\label{sec:usagenotes}
SCARFACE is an open access harmonized spatio-temporal dataset that integrates several domain, such as climate, air quality, pollution emissions, land cover, soil properties, agro-industry dynamics and socio-economic indicators, to jointly investigate interconnected processes linking agricultural systems, atmospheric dynamics, emissions, and socioeconomic conditions in the Po Valley, Northern Italy. The spatial reference unit is the Agrarian Sub-Region (ASR), that is, groups of contiguous municipalities that are considered relatively homogeneous with respect to natural conditions, agronomic characteristics, and agricultural production systems. The SCARFACE dataset adopts an annual panel structure from 2011 to 2024 defined over the 256 ASRs partitioning the Po Valley and comprises more than 2,700 indicators sourced from national and international public institutions.

The dataset is ready to be used ‘as is’ by any user interested in studying the complex interactions among agricultural systems, air quality management, environmental processes, and human activities in a critical area of Europe. However, users should take into account some potential caveats.

\paragraph{Potential fields of application}
As stated in Section \ref{sec:intro}, the dataset is designed for broad reuse across a range of analytical settings, including panel data analysis at moderate spatial and temporal resolutions, high-dimensional spatio-temporal modeling, spatial and spatio-temporal clustering, and policy-oriented applications. Among the methodological approaches particularly well suited for SCARFACE, area-level small area estimation (SAE) models represent a promising direction \citep{MoralesEtAl2021}. Indeed, key variables such as agronomic practices and the techno-productive structure of farms are derived from representative surveys, whose direct estimates may lack precision at fine spatial and temporal resolutions. SAE approaches allow for improving the reliability of such estimates by leveraging auxiliary information available within SCARFACE, thereby enhancing their statistical robustness. In environmental and agricultural contexts, where reliable and disaggregated information is essential, classical models such as the \cite{FayHerriot1979} model provide a natural framework to combine survey-based estimates with auxiliary covariates, yielding more precise and policy-relevant indicators at finer spatial scales. However, a typical requirement for the implementation of area-level SAE models is the availability of sampling variances associated with each domain or small area, which act as known inputs to properly specify the underlying hierarchical structure. In this respect, SCARFACE provides these quantities in a systematic way, ensuring full compatibility with standard SAE frameworks and enabling their straightforward application without the need to access additional data sources or perform ad hoc variance estimation procedures, thereby substantially reducing the methodological burden for applied researchers and policy analysts.

\paragraph{Potential missing values}
Due to structural heterogeneity of the data sources (i.e., different spatial and temporal coverages) several missing values could arise (e.g., socio-economic data cover the period 2014--2023, while emissions data cover the whole period 2011--2024), especially across the temporal domain. We recall that no missing data treatment was enforced during the dataset building. In this regard, users should keep in mind that the institutional nature of the data (e.g., socio-economic information) can induce interpretative biases in the case of imputation of missing data through statistical approaches. For transparency purposes, among the output made available to the public we provide a descriptive analysis about the missing values detected for each individual dataset, year, variable and ASR.

\paragraph{Uncertainty quantification and interpolation diagnostics}
In addition to the main data products, the SCARFACE framework provides a comprehensive set of uncertainty-related information, including metrics associated with temporal aggregation, spatial interpolation, and the tuning of interpolation hyperparameters.

Temporal aggregation uncertainty arises from the harmonization of input datasets originally available at monthly or higher temporal resolution (e.g., meteorological and drought indicators). Prior to spatial alignment via BK, these data were summarized at the grid-cell level using annual and seasonal statistics (mean, minimum, maximum, and standard deviation). The standard deviation captures infra-annual variability and provides an indirect measure of temporal aggregation uncertainty.

For indicators subject to spatial alignment and interpolation using spatial BK (i.e., gridded data on meteorology, emissions and concentrations) we provide the predicted kriging average and the associated kriging prediction variance associated with each ASR, as well as several metadata describing the interpolation procedure. Specifically, the latter includes the selected covariance model (specified by year, variable and sector), the number of neighboring observations used in the interpolation, the estimated regression parameters and the cross-validation diagnostics. Similarly, for survey-based indicators derived from the FADN, we provide the Horvitz-Thompson expansion estimator of the average value and the total value along with the corresponding estimated variances. In this case too, we provide users with complete information about the regularization process applied to the direct estimates of the variance and in particular we provide both the original non-smoothed variances and the GVF smoothed variances, as well as details on the empirical smoothing model, allowing a fair comparison of the pre-and-post-regularization distributions. In both cases, the estimated area-level variances quantify the interpolation uncertainty and are intended to support downstream analyses in which estimation precision is critical, such as weighted regression models, uncertainty propagation, sensitivity analyses, and the identification of domains with lower inferential reliability. In this regard, for modeling purposes, such uncertainty should be carefully taken into account through specific model-based adjustments, such as multi-stage bootstrap inference, in order to correct estimates and inference for the prior knowledge of the underlying data interpolation process.

In summary, by providing both point estimates and their associated uncertainty metrics, together with detailed information on the spatial interpolation process, the SCARFACE dataset enhances transparency and facilitates robust and reproducible reuse across a wide range of empirical applications.

\paragraph{Dataset update}
This dataset was initially compiled for the \textit{"Sequestering CARbon through Forests, AgriCulture, and land usE (SCARFACE)"} research initiative funded by the University of Milano-Bicocca and will be systematically updated whenever new information, for instance new releases of the underlying raw dataset, will become available.

\section{Code Availability}\label{sec:codedata}
The SCARFACE dataset, the replication code and the companion materials can be publicly accessed either at the SCARFACE project's Zenodo repository (\href{https://doi.org/10.5281/zenodo.19600020}{https://doi.org/10.5281/zenodo.19600020}, Version v2) or GitHub repository (\href{https://github.com/ScarfaceSeqCARForAgriCultLandusE}{https://github.com/ScarfaceSeqCARForAgriCultLandusE}). Both repositories contain the harmonized dataset (in both Excel and R formats), metadata and geometries, supplementary materials (i.e., methodological notes, extended data descriptors and synthesis files about missing data patterns and dataset structure), and the code developed to replicate the harmonization workflow presented in this paper. User instructions are provided in the corresponding description/README files. The code used to extract and process all the datasets were developed by the authors using \texttt{R} and \texttt{Python} programming languages. 

\section*{Funding}
This work is part of the \textit{"Sequestering CARbon through Forests, AgriCulture, and land usE (SCARFACE)"} research project, funded by the University of Milano-Bicocca, under grant number 2024-ATEQC-0048. Further information about the project can be found at the link \href{https://www.paolomaranzano.net/scarface}{https://www.paolomaranzano.net/scarface}. 

\section*{Acknowledgements}
We acknowledge the Italian Council for Agricultural Research and Economics -- Research Centre for Agricultural Policies and Bioeconomy (CREA-PB) for providing the research team with access to the RICA-FADN database within the \textit{AgroGeoStat} research agreement. We also acknowledge the GEMMA center in the framework of project MUR \textit{"Dipartimenti di eccellenza 2023-2027"}. We also acknowledge researchers from Associazione Economia e Sostenibilità (EStà) and Terra! for the feedback provided within the joint research projects \textit{"Allevamenti intensivi e sistemi alimentari sostenibili"} and \textit{"Per il lavoro dignitoso e la transizione giusta: verso l’Osservatorio Lavoro e Ambiente nei sistemi alimentari"}. We also acknowledge and thank all the colleagues that provided the research team with comments and suggestions, in particular Laura Marcis (University of Valle d'Aosta, IT), Renato Salvatore (University of Cassino and Southern Lazio, IT), Paul Parker (UCSC, USA) and Scott Holan (University of Missouri, USA) for the survey data integration.

\section*{Author Contributions}
PM: conceptualization, data acquisition and processing, harmonization design, code, writing, supervision.\\
PC: conceptualization, data processing, geostatistical workflow, harmonization design, code, writing.\\
RB: conceptualization, survey data processing, satellite-derived variables processing, supervision.\\
RP: conceptualization, survey data processing, satellite-derived variables processing, supervision.\\
MB: conceptualization, supervision.\\
FC: conceptualization, survey data collection and processing, supervision.\\
LF: conceptualization, atmospheric variables (air quality, emissions, meteorology) processing, supervision.\\
EB: conceptualization, atmospheric variables (air quality, emissions, meteorology) processing, supervision.\\

\section*{Competing Interests}
The authors declare no competing interests. 

\clearpage
\bibliographystyle{plainnat}
\bibliography{biblio}

@article{Agrimonia2023,
   author = {Fassò, Alessandro and Rodeschini, Jacopo and Moro, Alessandro Fusta and Shaboviq, Qendrim and Maranzano, Paolo and Cameletti, Michela and Finazzi, Francesco and Golini, Natalia and Ignaccolo, Rosaria and Otto, Philipp},
   title = {Agrimonia: a dataset on livestock, meteorology and air quality in the Lombardy region, Italy},
   journal = {Scientific Data},
   volume = {10},
   number = {1},
   pages = {143},
   ISSN = {2052-4463},
   DOI = {10.1038/s41597-023-02034-0},
   url = {https://doi.org/10.1038/s41597-023-02034-0},
   year = {2023},
   type = {Journal Article}
}

@book{chiles2012geostatistics,
  author    = {Jean-Paul Chiles and Pierre Delfiner},
  title     = {{Geostatistics: Modeling Spatial Uncertainty}},
  edition   = {2nd},
  year      = {2012},
  publisher = {Wiley},
  address   = {Hoboken, NJ}
}

@techreport{ASRdef,
   author = {ISTAT, .},
   title = {Regione Agraria},
   booktitle = {Statistiche dell'agricoltura. Anni 2001-2002.},
   publisher = {ISTAT, .},
   institution = {ISTAT, .},
   address = {Roma, Italy},
   ISBN = {88-458-1334-7.},
   year = {2006},
   type = {Book Section}
}

@article{MunozSabater2021,
  title={ERA5-Land: a state-of-the-art global reanalysis dataset for land applications},
  author={Mu{\~n}oz Sabater, Joaqu{\'i}n and others},
  journal={Earth System Science Data},
  volume={13},
  pages={4349--4383},
  year={2021},
  doi={10.5194/essd-13-4349-2021}
}

@article{Hersbach2020,
  title={The ERA5 global reanalysis},
  author={Hersbach, Hans and Bell, Bill and Berrisford, Paul and others},
  journal={Quarterly Journal of the Royal Meteorological Society},
  volume={146},
  pages={1999--2049},
  year={2020},
  doi={10.1002/qj.3803}
}

@article{Thompson2023,
  title={A global dataset of vapor pressure deficit at high spatial resolution},
  author={Thompson, Jonathan A. and others},
  journal={Scientific Data},
  volume={10},
  pages={167},
  year={2023},
  doi={10.1038/s41597-023-02034-0}
}

@article{Lawrence2005,
  title={The relationship between relative humidity and the dewpoint temperature in moist air},
  author={Lawrence, Mark G.},
  journal={Bulletin of the American Meteorological Society},
  volume={86},
  number={2},
  pages={225--233},
  year={2005},
  doi={10.1175/BAMS-86-2-225}
}

@book{Stull1988,
  title={An Introduction to Boundary Layer Meteorology},
  author={Stull, Roland B.},
  year={1988},
  publisher={Springer},
  doi={10.1007/978-94-009-3027-8}
}

@techreport{EDGAR,
   author = {Crippa, Monica and Guizzardi, Diego and Solazzo, Efisio and Muntean, Marilena and Schaaf, Edwin and Monforti-Ferrario, Fabio and Banja, Manola and Olivier, Jos and Grassi, Giacomo and Rossi, Simone},
   title = {GHG emissions of all world countries - 2024},
   institution = {Publications Office of the European Union},
   DOI = {10.2760/4002897, JRC138862.},
   url = {https://publications.jrc.ec.europa.eu/repository/handle/JRC138862},
   year = {2024},
   type = {Report}
}

@misc{EEA2023,
  title={European air quality data (interpolated data) – annual concentration maps},
  author={European Environment Agency (EEA),.},
  year={2023},
  url={https://www.eea.europa.eu/en/datahub/datahubitem-view/82700fbd-2953-467b-be0a-78a520c3a7ef}
}

@misc{TilezenTerrain,
  author = {{Tilezen}},
  title = {Joerd Terrain Tiles: Data Sources and Ground Resolution},
  year = {2024},
  url = {https://github.com/tilezen/joerd/blob/master/docs/data-sources.md#what-is-the-ground-resolution}
}

@misc{NUTS,
   author = {Eurostat,.},
   title = {NUTS 2021 - Nomenclature of territorial units for statistics},
   month = {2026/03/14},
   url = {https://ec.europa.eu/eurostat/web/nuts},
   year = {2024},
   type = {Online Database}
}

@article{pebesma2004multivariable,
  title={Multivariable geostatistics in S: the gstat package},
  author={Pebesma, Edzer J},
  journal={Computers \& geosciences},
  volume={30},
  number={7},
  pages={683--691},
  year={2004},
  publisher={Elsevier}
}

@article{gotway2002combining,
  title={Combining incompatible spatial data},
  author={Gotway, Carol A and Young, Linda J},
  journal={Journal of the American Statistical Association},
  volume={97},
  number={458},
  pages={632--648},
  year={2002},
  publisher={Taylor \& Francis}
}

@article{BigiGhermandi2014,
   author = {Bigi, A. and Ghermandi, G.},
   title = {Long-term trend and variability of atmospheric PM<sub>10</sub> concentration in the Po Valley},
   journal = {Atmos. Chem. Phys.},
   volume = {14},
   number = {10},
   pages = {4895-4907},
   ISSN = {1680-7324},
   DOI = {10.5194/acp-14-4895-2014},
   url = {https://acp.copernicus.org/articles/14/4895/2014/},
   year = {2014},
   type = {Journal Article}
}

@article{BigiGhermandiHarrison2012,
   author = {Bigi, Alessandro and Ghermandi, Grazia and Harrison, Roy M},
   title = {Analysis of the air pollution climate at a background site in the Po valley},
   journal = {Journal of Environmental Monitoring},
   volume = {14},
   number = {2},
   pages = {552-563},
   year = {2012},
   type = {Journal Article}
}

@article{PernigottiEtAl2012a,
   author = {Pernigotti, Denise and Georgieva, Emilia and Thunis, Philippe and Bessagnet, Bertrand},
   title = {Impact of meteorological modelling on air quality: summer and winter episodes in the Po valley (Northern Italy)},
   journal = {International Journal of Environment and Pollution},
   volume = {50},
   number = {1-4},
   pages = {111-119},
   DOI = {10.1504/ijep.2012.051185},
   url = {https://www.inderscienceonline.com/doi/abs/10.1504/IJEP.2012.051185},
   year = {2012},
   type = {Journal Article}
}

@article{PernigottiEtAl2012b,
   author = {Pernigotti, Denise and Georgieva, Emilia and Thunis, Philippe and Bessagnet, Bertrand},
   title = {Impact of meteorology on air quality modeling over the Po valley in northern Italy},
   journal = {Atmospheric environment},
   volume = {51},
   pages = {303-310},
   ISSN = {1352-2310},
   year = {2012},
   type = {Journal Article}
}

@article{OttoEtAl2024,
   author = {Otto, Philipp and Fassò, Alessandro and Maranzano, Paolo},
   title = {A review of regularised estimation methods and cross-validation in spatiotemporal statistics},
   journal = {Statistics Surveys},
   volume = {18},
   number = {none},
   pages = {299-340, 42},
   url = {https://doi.org/10.1214/24-SS150},
   year = {2024},
   type = {Journal Article}
}

@article{HorvitzThompson1952,
   author = {Horvitz, D. G. and Thompson, D. J.},
   title = {A Generalization of Sampling Without Replacement from a Finite Universe},
   journal = {Journal of the American Statistical Association},
   volume = {47},
   number = {260},
   pages = {663-685},
   ISSN = {0162-1459},
   DOI = {10.1080/01621459.1952.10483446},
   url = {https://doi.org/10.1080/01621459.1952.10483446},
   year = {1952},
   type = {Journal Article}
}

@article{Kolenikov2014,
   author = {Kolenikov, Stanislav},
   title = {Calibrating Survey Data using Iterative Proportional Fitting (Raking)},
   journal = {The Stata Journal},
   volume = {14},
   number = {1},
   pages = {22-59},
   DOI = {10.1177/1536867x1401400104},
   url = {https://journals.sagepub.com/doi/abs/10.1177/1536867X1401400104},
   year = {2014},
   type = {Journal Article}
}

@article{LomaxNorman2016,
   author = {Lomax, Nik and Norman, Paul},
   title = {Estimating Population Attribute Values in a Table: “Get Me Started in” Iterative Proportional Fitting},
   journal = {The Professional Geographer},
   volume = {68},
   number = {3},
   pages = {451-461},
   ISSN = {0033-0124},
   DOI = {10.1080/00330124.2015.1099449},
   url = {https://doi.org/10.1080/00330124.2015.1099449},
   year = {2016},
   type = {Journal Article}
}

@article{BeheraEtAl2013,
   author = {Behera, Sailesh N. and Sharma, Mukesh and Aneja, Viney P. and Balasubramanian, Rajasekhar},
   title = {Ammonia in the atmosphere: a review on emission sources, atmospheric chemistry and deposition on terrestrial bodies},
   journal = {Environmental Science and Pollution Research},
   volume = {20},
   number = {11},
   pages = {8092-8131},
   ISSN = {1614-7499},
   DOI = {10.1007/s11356-013-2051-9},
   url = {https://doi.org/10.1007/s11356-013-2051-9},
   year = {2013},
   type = {Journal Article}
}

@article{SuttonEtAl2013,
   author = {Sutton, Mark A. and Reis, Stefan and Riddick, Stuart N. and Dragosits, Ulrike and Nemitz, Eiko and Theobald, Mark R. and Tang, Y. Sim and Braban, Christine F. and Vieno, Massimo and Dore, Anthony J. and Mitchell, Robert F. and Wanless, Sarah and Daunt, Francis and Fowler, David and Blackall, Trevor D. and Milford, Celia and Flechard, Chris R. and Loubet, Benjamin and Massad, Raia and Cellier, Pierre and Personne, Erwan and Coheur, Pierre F. and Clarisse, Lieven and Van Damme, Martin and Ngadi, Yasmine and Clerbaux, Cathy and Skjøth, Carsten Ambelas and Geels, Camilla and Hertel, Ole and Wichink Kruit, Roy J. and Pinder, Robert W. and Bash, Jesse O. and Walker, John T. and Simpson, David and Horváth, László and Misselbrook, Tom H. and Bleeker, Albert and Dentener, Frank and de Vries, Wim},
   title = {Towards a climate-dependent paradigm of ammonia emission and deposition},
   journal = {Philosophical Transactions of the Royal Society B: Biological Sciences},
   volume = {368},
   number = {1621},
   ISSN = {0962-8436},
   DOI = {10.1098/rstb.2013.0166},
   url = {https://doi.org/10.1098/rstb.2013.0166},
   year = {2013},
   type = {Journal Article}
}

@misc{CAMS,
   author = {Copernicus Atmosphere Monitoring Service (CAMS) Atmosphere Data Store, .},
   title = {CAMS European air quality reanalyses (accessed on April 2nd, 2026)},
   DOI= {10.24381/7cc0465a},
   url = {https://ads.atmosphere.copernicus.eu/datasets/cams-europe-air-quality-reanalyses?tab=overview},
   year = {2021},
   type = {Dataset}
}

@article{raffaelli2020improving,
  title={Improving air quality in the Po Valley, Italy: Some results by the LIFE-IP-PREPAIR project},
  author={Raffaelli, Katia and Deserti, Marco and Stortini, Michele and Amorati, Roberta and Vasconi, Matteo and Giovannini, Giulia},
  journal={Atmosphere},
  volume={11},
  number={4},
  pages={429},
  year={2020},
  publisher={MDPI}
}

@article{bigi2012analysis,
  title={Analysis of the air pollution climate at a background site in the Po valley},
  author={Bigi, Alessandro and Ghermandi, Grazia and Harrison, Roy M},
  journal={Journal of Environmental Monitoring},
  volume={14},
  number={2},
  pages={552--563},
  year={2012},
  publisher={Royal Society of Chemistry}
}

@article{granella2024formation,
  title={The formation of secondary inorganic aerosols: A data-driven investigation of Lombardy's secondary inorganic aerosol problem},
  author={Granella, Francesco and Renna, Stefania and Reis, Lara Aleluia},
  journal={Atmospheric Environment},
  volume={327},
  pages={120480},
  year={2024},
  publisher={Elsevier}
}

@article{vicente2010new,
  title={A new global 0.5 gridded dataset (1901--2006) of a multiscalar drought index: comparison with current drought index datasets based on the Palmer Drought Severity Index},
  author={Vicente-Serrano, Sergio M and Beguer{\'\i}a, Santiago and L{\'o}pez-Moreno, Juan I and Angulo, Marta and El Kenawy, Ahmed},
  journal={Journal of Hydrometeorology},
  volume={11},
  number={4},
  pages={1033--1043},
  year={2010}
}

@article{begueria2014standardized,
  title={Standardized precipitation evapotranspiration index (SPEI) revisited: parameter fitting, evapotranspiration models, tools, datasets and drought monitoring},
  author={Beguer{\'\i}a, Santiago and Vicente-Serrano, Sergio M and Reig, Fergus and Latorre, Borja},
  journal={International journal of climatology},
  volume={34},
  number={10},
  pages={3001--3023},
  year={2014},
  publisher={Wiley Online Library}
}

@article{zscheischler2020record,
  title={The record-breaking compound hot and dry 2018 growing season in Germany},
  author={Zscheischler, Jakob and Fischer, Erich M},
  journal={Weather and Climate Extremes},
  volume={29},
  pages={100270},
  year={2020},
  publisher={Elsevier}
}

@article{lesk2016influence,
  title={Influence of extreme weather disasters on global crop production},
  author={Lesk, Corey and Rowhani, Pedram and Ramankutty, Navin},
  journal={Nature},
  volume={529},
  number={7584},
  pages={84--87},
  year={2016},
  publisher={Nature Publishing Group UK London}
}

@misc{willetts2022review,
  title={Review of IPCC evidence 2022},
  author={Willetts, Elizabeth and Thank Carlos, DC and Corvalan, Marina Maiero and Neville, Tara},
  year={2022},
  publisher={Geneva: World Health Organization}
}

@article{FriedlEtAl2010,
   author = {Friedl, Mark A. and Sulla-Menashe, Damien and Tan, Bin and Schneider, Annemarie and Ramankutty, Navin and Sibley, Adam and Huang, Xiaoman},
   title = {MODIS Collection 5 global land cover: Algorithm refinements and characterization of new datasets},
   journal = {Remote Sensing of Environment},
   volume = {114},
   number = {1},
   pages = {168-182},
   ISSN = {0034-4257},
   DOI = {https://doi.org/10.1016/j.rse.2009.08.016},
   url = {https://www.sciencedirect.com/science/article/pii/S0034425709002673},
   year = {2010},
   type = {Journal Article}
}

@article{GuptaEtAl2022,
   author = {Gupta, Surya and Papritz, Andreas and Lehmann, Peter and Hengl, Tomislav and Bonetti, Sara and Or, Dani},
   title = {Global Mapping of Soil Water Characteristics Parameters— Fusing Curated Data with Machine Learning and Environmental Covariates},
   journal = {Remote Sensing},
   volume = {14},
   number = {8},
   pages = {1947},
   ISSN = {2072-4292},
   url = {https://www.mdpi.com/2072-4292/14/8/1947},
   year = {2022},
   type = {Journal Article}
}

@article{HenglEtAl2026,
   author = {Hengl, T. and Consoli, D. and Tian, X. and Nauman, T. W. and Nussbaum, M. and Isik, M. S. and Parente, L. and Ho, Y. F. and Simoes, R. and Gupta, S. and Samuel-Rosa, A. and Zborowski Horst, T. and Safanelli, J. L. and Harris, N.},
   title = {OpenLandMap-soildb: global soil information at 30m spatial resolution for 2000–2022+ based on spatiotemporal Machine Learning and harmonized legacy soil samples and observations},
   journal = {Earth Syst. Sci. Data},
   volume = {18},
   number = {2},
   pages = {989-1036},
   ISSN = {1866-3516},
   DOI = {10.5194/essd-18-989-2026},
   url = {https://essd.copernicus.org/articles/18/989/2026/},
   year = {2026},
   type = {Journal Article}
}

@article{HenglEtAl2017,
   author = {Hengl, Tomislav and Mendes de Jesus, Jorge and Heuvelink, Gerard B. M. and Ruiperez Gonzalez, Maria and Kilibarda, Milan and Blagotić, Aleksandar and Shangguan, Wei and Wright, Marvin N. and Geng, Xiaoyuan and Bauer-Marschallinger, Bernhard and Guevara, Mario Antonio and Vargas, Rodrigo and MacMillan, Robert A. and Batjes, Niels H. and Leenaars, Johan G. B. and Ribeiro, Eloi and Wheeler, Ichsani and Mantel, Stephan and Kempen, Bas},
   title = {SoilGrids250m: Global gridded soil information based on machine learning},
   journal = {PLOS ONE},
   volume = {12},
   number = {2},
   pages = {e0169748},
   DOI = {10.1371/journal.pone.0169748},
   url = {https://doi.org/10.1371/journal.pone.0169748},
   year = {2017},
   type = {Journal Article}
}

@article{LundstadEtAl2023,
   author = {Lundstad, Elin and Brugnara, Yuri and Pappert, Duncan and Kopp, Jérôme and Samakinwa, Eric and Hürzeler, André and Andersson, Axel and Chimani, Barbara and Cornes, Richard and Demarée, Gaston and Filipiak, Janusz and Gates, Lydia and Ives, Gemma L. and Jones, Julie M. and Jourdain, Sylvie and Kiss, Andrea and Nicholson, Sharon E. and Przybylak, Rajmund and Jones, Philip and Rousseau, Daniel and Tinz, Birger and Rodrigo, Fernando S. and Grab, Stefan and Domínguez-Castro, Fernando and Slonosky, Victoria and Cooper, Jason and Brunet, Manola and Brönnimann, Stefan},
   title = {The global historical climate database HCLIM},
   journal = {Scientific Data},
   volume = {10},
   number = {1},
   pages = {44},
   ISSN = {2052-4463},
   DOI = {10.1038/s41597-022-01919-w},
   url = {https://doi.org/10.1038/s41597-022-01919-w},
   year = {2023},
   type = {Journal Article}
}

@article{MistryEtAl2019,
   author = {Mistry, Malcolm N.},
   title = {A High-Resolution Global Gridded Historical Dataset of Climate Extreme Indices},
   journal = {Data},
   volume = {4},
   number = {1},
   pages = {41},
   ISSN = {2306-5729},
   url = {https://www.mdpi.com/2306-5729/4/1/41},
   year = {2019},
   type = {Journal Article}
}

@article{ParenteEtAl2024,
   author = {Parente, Leandro and Sloat, Lindsey and Mesquita, Vinicius and Consoli, Davide and Stanimirova, Radost and Hengl, Tomislav and Bonannella, Carmelo and Teles, Nathália and Wheeler, Ichsani and Hunter, Maria and Ehrmann, Steffen and Ferreira, Laerte and Mattos, Ana Paula and Oliveira, Bernard and Meyer, Carsten and Şahin, Murat and Witjes, Martijn and Fritz, Steffen and Malek, Ziga and Stolle, Fred},
   title = {Annual 30-m maps of global grassland class and extent (2000–2022) based on spatiotemporal Machine Learning},
   journal = {Scientific Data},
   volume = {11},
   number = {1},
   pages = {1303},
   ISSN = {2052-4463},
   DOI = {10.1038/s41597-024-04139-6},
   url = {https://doi.org/10.1038/s41597-024-04139-6},
   year = {2024},
   type = {Journal Article}
}

@article{PesaresiEtAl2013,
   author = {Pesaresi, M. and Huadong, G. and Blaes, X. and Ehrlich, D. and Ferri, S. and Gueguen, L. and Halkia, M. and Kauffmann, M. and Kemper, T. and Lu, L. and Marin-Herrera, M. A. and Ouzounis, G. K. and Scavazzon, M. and Soille, P. and Syrris, V. and Zanchetta, L.},
   title = {A Global Human Settlement Layer From Optical HR/VHR RS Data: Concept and First Results},
   journal = {IEEE Journal of Selected Topics in Applied Earth Observations and Remote Sensing},
   volume = {6},
   number = {5},
   pages = {2102-2131},
   ISSN = {2151-1535},
   DOI = {10.1109/JSTARS.2013.2271445},
   year = {2013},
   type = {Journal Article}
}

@article{TatemEtAl2017,
   author = {Tatem, Andrew J.},
   title = {WorldPop, open data for spatial demography},
   journal = {Scientific Data},
   volume = {4},
   number = {1},
   pages = {170004},
   ISSN = {2052-4463},
   DOI = {10.1038/sdata.2017.4},
   url = {https://doi.org/10.1038/sdata.2017.4},
   year = {2017},
   type = {Journal Article}
}

@article{LiangEtAl2026,
  author = {Liang, Qixiang and Di, Yanfeng and Hao, Xingming and Zhang, Jingjing and others},
  title = {A 30 m Multi-Year Dataset of Major Crop Distributions in Xinjiang, China (2013--2024) Based on Harmonized Landsat--Sentinel-2 Data},
  journal = {Scientific Data},
  year = {2026},
  volume = {13},
  doi = {10.1038/s41597-026-07082-w}
}

@article{ChongEtAl2026,
  author = {Chong, Yue Linn and Kr{\"a}mer, Julie and Chakhvashvili, Erekle and Marks, Elias and others},
  title = {The Multi-Sensor and Multi-Temporal Dataset of Multiple Crops for In-Field Phenotyping and Monitoring},
  journal = {Scientific Data},
  year = {2026},
  volume = {13},
  pages = {17},
  doi = {10.1038/s41597-025-06462-y}
}

@article{DiEtAl2026,
  author = {Di, Yuanyuan and Dong, Jinwei and You, Nanshan and Li, Zhichao and others},
  title = {30 m-resolution annual crop type maps in Northeast China from 2001 to 2022},
  journal = {Scientific Data},
  year = {2026},
  volume = {13},
  pages = {197},
  doi = {10.1038/s41597-025-06516-1}
}

@article{CatarinoEtAl2025,
  author = {Catarino, Rui and Klinnert, Ana and Barbosa, Ana Luisa and d'Andrimont, Raphael and others},
  title = {Mapping the potential of Natural Pest Control services in pan-European landscapes at 50 m resolution},
  journal = {Scientific Data},
  year = {2025},
  volume = {12},
  pages = {1865},
  doi = {10.1038/s41597-025-06138-7}
}

@article{ReesEtAl2025,
  author = {Rees, Charles and Ineichen, Lorin and Finger, Robert and Grovermann, Christian},
  title = {Data covering soil management practices and farm characteristics on Swiss arable farms},
  journal = {Scientific Data},
  year = {2025},
  volume = {12},
  pages = {1471},
  doi = {10.1038/s41597-025-05731-0}
}

@book{Yearbook2024,
   author = {Eurostat, .},
   publisher = {Eurostat, .},
   title = {Eurostat regional yearbook – 2024 edition},
   ISBN = {978-92-68-14307-0 },
   DOI = {10.2785/174221},
   url = {https://ec.europa.eu/eurostat/en/web/products-flagship-publications/w/ks-ha-24-001},
   year = {2024},
   type = {Book}
}

@article{ColomboEtAl2023,
   author = {Colombo, Loris and Marongiu, Alessandro and Malvestiti, Giulia and Fossati, Giuseppe and Angelino, Elisabetta and Lazzarini, Matteo and Gurrieri, Gian Luca and Pillon, Silvia and Lanzani, Guido Giuseppe},
   title = {Assessing the impacts and feasibility of emissions reduction scenarios in the Po Valley},
   journal = {Frontiers in Environmental Science},
   volume = {11},
   ISSN = {2296-665X},
   DOI = {10.3389/fenvs.2023.1240816},
   url = {https://www.frontiersin.org/articles/10.3389/fenvs.2023.1240816},
   year = {2023},
   type = {Journal Article}
}

@article{MarongiuEtAl2022,
   author = {Marongiu, Alessandro and Angelino, Elisabetta and Moretti, Marco and Malvestiti, Giulia and Fossati, Giuseppe},
   title = {Atmospheric emission sources in the po-basin from the LIFE-IP PREPAIR project},
   journal = {Open Journal of Air Pollution},
   volume = {11},
   number = {3},
   pages = {70-83},
   year = {2022},
   type = {Journal Article}
}

@article{MarongiuEtAl2024,
   author = {Marongiu, Alessandro and Collalto, Anna Gilia and Distefano, Gabriele Giuseppe and Angelino, Elisabetta},
   title = {Application of Machine Learning to Estimate Ammonia Atmospheric Emissions and Concentrations},
   journal = {Air},
   volume = {2},
   number = {1},
   pages = {38-60},
   ISSN = {2813-4168},
   url = {https://www.mdpi.com/2813-4168/2/1/3},
   year = {2024},
   type = {Journal Article}
}

@article{MarongiuEtAl2025,
   author = {Marongiu, Alessandro and Colombo, Loris and Collalto, Anna Gilia},
   title = {Dynamic emission inventory of ammonia in northern Italy by machine learning},
   journal = {Air Quality, Atmosphere \& Health},
   ISSN = {1873-9326},
   DOI = {10.1007/s11869-025-01779-4},
   url = {https://doi.org/10.1007/s11869-025-01779-4},
   year = {2025},
   type = {Journal Article}
}

@article{MoralesEtAl2021,
   author = {Morales, Domingo and Esteban, María Dolores and Pérez, Agustín and Hobza, Tomáš},
   title = {A course on small area estimation and mixed models},
   journal = {Methods, theory and applications in R},
   year = {2021},
   type = {Journal Article}
}

@article{FayHerriot1979,
   author = {Fay, Robert E. III and Herriot, Roger A.},
   title = {Estimates of Income for Small Places: An Application of James-Stein Procedures to Census Data},
   journal = {Journal of the American Statistical Association},
   volume = {74},
   number = {366a},
   pages = {269-277},
   ISSN = {0162-1459},
   DOI = {10.1080/01621459.1979.10482505},
   url = {https://doi.org/10.1080/01621459.1979.10482505},
   year = {1979},
   type = {Journal Article}
}

@book{CressieWikle2015,
   author = {Cressie, Noel and Wikle, Christopher K},
   title = {Statistics for spatio-temporal data},
   publisher = {John Wiley \& Sons},
   ISBN = {1119243041},
   year = {2015},
   type = {Book}
}

@article{Wikle2015,
   author = {Wikle, Christopher K},
   title = {Modern perspectives on statistics for spatio‐temporal data},
   journal = {Wiley Interdisciplinary Reviews: Computational Statistics},
   volume = {7},
   number = {1},
   pages = {86-98},
   ISSN = {1939-5108},
   year = {2015},
   type = {Journal Article}
}

@article{Elhorst2010,
   author = {Elhorst, J. Paul},
   title = {Applied Spatial Econometrics: Raising the Bar},
   journal = {Spatial Economic Analysis},
   volume = {5},
   number = {1},
   pages = {9-28},
   ISSN = {1742-1772},
   DOI = {10.1080/17421770903541772},
   url = {https://doi.org/10.1080/17421770903541772},
   year = {2010},
   type = {Journal Article}
}

@article{ParentLeSage2012,
   author = {Parent, Olivier and LeSage, James P.},
   title = {Spatial dynamic panel data models with random effects},
   journal = {Regional Science and Urban Economics},
   volume = {42},
   number = {4},
   pages = {727-738},
   ISSN = {0166-0462},
   DOI = {https://doi.org/10.1016/j.regsciurbeco.2012.04.008},
   url = {https://www.sciencedirect.com/science/article/pii/S0166046212000361},
   year = {2012},
   type = {Journal Article}
}

@article{CerquetiMattera2025,
   author = {Cerqueti, Roy and Mattera, Raffaele},
   title = {Measuring unit relevance and stability in hierarchical spatio-temporal clustering},
   journal = {Spatial Statistics},
   volume = {66},
   pages = {100880},
   ISSN = {2211-6753},
   DOI = {https://doi.org/10.1016/j.spasta.2025.100880},
   url = {https://www.sciencedirect.com/science/article/pii/S2211675325000028},
   year = {2025},
   type = {Journal Article}
}

@article{MorelliEtAl2026,
   author = {Morelli, Caterina and Maranzano, Paolo and Otto, Philipp},
   title = {Spatiotemporal clustering of GHGs emissions in Europe: Exploring the role of spatial component},
   journal = {Spatial Statistics},
   pages = {100960},
   ISSN = {2211-6753},
   DOI = {https://doi.org/10.1016/j.spasta.2026.100960},
   url = {https://www.sciencedirect.com/science/article/pii/S2211675326000084},
   year = {2026},
   type = {Journal Article}
}

@misc{Kim2026,
      title={Statistics in Survey Sampling}, 
      author={Jae Kwang Kim},
      year={2024},
      eprint={2401.07625},
      archivePrefix={arXiv},
      primaryClass={stat.ME},
      url={https://arxiv.org/abs/2401.07625}, 
}

@inbook{DasEtAl2021,
   author = {Das, Sujit and Pal, Debanjana and Sarkar, Abhijit},
   title = {Particulate Matter Pollution and Global Agricultural Productivity},
   booktitle = {Sustainable Agriculture Reviews 50: Emerging Contaminants in Agriculture},
   editor = {Kumar Singh, Vipin and Singh, Rishikesh and Lichtfouse, Eric},
   publisher = {Springer International Publishing},
   address = {Cham},
   pages = {79-107},
   ISBN = {978-3-030-63249-6},
   DOI = {10.1007/978-3-030-63249-6_4},
   url = {https://doi.org/10.1007/978-3-030-63249-6_4},
   year = {2021},
   type = {Book Section}
}

@article{CrovaEtAl2021,
   author = {Crova, F. and Valli, G. and Bernardoni, V. and Forello, A. C. and Valentini, S. and Vecchi, R.},
   title = {Effectiveness of airborne radon progeny assessment for atmospheric studies},
   journal = {Atmospheric Research},
   volume = {250},
   pages = {105390},
   ISSN = {0169-8095},
   DOI = {https://doi.org/10.1016/j.atmosres.2020.105390},
   url = {https://www.sciencedirect.com/science/article/pii/S0169809520313272},
   year = {2021},
   type = {Journal Article}
}

@inproceedings{DiNicolantonioEtAl2007,
   author = {Di Nicolantonio, Walter and Cacciari, Alessandra and Bolzacchini, Ezio and Ferrero, Luca and Volta, Marialuisa and Pisoni, Enrico},
   title = {MODIS aerosol optical properties over north Italy for estimating surface-level PM2. 5},
   booktitle = {Proceedings of Envisat Symposium},
   publisher = {ESA Publications Division Montreux, Switzerland},
   volume = {636},
   type = {Conference Proceedings},
   year = {2007}
}

@article{DiNicolantonio2009,
   author = {Di Nicolantonio, W. and Cacciari, A. and Petritoli, A. and Carnevale, C. and Pisoni, E. and Volta, M. L. and Stocchi, P. and Curci, G. and Bolzacchini, E. and Ferrero, L. and Ananasso, C. and Tomasi, C.},
   title = {MODIS and OMI satellite observations supporting air quality monitoring},
   journal = {Radiation Protection Dosimetry},
   volume = {137},
   number = {3-4},
   pages = {280-287},
   ISSN = {0144-8420},
   DOI = {10.1093/rpd/ncp231},
   url = {https://doi.org/10.1093/rpd/ncp231},
   year = {2009},
   type = {Journal Article}
}

@article{DiemozEtAl2019,
   author = {Diémoz, H. and Barnaba, F. and Magri, T. and Pession, G. and Dionisi, D. and Pittavino, S. and Tombolato, I. K. F. and Campanelli, M. and Della Ceca, L. S. and Hervo, M. and Di Liberto, L. and Ferrero, L. and Gobbi, G. P.},
   title = {Transport of Po Valley aerosol pollution to the northwestern Alps – Part 1: Phenomenology},
   journal = {Atmos. Chem. Phys.},
   volume = {19},
   number = {5},
   pages = {3065-3095},
   ISSN = {1680-7324},
   DOI = {10.5194/acp-19-3065-2019},
   url = {https://acp.copernicus.org/articles/19/3065/2019/},
   year = {2019},
   type = {Journal Article}
}

@article{FerreroEtAl2012,
   author = {Ferrero, L. and Cappelletti, D. and Moroni, B. and Sangiorgi, G. and Perrone, M. G. and Crocchianti, S. and Bolzacchini, E.},
   title = {Wintertime aerosol dynamics and chemical composition across the mixing layer over basin valleys},
   journal = {Atmospheric Environment},
   volume = {56},
   pages = {143-153},
   ISSN = {1352-2310},
   DOI = {https://doi.org/10.1016/j.atmosenv.2012.03.071},
   url = {https://www.sciencedirect.com/science/article/pii/S1352231012003123},
   year = {2012},
   type = {Journal Article}
}

@article{FerreroEtAl2014,
   author = {Ferrero, L. and Castelli, M. and Ferrini, B. S. and Moscatelli, M. and Perrone, M. G. and Sangiorgi, G. and D'Angelo, L. and Rovelli, G. and Moroni, B. and Scardazza, F. and Močnik, G. and Bolzacchini, E. and Petitta, M. and Cappelletti, D.},
   title = {Impact of black carbon aerosol over Italian basin valleys: high-resolution measurements along vertical profiles, radiative forcing and heating rate},
   journal = {Atmos. Chem. Phys.},
   volume = {14},
   number = {18},
   pages = {9641-9664},
   ISSN = {1680-7324},
   DOI = {10.5194/acp-14-9641-2014},
   url = {https://acp.copernicus.org/articles/14/9641/2014/},
   year = {2014},
   type = {Journal Article}
}

@article{FerreroEtAl2019,
   author = {Ferrero, L. and Riccio, A. and Ferrini, B. S. and D'Angelo, L. and Rovelli, G. and Casati, M. and Angelini, F. and Barnaba, F. and Gobbi, G. P. and Cataldi, M. and Bolzacchini, E.},
   title = {Satellite AOD conversion into ground PM10, PM2.5 and PM1 over the Po valley (Milan, Italy) exploiting information on aerosol vertical profiles, chemistry, hygroscopicity and meteorology},
   journal = {Atmospheric Pollution Research},
   volume = {10},
   number = {6},
   pages = {1895-1912},
   ISSN = {1309-1042},
   DOI = {https://doi.org/10.1016/j.apr.2019.08.003},
   url = {https://www.sciencedirect.com/science/article/pii/S1309104219304416},
   year = {2019},
   type = {Journal Article}
}

@article{FerreroEtAl2011,
   author = {Ferrero, L. and Riccio, A. and Perrone, M. G. and Sangiorgi, G. and Ferrini, B. S. and Bolzacchini, E.},
   title = {Mixing height determination by tethered balloon-based particle soundings and modeling simulations},
   journal = {Atmospheric Research},
   volume = {102},
   number = {1},
   pages = {145-156},
   ISSN = {0169-8095},
   DOI = {https://doi.org/10.1016/j.atmosres.2011.06.016},
   url = {https://www.sciencedirect.com/science/article/pii/S0169809511002079},
   year = {2011},
   type = {Journal Article}
}

@article{WyerEtAl2022,
   author = {Wyer, Katie E. and Kelleghan, David B. and Blanes-Vidal, Victoria and Schauberger, Günther and Curran, Thomas P.},
   title = {Ammonia emissions from agriculture and their contribution to fine particulate matter: A review of implications for human health},
   journal = {Journal of Environmental Management},
   volume = {323},
   pages = {116285},
   ISSN = {0301-4797},
   DOI = {https://doi.org/10.1016/j.jenvman.2022.116285},
   url = {https://www.sciencedirect.com/science/article/pii/S0301479722018588},
   year = {2022},
   type = {Journal Article}
}

@article{BhattacharyyaEtAl2022,
title = {Soil carbon sequestration, greenhouse gas emissions, and water pollution under different tillage practices},
journal = {Science of The Total Environment},
volume = {826},
pages = {154161},
year = {2022},
issn = {0048-9697},
doi = {https://doi.org/10.1016/j.scitotenv.2022.154161},
url = {https://www.sciencedirect.com/science/article/pii/S0048969722012530},
author = {Siddhartha Shankar Bhattacharyya and Fernanda Figueiredo Granja Dorilêo Leite and Casey L. France and Adetomi O. Adekoya and Gerard H. Ros and Wim {de Vries} and Elda M. Melchor-Martínez and Hafiz M.N. Iqbal and Roberto Parra-Saldívar}
}

@article{AbbasEtAl2020,
title = {A review of soil carbon dynamics resulting from agricultural practices},
journal = {Journal of Environmental Management},
volume = {268},
pages = {110319},
year = {2020},
issn = {0301-4797},
doi = {https://doi.org/10.1016/j.jenvman.2020.110319},
url = {https://www.sciencedirect.com/science/article/pii/S0301479720302541},
author = {Farhat Abbas and Hafiz Mohkum Hammad and Wajid Ishaq and Aitazaz Ahsan Farooque and Hafiz Faiq Bakhat and Zahida Zia and Shah Fahad and Wajid Farhad and Artemi Cerdà}
}

@article{KauerEtAl2015,
title = {Soil carbon dynamics estimation and dependence on farming system in a temperate climate},
journal = {Soil and Tillage Research},
volume = {154},
pages = {53-63},
year = {2015},
issn = {0167-1987},
doi = {https://doi.org/10.1016/j.still.2015.06.010},
url = {https://www.sciencedirect.com/science/article/pii/S0167198715001270},
author = {Karin Kauer and Berit Tein and Diego {Sanchez de Cima} and Liina Talgre and Vyacheslav Eremeev and Evelin Loit and Anne Luik}
}

@article{PapadopoulosEtAl2014,
title = {Does organic management lead to enhanced soil physical quality?},
journal = {Geoderma},
volume = {213},
pages = {435-443},
year = {2014},
issn = {0016-7061},
doi = {https://doi.org/10.1016/j.geoderma.2013.08.033},
url = {https://www.sciencedirect.com/science/article/pii/S0016706113003157},
author = {A. Papadopoulos and N.R.A. Bird and A.P. Whitmore and S.J. Mooney}
}

@article{Plieninger2011,
   author = {Plieninger, Tobias},
   title = {Capitalizing on the Carbon Sequestration Potential of Agroforestry in Germany's Agricultural Landscapes: Realigning the Climate Change Mitigation and Landscape Conservation Agendas},
   journal = {Landscape Research},
   volume = {36},
   number = {4},
   pages = {435-454},
   ISSN = {0142-6397},
   DOI = {10.1080/01426397.2011.582943},
   url = {https://doi.org/10.1080/01426397.2011.582943},
   year = {2011},
   type = {Journal Article}
}

@article{CrippaEtAl2024_ESSD,
   author = {Crippa, M. and Guizzardi, D. and Pagani, F. and Schiavina, M. and Melchiorri, M. and Pisoni, E. and Graziosi, F. and Muntean, M. and Maes, J. and Dijkstra, L. and Van Damme, M. and Clarisse, L. and Coheur, P.},
   title = {Insights into the spatial distribution of global, national, and subnational greenhouse gas emissions in the Emissions Database for Global Atmospheric Research (EDGAR v8.0)},
   journal = {Earth Syst. Sci. Data},
   volume = {16},
   number = {6},
   pages = {2811-2830},
   ISSN = {1866-3516},
   DOI = {10.5194/essd-16-2811-2024},
   url = {https://essd.copernicus.org/articles/16/2811/2024/},
   year = {2024},
   type = {Journal Article}
}

\clearpage
\appendix

\renewcommand{\thesection}{S\arabic{section}}
\renewcommand{\thesubsection}{S\arabic{section}.\arabic{subsection}}

\setcounter{section}{0}
\setcounter{subsection}{0}

\clearpage

\section{Spatial alignment of gridded datasets via spatial block kriging}\label{sec:spatial_alignment_appendix}
Several environmental datasets included in SCARFACE, such as meteorological variables, atmospheric concentrations, and emission inventories, are originally provided on regular spatial grids and need to be realigned to the ASR spatial support to be consistent with the final structure of the dataset.

The need to integrate information collected on different spatial supports is known in spatial statistics as the \textit{Change of Support Problem}, extensively discussed in \cite{gotway2002combining}. In our case, the original data consist of point or grid-cell observations, whereas the target quantities correspond to averages over irregular polygonal domains. When the objective is to obtain spatial averages over areas, spatial block kriging (BK) represents a natural and statistically principled solution \citep{chiles2012geostatistics}.

Let $Y = \{Y(s),\, s \in D\}$ denote a real-valued spatial random field defined over the geographical region $D$, and let $V \subset D$ be a polygonal subregion corresponding to one ASR with area $|V|$. Block kriging is used to predict the spatial average of the random field over a target polygon $V$, defined as
\begin{equation}
Y_V = \frac{1}{|V|}\int_V Y(s)\,ds .
\end{equation}

The ordinary block kriging predictor is obtained as the best linear unbiased predictor (BLUP) of the spatial average $Y_V$ using a linear combination of observations at nearby sampling locations:
\[
\hat{Y}_V=\sum_{i=1}^{k}\alpha_i\,Y(s_i) = \boldsymbol{\alpha}^\top \mathbf{Y},
\]
where $\mathbf{Y} = (Y(s_1),\ldots,Y(s_k))^\top$ is the vector of observations (i.e., values at grid-cell centers) selected in the local neighborhood and $\boldsymbol{\alpha} = (\alpha_1,\ldots,\alpha_k)^\top$ is the vector of kriging weights.

The vector of kriging weights is obtained as the solution to the constrained optimization problem
\begin{equation}
(\alpha_1,\ldots,\alpha_k) = \arg\min_{\alpha_1,\ldots,\alpha_k}\operatorname{Var}\!\left(Y_V - \sum_{i=1}^{k}\alpha_i Y(s_i)\right)
\end{equation}
subject to the unbiasedness constraint
\begin{equation}
\sum_{i=1}^{k}\alpha_i = 1.
\end{equation}

This optimization depends on the spatial covariance structure of the process. In point kriging, the covariance vector is composed of point-to-point covariances between each observation location $s_i$ and the prediction location $u$. In block kriging, the relevant covariance terms are defined with respect to the target block support rather than to a single prediction location, and are evaluated through block-to-point and block-to-block covariance integrals, which in practice are approximated numerically using discretization of the polygonal support \citep[detailed derivations can be found in][]{chiles2012geostatistics}. Specifically, if the block $V$ is discretized by $N_V$ representative points $u_1^{(V)},\ldots,u_{N_V}^{(V)}$, the point-to-block covariance between an observation location $s_i$ and the target block $V$ is approximated as the arithmetic mean of the point-to-point covariances between $s_i$ and all discretization points of the block:
\begin{equation}
C(s_i,V)\approx \frac{1}{N_V}\sum_{j=1}^{N_V} C\!\left(s_i,u_j^{(V)}\right).    
\end{equation}
Hence, $C(s_i,V)$ measures how strongly the observation at location $s_i$ co-varies with the \emph{average value} of the process over the whole support $V$, rather than with the process at a single location. Geometrically, this quantity is obtained by connecting the point $s_i$ to every discretization point inside the target block and averaging the corresponding point-to-point covariances.

More generally, if two blocks $V$ and $V'$ are discretized by $\{u_1^{(V)},\ldots,u_{N_V}^{(V)}\}$ and $\{u_1^{(V')},\ldots,u_{N_{V'}}^{(V')}\}$, respectively, the block-to-block covariance is approximated by the average of all pairwise point-to-point covariances between discretization points in the two blocks:
\begin{equation}
C(V,V')\approx \frac{1}{N_V N_{V'}}\sum_{i=1}^{N_V}\sum_{j=1}^{N_{V'}} C\!\left(u_i^{(V)},u_j^{(V')}\right).
\end{equation}
This quantity expresses the covariance between the block averages over $V$ and $V'$. As a special case, when $V'=V$, one obtains the block-to-block covariance
\begin{equation}
C(V,V)\approx \frac{1}{N_V^2}\sum_{i=1}^{N_V}\sum_{j=1}^{N_V} C\!\left(u_i^{(V)},u_j^{(V)}\right),
\end{equation}
which corresponds to the covariance of the block average with itself and enters the expression of the block kriging variance.

The above optimization problem can also be recast into the following ordinary kriging system
\begin{equation}
\begin{bmatrix}
\mathbf{C} & \mathbf{1}\\
\mathbf{1}^\top & 0
\end{bmatrix}
\begin{bmatrix}
\boldsymbol{\alpha}\\
\lambda
\end{bmatrix}
=
\begin{bmatrix}
\mathbf{c}_V\\
1
\end{bmatrix},
\end{equation}
where $\mathbf{C}$ is the covariance matrix among the selected observations, $\mathbf{c}_V=\left(C(s_1,V),\ldots,C(s_k,V)\right)^\top$ is the point-to-block covariance vector, and $\lambda$ is the Lagrange multiplier enforcing the unbiasedness constraint $\sum_{i=1}^k\alpha_i=1$. The resulting predictor is the best linear unbiased predictor of the block mean, and the associated block kriging prediction variance is
\begin{equation}
\widehat{\mathrm{Var}}(\hat{Y}_V - Y_V) = C(V,V)-\boldsymbol{\alpha}^\top \mathbf{c}_V+\lambda.
\end{equation}

\begin{figure}[htbp]
\centering
\begin{subfigure}[b]{0.48\textwidth}
\centering
\begin{tikzpicture}[scale=0.95]
\filldraw[black] (0,0) circle (3pt);
\node[above left] at (0,0.05) {$u$};
\draw[gray] (3,-1.4) rectangle (6.2,1.4);
\node[above] at (4.6,1.45) {$V$};
\foreach \x/\y in {3.4/0.9,4.5/0.9,5.7/0.9,3.4/0,4.5/0,5.7/0,3.4/-0.9,4.5/-0.9,5.7/-0.9}{
  \filldraw[black] (\x,\y) circle (2.8pt);
  \draw[black!60] (0,0) -- (\x,\y);
}
\end{tikzpicture}
\caption{Point-to-block covariance. Each segment represents one point-to-point covariance $C(u,u_j^{(V)})$ contributing to the average $C(u,V)\approx \frac{1}{N_V}\sum_{j=1}^{N_V} C(u,u_j^{(V)})$.}
\end{subfigure}
\hfill
\begin{subfigure}[b]{0.48\textwidth}
\centering
\begin{tikzpicture}[scale=0.95]
\draw[gray] (0,-1.2) rectangle (2.4,1.2);
\node[above] at (1.2,1.25) {$V$};
\draw[gray] (4,-1.2) rectangle (6.4,1.2);
\node[above] at (5.2,1.25) {$V'$};
\foreach \x/\y in {0.6/0.6,1.8/0.6,0.6/-0.6,1.8/-0.6}{
  \filldraw[black] (\x,\y) circle (2.8pt);
}
\foreach \x/\y in {4.6/0.8,5.8/0.8,4.6/-0.2,5.8/-0.2}{
  \filldraw[black] (\x,\y) circle (2.8pt);
}
\foreach \xa/\ya in {0.6/0.6,1.8/0.6,0.6/-0.6,1.8/-0.6}{
  \foreach \xb/\yb in {4.6/0.8,5.8/0.8,4.6/-0.2,5.8/-0.2}{
    \draw[black!55] (\xa,\ya) -- (\xb,\yb);
  }
}
\end{tikzpicture}
\caption{Block-to-block covariance. Each segment represents one point-to-point covariance between discretization points in $V$ and $V'$, contributing to $C(V,V')\approx \frac{1}{N_VN_{V'}}\sum_{i=1}^{N_V}\sum_{j=1}^{N_{V'}} C(u_i^{(V)},u_j^{(V')})$.}
\end{subfigure}
\caption{Schematic representation of the covariance terms involved in block kriging. Left: point-to-block covariance between a point support and a target block. Right: block-to-block covariance between two block supports.}
\label{fig:block_covariances}
\end{figure}
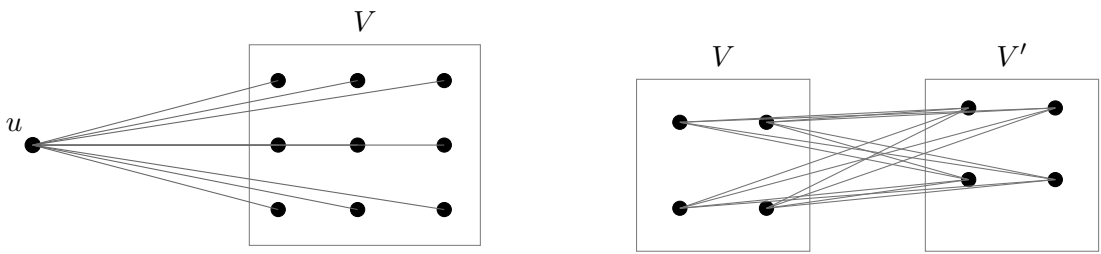

The spatial prediction is implemented using a local ordinary block kriging algorithm \citep{pebesma2004multivariable} with constant but unknown mean (i.e., ordinary kriging with intercept-only trend specification) and depending on the spatial covariance structure of the process, typically expressed through a semi-variogram function. The block kriging strategy is applied for each combination of variable, year and sector (when applicable), such that only the spatial field observed at grid-cell location and a subset of nearby observations is used in to predict the average value over each ASR polygon. The kriging predictor is constructed as a weighted linear combination of nearby observations, where the set of contributing observations is restricted to the closest sampling locations around the target block. The locality of the prediction is controlled by the size of the kriging neighborhood (i.e., $n_{max}$), that is, the maximum number of neighboring observations included in the predictor. Smaller neighborhoods produce more localized predictions that reflect short-range spatial variability, whereas larger neighborhoods yield smoother predictions that capture broader spatial structures but increase computational burden. This local formulation reduces computational cost (the kriging system scales approximately as $O(n_{\max}^3)$ with respect to the number of neighboring observations) and improves numerical stability when working with large gridded datasets.

In practice, the up-scaling procedure consists of three main steps. First, a spatial covariance model is estimated. Several isotropic covariance families are considered, nmely spherical, exponential, Gaussian, and Matérn models. For each candidate covariance model, prediction performance is evaluated using repeated out-of-sample predictions obtained through random $5$-fold cross-validation (CV) \citep{OttoEtAl2024}. The covariance model producing the lowest root mean squared prediction error (RMSE) is selected. Second, for the selected covariance model, the a grid of candidate neighborhood size is tuned through CV, and the value minimizing the RMSE is retained. Third, using the selected covariance structure and neighborhood configuration, ordinary block kriging is performed over the ASR polygons. The procedure yields both the predicted spatial mean for each ASR and the corresponding kriging prediction variance, which quantifies the interpolation uncertainty associated with the area-level spatial prediction. The spatial alignment procedure based on local block kriging is summarized in Algorithm~\ref{alg:block_kriging}.

\begin{algorithm}[htbp]
\caption{Local block kriging procedure for ASR spatial alignment}
\label{alg:block_kriging}
\begin{algorithmic}[1]
\Require Gridded observations $\{(s_i, Y(s_i)) : i=1,\dots,n\}$ for a given variable, year, and sector; ASR polygons $\{V_d : d=1,\dots,D\}$; candidate covariance families $\mathcal{C}$; candidate neighborhood sizes $\mathcal{N}$
\Ensure For each ASR $V_d$, block kriging estimate $\hat{Y}_{V_d}$ and prediction variance $\widehat{\mathrm{Var}}(Y_{V_d} - \hat{Y}_{V_d})$

\State \textbf{Covariance-model selection}
\State Fix an initial neighborhood size $n_{\max}^{(0)}$
\ForAll{$C \in \mathcal{C}$}
    \State Fit the isotropic variogram model corresponding to covariance family $C$
    \State Perform random 5-fold cross-validation using local ordinary kriging with the $n_{\max}^{(0)}$ closest observations
    \State Compute the cross-validated root mean squared error $RMSE(C \mid n_{\max}^{(0)})$
\EndFor
\State Select
\[
C^\star = \arg\min_{C \in \mathcal{C}} RMSE(C \mid n_{\max}^{(0)})
\]

\Statex

\State \textbf{Neighborhood-size tuning}
\ForAll{$n_{\max} \in \mathcal{N}$}
    \State Perform random 5-fold cross-validation using covariance family $C^\star$
    \State Compute the cross-validated root mean squared error $RMSE(n_{\max} \mid C^\star)$
\EndFor
\State Select
\[
n_{\max}^\star = \arg\min_{n_{\max} \in \mathcal{N}} RMSE(n_{\max} \mid C^\star)
\]

\Statex

\State \textbf{Block kriging over ASR polygons}
\For{$d=1,\dots,D$}
    \State Use covariance family $C^\star$ and neighborhood size $n_{\max}^\star$
    \State Perform local ordinary block kriging to predict the block mean $Y_{V_d}$
    \State Compute the predicted block mean $\hat{Y}_{V_d}$
    \State Compute the block kriging prediction variance $\widehat{\mathrm{Var}}(Y_{V_d} - \hat{Y}_{V_d})$
\EndFor

\State \Return $\left\{\hat{Y}_{V_d},\widehat{\mathrm{Var}}(Y_{V_d} - \hat{Y}_{V_d}) : d=1,\dots,D \right\}$
\end{algorithmic}
\end{algorithm}

In the implementation adopted in SCARFACE, block kriging predictions and associated uncertainties are computed using the \texttt{krige} function from the \texttt{gstat} package \citep{pebesma2004multivariable} in \textsf{R}. In \texttt{gstat}, the point-to-block and block-to-block covariances are approximated numerically by discretizing the target polygonal support $V$ into a finite set of representative points and averaging the implied point-to-point covariances. When prediction targets are supplied as polygons, the package automatically computes the block average for each polygon and returns both the block kriging prediction and the corresponding prediction variance. In our implementation, prediction is further localized through the parameter \texttt{nmax}, which restricts the kriging system to the closest observations in the neighborhood, and the intercept-only specification \texttt{~1} is used throughout the fitting and prediction steps \citep{pebesma2004multivariable}.

\clearpage
\section{Post-stratification procedure adopted to generate the spatio-temporal weighting system for FADN survey data}\label{sec:appendix_raking}

This appendix provides a detailed description of the calibration procedure used to construct the spatio-temporal weighting system linking the FADN sample to the reconstructed population of farms at the ASR level. The procedure combines direct post-stratification via iterative proportional fitting (IPF) or raking \citep{Kolenikov2014,LomaxNorman2016} with a hierarchical calibration strategy designed to address sparse sample coverage across strata.

The final weighting system is obtained by applying the following hierarchy within each ASR--year domain:
\begin{enumerate}
\item direct cell post-stratification when all strata are observed;
\item two-dimensional raking calibration on both margins;
\item one-dimensional calibration on a single margin;
\item temporal donor reconstruction using nearby years;
\item uniform domain weighting as a final fallback.
\end{enumerate}

This hierarchical strategy ensures that the resulting weights remain stable and well defined even under sparse sampling conditions while preserving consistency with the reconstructed population totals at the ASR level.

\subsection{Notation}
Let $d=1,\ldots,D$ denote the ASRs, $s\in\{1,2\}$ the economic size classes (i.e., small, large), $t\in\{1,2,3\}$ the technical specialization classes (i.e., crop, livestock, mixed), and $y$ the year.

Let $N_{dsty}$ denote the number of farms in the population and $n_{dsty}$ the number of sampled farms in the FADN survey for the corresponding stratum. 

Population margins are defined as
\[
N_{dsy}=\sum_{t}N_{dsty}, 
\qquad
N_{dty}=\sum_{s}N_{dsty}.
\]

Similarly, sample margins are
\[
n_{dsy}=\sum_{t}n_{dsty}, 
\qquad
n_{dty}=\sum_{s}n_{dsty}.
\]

A schematic representation of the post-stratification workflow is reported in Figure~\ref{fig:new_poststratification}.

\begin{figure}
\centering
\includegraphics[height=5cm]{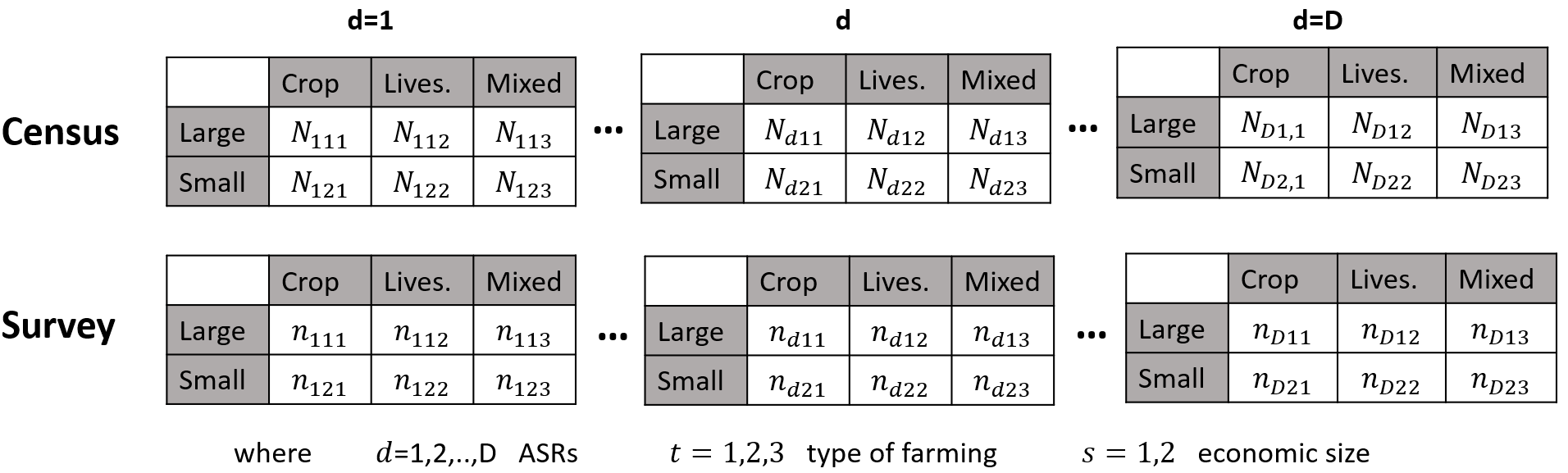}
\caption{Census-based post-stratification quantities used to construct the spatio-temporal weighting system for the FADN sample. Annual population counts of farms by economic size and technical specialization ($N_{dsty}$) are reconstructed at the ASR level using the 2010 and 2020 Italian Agricultural Censuses and interpolated for intermediate years. These reconstructed population totals are matched with the FADN sample counts ($n_{dsty}$) to derive post-stratification weights through the iterative proportional fitting procedure.}
\label{fig:new_poststratification_appendix}
\end{figure}

The objective is to compute weights $w_{dsty}$ such that the weighted sample reproduces the known population margins while remaining well defined even when some strata are not represented in the survey sample.

\subsection{Direct post-stratification}
When the survey sample covers all strata $(s,t)$ within a given ASR--year domain $(d,y)$, weights are computed directly through post-stratification:
\[
w_{dsty}=\frac{N_{dsty}}{n_{dsty}}.
\]
This corresponds to the classical cell-weighting estimator and guarantees exact reproduction of the population counts within each stratum.

\subsection{Two-dimensional raking calibration}
When some strata are missing from the sample but the support of the sample allows identification of both margins, weights are calibrated through IPF. The calibration weights are obtained by solving the system
\[
\sum_{t} w_{dsty} n_{dsty} = N_{dsy}, \qquad
\sum_{s} w_{dsty} n_{dsty} = N_{dty}.
\]
Starting from initial design weights $w_{dsty}^{(0)}$, the raking algorithm iteratively rescales the weights along the two margins until convergence is achieved. The resulting weights reproduce the population totals by economic size and by technical specialization within each ASR--year domain.

\subsection{One-dimensional calibration}
If two-dimensional calibration cannot be performed because the sample lacks support for one of the margins (for example, all sampled farms belong to the same specialization class), calibration is attempted along a single margin.

Two alternative calibrations are considered:
\begin{enumerate}
\item calibration on economic size:
\[
\sum_{t} w_{dsty} n_{dsty}=N_{dsy},
\]
\item calibration on technical specialization:
\[
\sum_{s} w_{dsty} n_{dsty}=N_{dty}.
\]
\end{enumerate}
When both calibrations are feasible, the solution minimizing the maximum absolute deviation between weighted sample counts and reconstructed population totals is selected.

\subsection{Temporal donor reconstruction}
In some ASR--year domains the survey sample is too sparse to support calibration. In these cases, information from adjacent years within the same ASR is used. A temporal donor year $y^\ast$ is selected according to the following criteria:
\begin{enumerate}
\item minimum temporal distance, that is, $|y-y^\ast|$,
\item best calibration quality in the donor year,
\item most recent donor year when ties occur.
\end{enumerate}

Let
\[
p_{dsty^\ast}=\frac{N_{dsty^\ast}}{\sum_{s,t}N_{dsty^\ast}}
\]
denote the population share of each stratum in the donor year. The reconstructed population totals for the target year are obtained by applying these shares to the total number of farms in the target domain:
\[
\tilde N_{dsty}=p_{dsty^\ast}\, N_{dy}.
\]
When sampled farms are present in the corresponding strata, weights are derived as
\[
w_{dsty}=\frac{\tilde N_{dsty}}{n_{dsty}}.
\]

\subsection{Uniform weighting}
As a final fallback, when neither calibration nor temporal reconstruction is feasible, a uniform domain weight is applied:
\[
w_{dsty}=\frac{N_{dy}}{n_{dy}},
\]
where
\[
N_{dy}=\sum_{s,t}N_{dsty}, 
\qquad
n_{dy}=\sum_{s,t}n_{dsty}.
\]
This ensures that the weighted sample reproduces the total number of farms within the ASR--year domain even when the internal stratification cannot be recovered.

\subsection{Area-level estimation of the total and the mean}
Using the calibrated weights, farm-level indicators from FADN are then aggregated to the ASR--year level through design-based \cite{HorvitzThompson1952} estimators (HT). For a generic farm-level variable $X_{dstyk}$ observed for farm $k$ belonging to stratum $(d,s,t,y)$, the HT estimator of the ASR total for year $y$ and area $d$ is
\begin{equation*}
\hat{T}_{dy}(X)=\sum_{s=1}^{2}\sum_{t=1}^{3}\sum_{k=1}^{n_{dsty}} w_{dsty}X_{dstyk}.
\end{equation*}
The corresponding estimator of the ASR mean for area $d$ in year $y$ is obtained as
\begin{equation*}
\hat{\mu}_{dy}(X)=\frac{\hat{T}_{dy}(X)}{N_{dy}}, \qquad N_{dy}=\sum_{s=1}^{2}\sum_{t=1}^{3} N_{dsty}.
\end{equation*}

\clearpage
\section{Generalized Variance Function (GVF) methodology adopted to regularize direct estimates of the variance for FADN survey data}
This appendix describes the \textit{Generalized Variance Function} (GVF) approach adopted to regularize the sampling variances of the \cite{HorvitzThompson1952} estimators (HT). The objective of the GVF procedure is to stabilize highly variable direct variance estimates, particularly in domains with small sample sizes, while preserving the design-based variance structure implied by the survey design (especially in well-sampled domains). Regularization is achieved by exploiting the systematic relationship between sampling variability and domain sample size. This allows for consistent uncertainty quantification across all domain--time observations.

\subsection{Model specification}
Let $\widehat{Y}_d$ denote the HT estimator for domain $d$, and let $\widehat{V}_d$ be the corresponding direct variance estimate. The variance structure is modeled using a GVF, defined on a logarithmic scale.

Two alternative formulations are considered:

\paragraph{Variance formulation}
\begin{equation}
\log(\widehat{V}_d) = \alpha + \beta \log(\widehat{Y}_d) + g(p_d) + u_{s(d)} + v_{t(d)} + \varepsilon_d
\end{equation}

\paragraph{Relative variance formulation}
\begin{equation}
\log(CV_d^2) = \alpha + g(p_d) + u_{s(d)} + v_{t(d)} + \varepsilon_d
\end{equation}
where
\begin{equation}
CV_d^2 = \frac{\widehat{V}_d}{\widehat{Y}_d^2}
\end{equation}

The first formulation models the variance directly, while the second focuses on the relative precision of the estimator. These complementary specifications allow the model to capture both scale-dependent and scale-invariant components of the variance structure.

\subsection{Precision modeling}
The function $g(p_d)$ captures the relationship between sampling precision and domain-level sampling characteristics. Three alternative specifications are considered:
\begin{equation}
g(p_d)=
\begin{cases}
\gamma_1 \log(n_d) \\
\gamma_1 \log\left(\frac{n_d}{N_d}\right) \\
\gamma_1 \log(n_d) + \gamma_2 \log(N_d)
\end{cases}
\end{equation}
where $n_d$ is the effective sample size and $N_d$ is the domain population size.

These specifications allow the GVF to account for both absolute sample size effects and relative sampling intensity. Collinear representations are avoided by not including equivalent transformations simultaneously.

\subsection{Estimation framework}
The GVF is estimated using domain--year observations with sufficiently reliable direct variance estimates. In particular, only domains satisfying
\begin{equation}
n_d \ge 3
\end{equation}
are included in the estimation sample.

Spatial and temporal heterogeneity in the variance structure are captured through random intercepts:
\begin{equation}
u_{s(d)} \sim N(0,\sigma_s^2), \qquad v_{t(d)} \sim N(0,\sigma_t^2)
\end{equation}
where $s(d)$ and $t(d)$ denote the spatial domain (ASR) and the time index, respectively.

\subsection{Model selection}
Several candidate GVF models are obtained by combining the alternative response formulations and precision specifications. Model selection is based on three complementary criteria reflecting the objectives of variance regularization.

\paragraph{Coherence with direct variances}
The agreement between GVF predictions and direct variance estimates is measured using the logarithmic root mean squared error (RMSE):
\begin{equation}
RMSE_{\log} =
\sqrt{
\frac{1}{m}
\sum_{d=1}^{m}
\left(
\log(\widehat{V}_d) -
\log(\widehat{V}^{GVF}_d)
\right)^2
}
\end{equation}
Lower values indicate that the GVF reproduces the structure of the direct variances more closely.

Only models satisfying
\begin{equation}
RMSE_{\log} \le RMSE_{\log}^{\min}(1+\tau)
\end{equation}
with $\tau = 0.05$ are retained for further comparison.

\paragraph{Reduction of extreme variances}
To evaluate the ability of the GVF to regularize unstable estimates, attention is restricted to the upper tail of the variance distribution, that is, preference is given to the specification that most effectively reduces excessively large variances. Let $Q_{0.75}(\widehat{V})$ denote the 75th percentile of the direct variances. For domains with $\widehat{V}_d \ge Q_{0.75}(\widehat{V})$, the reduction factor is defined as
\begin{equation}
R_d = \frac{\widehat{V}_d^{final}}{\widehat{V}_d}
\end{equation}
and summarized through
\begin{equation}
Reduction_{upper} = \operatorname{median}(R_d)
\end{equation}

\paragraph{Control of variance inflation}
The GVF may increase some variance estimates. This behavior is quantified by
\begin{equation}
Increase\_share = \frac{1}{m}\sum_{d=1}^{m}\mathbf{1}\left(\frac{\widehat{V}_d^{final}}{\widehat{V}_d} > 1.05\right)
\end{equation}
This indicator measures the proportion of domains where the final variance exceeds the direct variance by more than 5\%. Lower values indicate fewer substantial increases.

\paragraph{Selection rule}
The final model is selected through a hierarchical rule:
\begin{enumerate}
\item retain models within the $RMSE_{\log}$ tolerance band;
\item among them, select the model minimizing $Reduction_{upper}$;
\item if necessary, select the model minimizing $Increase\_share$;
\item break remaining ties using $RMSE_{\log}$.
\end{enumerate}

\subsection{Variance prediction}
After model estimation, predicted squared coefficients of variation are obtained for all domains. The corresponding variance estimates are reconstructed as
\begin{equation}
\widehat{V}_d^{GVF} = \widehat{CV}_d^{\,2}\,\widehat{Y}_d^{\,2}
\end{equation}
No bias correction is applied when transforming predictions from the logarithmic scale.

\subsection{Blending of variance estimates}
The final variance estimate combines the direct estimator and the GVF prediction:
\begin{equation}
\widehat{V}_d^{final} = w(n_d)\widehat{V}_d^{GVF} + \left(1-w(n_d)\right)\widehat{V}_d
\end{equation}
where the weight depends on the domain sample size:
\begin{equation}
w(n_d)=
\begin{cases}
1 & n_d \le 1 \\
0.5 & n_d = 2 \\
0 & n_d \ge 3
\end{cases}
\end{equation}

This implies that domains with a single observation rely entirely on the GVF prediction; domains with two observations combine the GVF prediction and the direct variance equally; and domains with three or more observations retain the direct variance estimate. This rule reflects the increasing reliability of direct variance estimates as the sample size grows.

\clearpage
\section{Core thematic domains integrated in the SCARFACE dataset}
\begin{table}[htbp]
\centering
\resizebox{\textwidth}{!}{
\begin{threeparttable}
\caption{Overview of core thematic domains integrated in the SCARFACE dataset}
\label{tab:core_domains_appendix}
\begin{tabular}{p{2.6cm} p{5.0cm} p{2.4cm} p{2.0cm} p{2.4cm} p{4.8cm} p{1.6cm}}
\toprule
\textbf{Domain} & \textbf{Data source} & \textbf{Spat. resolution} & \textbf{Temp. resolution} & \textbf{Period} & \textbf{ASR method} & \textbf{Num. indic.} \\
\midrule
Farm activities & FADN (CREA) & Farm-level microdata & Annual & 2011--2023 & Population-weighted direct aggregation (sum/mean) & 436 \\
Emissions & EDGAR GHG v2025; EDGAR AP v8.1 (JRC, European Commission) & $0.1^{\circ}$ grid & Annual & 2011--2024 (GHG); 2011--2022 (AP) & Spatial block kriging & 96 \\
Air quality (1) & European air quality interpolated data (EEA Datahub) & 1 km grid & Annual & 2011--2022 & Spatial block kriging & 8 \\
Air quality (2) & CAMS European air quality reanalyses (CAMS Datahub) & $0.1^{\circ}$ grid & Hourly & 2013--2024 & Temporal aggregation + spatial block kriging & 680 \\
Meteorology & ERA5-Land reanalysis (ECMWF / Copernicus CDS) & $0.1^{\circ}$ hourly grid & Hourly & 2011--2024 & Temporal aggregation + spatial block kriging &  1170 \\
Extreme weather & European Drought Observatory (EDO) & $0.1^{\circ}$ grid & Infra-annual & 2011--2024 & Temporal aggregation + spatial block kriging &  208 \\
Land cover (1) & CORINE Land Cover (Copernicus) & 100 m grid & Multi-year snapshots & 2012 and 2018 & Reclassification + piecewise constant + area shares & 9 \\
Land cover (2) & GDLC dataset (Global Dynamic Land Cover) & 100 m grid & Multi-year snapshots & 2015--2019 & Reclassification + piecewise constant + area shares &  6 \\
Livestock & BDN livestock registry (Italian Ministry of Health) & Municipality (LAU) & Annual & 2011--2024 & Direct aggregation (sum) &  16 \\
Socio-economic indicators & Sistema informativo ``A misura di comune'' (ISTAT) & Municipality (LAU) & Annual & 2014--2023 & Direct aggregation (sum/mean) & 95 \\
Geographical features & Terrain elevation and morphology indicators (AWS Terrain Tiles; ISTAT) & Raster / administrative polygons & Static & Time-invariant & Spatial extraction + aggregation & 10 \\
Administrative metadata & NUTS territorial attributes (Eurostat GISCO) & Administrative regions & Static & Time-invariant & Spatial join to ASR & 16 \\
\bottomrule
\end{tabular}
\begin{tablenotes}[flushleft]
\footnotesize
\item \textit{Note:} All variables are harmonized at the ASR level through temporal aggregation, spatial block kriging, or direct aggregation depending on the nature of the original data source. The number of indicators (Num. Indic.) is given by the sum of annual and seasonal indicators (for CAMS, EDO and ERA5) or the sum of mean and total indicators plus the corresponding variances (for FADN data).
\end{tablenotes}
\end{threeparttable}
}
\end{table}

\clearpage
\section{Pollution emissions by sector and pollutant from EDGAR}
\begin{table}[htbp]
\centering
\caption{Agricultural emission sectors included in the SCARFACE dataset based on the EDGAR inventories. The table reports the corresponding IPCC classification codes and the GHGs and air pollutants considered for each sector.}
\label{tab:edgar_sectors_appendix}
\begin{tabular}{llllp{5.5cm}}
\toprule
Sector & Description & IPCC 1996 & IPCC 2006 & Emitted species \\
\midrule
AGS & Agricultural soils & 4C+4D1  & 3C2+3C3 & N$_2$O, NH$_3$, NO$_x$ \\
    &                    & 4D2+4D4 & 3C4+3C7 & NMVOC, PM$_{10}$, \\
AWB & Agricultural waste & 4F & 3C1b & PM$_{2.5}$, CO$_2$, CH$_4$, N$_2$O, \\
    & burning            & & & CO, NO$_x$, NMVOC, PM$_{10}$, PM$_{2.5}$, BC, OC \\
ENF & Enteric fermentation & 4A & 3A1 & CH$_4$, NMVOC, NH$_3$ \\
MNM & Manure management & 4B & 3A2 & CH$_4$, N$_2$O, NH$_3$, \\
 & & & & NMVOC, NO$_x$, PM$_{10}$, PM$_{2.5}$ \\
N2O & Indirect N$_2$O emiss. & 4D3 & 3C5+3C6 & N$_2$O\\
& from agriculture & & & \\
\bottomrule
\end{tabular}
\end{table}

\clearpage
\section{Air quality indicators from EEA and CAMS}
\begin{table}[htbp]
\centering
\caption{Air quality datasets integrated in the SCARFACE database.}
\label{tab:air_quality_datasets_appendix}
\begin{tabular}{p{4cm} p{3.3cm} p{2.5cm} p{5.0cm}}
\toprule
\textbf{Dataset and source} & \textbf{Spatial resolution} (grid) & \textbf{Temporal resolution} & \textbf{Pollutants (unit: $\mu g\,/m^{3}$)} \\
\midrule
EEA interpolated air quality (European Environment Agency) 
& 1\,km $\times$ 1\,km 
& Annual 
& PM$_{10}$, PM$_{2.5}$, NO$_2$, NO$_x$ \\[0.4em]
\midrule
CAMS air quality reanalysis (Copernicus Atmosphere Monitoring Service) 
& 0.1$^{\circ}$$\times$0.1$^{\circ}$ 
& Hourly $\rightarrow$ annual \& seasonal 
& NH$_3$, PM$_{2.5}$, PM$_{10}$, NO$_2$, NO, SO$_2$, O$_3$, CO; secondary PM$_{2.5}$ components (ammonium, nitrate, sulfate, organic matter, elemental carbon); dust and sea salt \\
\bottomrule
\end{tabular}
\end{table}

\clearpage
\begin{table}[htbp]
\centering
\begin{threeparttable}
\caption{Air pollutant variables included in the SCARFACE dataset}
\label{tab:air_quality_variables}
\begin{tabular}{p{7.0cm} p{2.5cm} p{4.0cm}}
\toprule
\textbf{Pollutant} & \textbf{Source} & \textbf{Temporal coverage} \\
\midrule
PM$_{10}$   & EEA & 2011--2022 \\
            & CAMS & 2013--2024 \\
PM$_{2.5}$  & EEA & 2011--2022 \\
            & CAMS & 2013--2024 \\
NO$_2$      & EEA & 2011--2022 \\
            & CAMS & 2013--2024 \\
NO$_x$ & EEA & 2011--2022 \\
NH$_3$ & CAMS & 2018--2024 \\
CO & CAMS & 2016--2024 \\
SO$_2$ & CAMS & 2016--2024 \\
O$_3$ & CAMS & 2013--2024 \\
NO & CAMS & 2018--2024 \\
Dust & CAMS & 2018--2024 \\
Secondary inorganic aerosol & CAMS & 2018--2024 \\
PM$_{2.5}$ ammonium & CAMS & 2023 \\
PM$_{2.5}$ nitrate & CAMS & 2023 \\
PM$_{2.5}$ sulphate & CAMS & 2023 \\
PM$_{2.5}$ total organic matter & CAMS & 2022--2024 \\
PM$_{2.5}$ total elementary carbon & CAMS & 2020--2024 \\
PM$_{2.5}$ residential elementary carbon & CAMS & 2020--2024 \\
PM$_{10}$ sea salt (dry) & CAMS & 2022--2024 \\
PM$_{10}$ wildfires & CAMS & 2019--2024 \\
\bottomrule
\end{tabular}
\begin{tablenotes}[flushleft]
\footnotesize
\item \textit{Note:} Concentrations are measured in $\mu g\,/m^{3}$. Data are obtained from the CAMS European air quality reanalyses service: \href{https://ads.atmosphere.copernicus.eu/datasets/cams-europe-air-quality-reanalyses?tab=overview}{https://ads.atmosphere.copernicus.eu/datasets/cams-europe-air-quality-reanalyses?tab=overview} (accessed on March 27th, 2026) or European Environmenta Agency (EEA) air quality data (interpolated data)- Series: \href{https://www.eea.europa.eu/en/datahub/datahubitem-view/82700fbd-2953-467b-be0a-78a520c3a7ef}{https://www.eea.europa.eu/en/datahub/datahubitem-view/82700fbd-2953-467b-be0a-78a520c3a7ef} (accessed on March 27th, 2026). CAMS data are originally provided ath hourly temporal resolutions; annual and seasonal (Winter, Spring, Summer, Fall) summary indicators (mean, minimum, maximum, and standard deviation) are computed prior to spatial aggregation. EEA data are already provided at the yearly frequency.
\end{tablenotes}
\end{threeparttable}
\end{table}

\clearpage
\section{Meteorological and land-surface variables from ERA5-Land and drought-related and extreme weather indicators from EDO}
\footnotesize
\setlength{\tabcolsep}{4pt}
\renewcommand{\arraystretch}{1.05}

\begin{longtable}{@{}>{\centering\arraybackslash}m{2cm}
                    >{\centering\arraybackslash}m{4cm}
                    >{\centering\arraybackslash}m{1.5cm}
                    >{\centering\arraybackslash}m{2cm}
                    p{5cm}@{}}
\caption{Meteorological and land-surface variables derived from ERA5-Land hourly data and aggregated annually for Agrarian Sub Regions (ASRs) in Northern Italy (2011--2024). Available aggregation functions are minimum, mean, maximum and standard deviation. Seasons are defined according to climatic calendar, that is Winter (December–February), Spring (March–May), Summer (June–August), and Fall (September–November).}
\label{tab:meteo_variables}\\
\toprule
Variable & Description & Unit & Class & Notes \\
\midrule
\endfirsthead
\toprule
Variable & Description & Unit & Class & Notes \\
\midrule
\endhead
\midrule
\multicolumn{5}{r}{Continued on next page} \\
\midrule
\endfoot
\bottomrule
\endlastfoot
t2m & 2m air temperature & $^\circ$C & Atmospheric & Converted from Kelvin to Celsius before aggregation. \\
d2m & 2m dew point temperature & $^\circ$C & Atmospheric & Converted from Kelvin to Celsius before aggregation. \\
rh & Relative humidity & \% & Atmospheric & Derived variable computed from air temperature ($t2m$) and dew point temperature ($d2m$) using the standard saturation vapor pressure formulation. \\
sp & Surface pressure & Pa & Atmospheric &  \\
u10 & 10m zonal wind component & m/s & Atmospheric &  \\
v10 & 10m meridional wind component & m/s & Atmospheric &  \\
ws & Wind speed & m/s & Atmospheric & Derived from the horizontal wind components as $ws=\sqrt{u10^{2}+v10^{2}}$. \\
wd & Wind direction & 0-360$^\circ$ & Atmospheric & Derived from wind components using $wd = 180 - \left(\mathrm{atan2}(u10/ws,v10/ws)\cdot180/\pi\right)$. \\
\midrule
ssr & Surface solar radiation & J/m$^{2}$ & Radiation &  \\
str & Surface thermal radiation & J/m$^{2}$ & Radiation &  \\
\midrule
tp & Total precipitation & mm & Hydrology & Converted from meters to millimeters before aggregation. \\
sf & Snowfall & mm & Hydrology & Converted from meters to millimeters before aggregation. \\
ro & Total runoff & mm & Hydrology & Converted from meters to millimeters before aggregation. \\
sro & Surface runoff & mm & Hydrology & Converted from meters to millimeters before aggregation. \\
ssro & Sub-surface runoff & mm & Hydrology & Converted from meters to millimeters before aggregation. \\
\midrule
skt & Skin temperature & $^\circ$C & Soil & Converted from Kelvin to Celsius before aggregation. \\
stl1 & Soil temperature layer 1 (0--7 cm) & $^\circ$C & Soil & Converted from Kelvin to Celsius before aggregation. \\
stl2 & Soil temperature layer 2 (7--28 cm) & $^\circ$C & Soil & Converted from Kelvin to Celsius before aggregation. \\
stl3 & Soil temperature layer 3 (28--100 cm) & $^\circ$C & Soil & Converted from Kelvin to Celsius before aggregation. \\
stl4 & Soil temperature layer 4 (100--289 cm) & $^\circ$C & Soil & Converted from Kelvin to Celsius before aggregation. \\
swvl1 & Soil volumetric water content layer 1 & m$^{3}$/m$^{3}$ & Soil &  \\
swvl2 & Soil volumetric water content layer 2 & m$^{3}$/m$^{3}$ & Soil &  \\
swvl3 & Soil volumetric water content layer 3 & m$^{3}$/m$^{3}$ & Soil &  \\
swvl4 & Soil volumetric water content layer 4 & m$^{3}$/m$^{3}$ & Soil &  \\
\midrule
cvh & High vegetation cover fraction & none & Vegetation &  \\
cvl & Low vegetation cover fraction & none & Vegetation &  \\
lai\_hv & Leaf area index (high vegetation) & m$^{2}$/m$^{2}$ & Vegetation &  \\
lai\_lv & Leaf area index (low vegetation) & m$^{2}$/m$^{2}$ & Vegetation &  \\
\end{longtable}

\clearpage
\begin{table}[htbp]
\centering
\begin{threeparttable}
\caption{Drought-related indicators and temperature extremes indicators included in the SCARFACE dataset}
\label{tab:edo_variables}
\begin{tabular}{p{2.5cm} p{6.0cm} p{2.5cm} p{2.8cm} p{2.5cm}}
\toprule
\textbf{Variable} & \textbf{Description} & \textbf{Unit} & \textbf{Original temporal resolution} & \textbf{Temporal coverage} \\
\midrule
\multicolumn{5}{l}{\textit{Soil moisture and drought indicators}} \\
SMI & Normalized soil moisture index describing wet/dry conditions & / & Monthly & 2011--2024 \\
SMIAN & Anomaly of soil moisture index relative to climatology & / & Monthly & 2011--2024 \\
SMIAN-1 & Short-term soil moisture anomaly (1-month scale) & / & Monthly & 2011--2024 \\
SMIAN-6 & Medium-term soil moisture anomaly (6-month scale) & / & Monthly & 2011--2024 \\
SMIAN-12 & Long-term soil moisture anomaly (12-month scale) & / & Monthly & 2011--2024 \\
CDI & Composite (combined) drought indicator integrating precipitation, soil moisture and vegetation response & Categorical index & Monthly & 2012--2024 \\
SPEI-1 & Standardized Precipitation-Evapotranspiration Index (1-month scale) & Standardized index & Monthly & 2011--2024 \\
SPEI-6 & Standardized Precipitation-Evapotranspiration Index (6-month scale) & Standardized index & Monthly & 2011--2024 \\
SPEI-12 & Standardized Precipitation-Evapotranspiration Index (12-month scale) & Standardized index & Monthly & 2011--2024 \\
\midrule
\multicolumn{5}{l}{\textit{Temperature extremes indicators}} \\
CWD & Cold wave duration: number of days with cold-wave conditions & Days & Daily & 2011--2024 \\
CWI & Cold Wave Intensity: intensity of cold-wave events & Index & Daily & 2011--2024 \\
HWD & Heat wave duration: number of days with heat-wave conditions & Days & Daily & 2011--2024 \\
HWI & Heat wave intensity: intensity of heat-wave events & Index & Daily & 2011--2024 \\
\bottomrule
\end{tabular}
\begin{tablenotes}[flushleft]
\footnotesize
\item \textit{Note:} Data are obtained from the European Drought Observatory (EDO), Copernicus Emergency Management Service: \href{https://drought.emergency.copernicus.eu/tumbo/edo/download/}{https://drought.emergency.copernicus.eu/tumbo/edo/download/} (accessed on March 27th, 2026). Original temporal resolutions vary across indicators (10-days, monthly, or daily) according to the specific product definition.
\end{tablenotes}
\end{threeparttable}
\end{table}

\clearpage
\section{Livestock consistency data from the National Livestock Registry (BDN)}

\begin{table}[htbp]
\centering
\caption{Livestock indicators derived from the National Livestock Registry (BDN) and included in the SCARFACE dataset.}
\label{tab:livestock_variables}
\begin{tabular}{p{3.2cm} p{3.0cm} p{5.8cm}}
\toprule
\textbf{Species} & \textbf{Indicator} & \textbf{Livestock mode} \\
\midrule
Bovine--buffalo & Number of farms & Stabled, SemiWild, Unknown \\
Bovine--buffalo & Number of heads & Stabled, SemiWild, Unknown \\
Swine & Number of farms & Stabled, SemiWild, Unknown \\
Swine & Number of heads & Stabled, SemiWild, Unknown \\
\bottomrule
\end{tabular}
\end{table}

\clearpage
\section{Socio-economic indicators from ISTAT}
\footnotesize
\setlength{\tabcolsep}{4pt}
\renewcommand{\arraystretch}{1.05}

\begin{longtable}{@{}>{\ttfamily}p{3.7cm}
                    p{2.2cm}
                    p{5.0cm}
                    p{2.6cm}
                    >{\centering\arraybackslash}p{1.8cm}@{}}
\caption{Socio-economic indicators included in the SCARFACE dataset.}
\label{tab:socioeconomic_variables}\\
\toprule
Variable & Domain & Description & Unit & Aggregation \\
\midrule
\endfirsthead
\toprule
Variable & Domain & Description & Unit & Aggregation \\
\midrule
\endhead
\midrule
\multicolumn{5}{r}{Continued on next page} \\
\midrule
\endfoot
\bottomrule
\bottomrule
\multicolumn{5}{p{\textwidth}}{
\footnotesize
\textit{Note:} data are obtained from the experimental statistical dataset \textit{Sistema informativo a misura di comune} (\href{https://www.istat.it/statistica-sperimentale/aggiornamento-degli-indicatori-del-sistema-informativo-a-misura-di-comune/}{https://www.istat.it/statistica-sperimentale/aggiornamento-degli-indicatori-del-sistema-informativo-a-misura-di-comune/}, accessed on April 3rd, 2026) produced by the Italian National Institute of Statistics (ISTAT). Available data cover the periodo 2014--2023. Variables ending with $\texttt{\_*}$ are further disaggregated by age group, gender, typology or sector (depending on the variable itself). ISTAT data are originally provided at the municipal level and then are aggregated at the ASR-level using the aggregation function reported in the last column.
}
\endlastfoot
Pop\_* & Demography & Resident population (by gender or age group) & persons & Sum \\
PopForeign\_* & Demography & Foreign resident population (by gender) & persons & Sum \\
BirthRate & Demography & Birth rate & per 1,000 inhabitants & Mean \\
MortalityRate & Demography & Mortality rate & per 1,000 inhabitants & Mean \\
NetMigrRate & Demography & Net migration rate & per 1,000 inhabitants & Mean \\
OldAgeIndex & Demography & Old-age index & ratio & Mean \\
OldAgeDepIndex & Demography & Old-age dependency index & ratio & Mean \\
StructDepIndex & Demography & Structural dependency index & ratio & Mean \\
\midrule
NrFam & Families & Number of families & households & Sum \\
NrFam\_* & Families & Number of families by type & households & Sum \\
NrFamWithForeigners & Families & Families with foreign members & households & Sum \\
AvgFamilySize & Families & Average family size & persons per household & Mean \\
NrSingleIncomeFamWith YoungChildren & Families & Single-income families with children under 6 & households & Sum \\
NrLowWorkIntensityFam & Families & Families with low work intensity & households & Sum \\
\midrule
EmpRate & Labor market & Employment rate & \% & Mean \\
UnempRate & Labor market & Unemployment rate & \% & Mean \\
InactivityRate & Labor market & Inactivity rate & \% & Mean \\
NEET1529 & Labor market & Youth (15--29) not in employment, education or training & \% & Mean \\
NrNonPermanentWorkers & Labor market & Non-permanent workers registered in October & persons & Sum \\
\midrule
EntrepreneurshipRate & Economy & Entrepreneurship rate & \% & Mean \\
LocalUnitsDensity & Economy & Density of local production units & units/km$^2$ & Mean \\
LocationQuotients\_* & Economy & Location quotient by economic sector & index & Mean \\
ShareLocalUnits\_* & Economy & Share of local production units by sector & \% & Mean \\
ShareWorkers\_* & Economy & Share of workers by sector & \% & Mean \\
HighTechSpecialization & Economy & High-tech production specialization index & index & Mean \\
TaxableIncomePer Taxpayer & Economy & Average taxable income per taxpayer & euros & Mean \\
\midrule
NrVehicles & Transport & Registered passenger vehicles & vehicles & Sum \\
NrMotorcycles & Transport & Registered motorcycles & vehicles & Sum \\
RoadAccidentRate & Transport & Road accident rate & accidents per 1,000 inhabitants & Mean \\
RoadAccidentMortality Index & Transport & Road accident mortality index & index & Mean \\
RoadAccidentInjury Index & Transport & Road accident injury index & index & Mean \\
\midrule
NrLibrariesPer100k & Culture & Libraries per 100,000 inhabitants & libraries per 100k inhabitants & Mean \\
CulturalSitesPer100k & Culture & Museums and cultural sites per 100,000 inhabitants & sites per 100k inhabitants & Mean \\
MuseumVisitorsPer100 & Culture & Museum visitors & visitors per 100 inhabitants & Mean \\
\midrule
SocialSpending PerCapita & Public finance & Social expenditure per capita & euros & Mean \\
RoadFinesRevenue PerCapita & Public finance & Road fines revenue per capita & euros & Mean \\
SpeedingFinesRevenue Share & Public finance & Share of fines from speed violations & \% & Mean \\
\midrule
WomenCouncilShare & Governance & Share of women in municipal councils & \% & Mean \\
WomenExecutiveShare & Governance & Share of women in municipal executive boards & \% & Mean \\
VoterTurnoutFirst Round & Governance & Voter turnout in municipal elections (first round) & \% & Mean \\
AvgCouncilAge & Governance & Average age of municipal councillors & years & Mean \\
AvgAdministratorAge & Governance & Average age of municipal administrators & years & Mean \\
\midrule
RecycleRate & Environment & Municipal waste recycling rate & \% & Mean \\
LandConsumption & Environment & Land consumption & \% of municipal area & Mean \\
\midrule
SecondEduShare & Education & Share of population with secondary education & \% & Mean \\
TertEduShare & Education & Share of population with tertiary education & \% & Mean \\
\midrule
NrChildcareChildren Served & Social services & Children enrolled in municipal childcare services & persons & Sum \\
\end{longtable}

\clearpage
\section{Land cover and land use variables from Copernicus}

\begin{table}[htbp]
\centering
\caption{Land cover datasets integrated in the SCARFACE database.}
\label{tab:lulc_datasets_appendix}
\begin{tabular}{p{4.2cm} p{3.6cm} p{2.5cm} p{6.3cm}}
\toprule
\textbf{Dataset} & \textbf{Source} & \textbf{Classes / Resolution} & \textbf{Note} \\
\midrule
CORINE Land Cover (CLC) 
& Copernicus Land Monitoring Service 
& 44 classes; 100m
& Data available for 2012 (CLC12) and 2018 (CLC18). Piecewise-constant assumption: CLC12 used for 2011–2017 and CLC18 for 2018–2024 under a piecewise-constant assumption \\[0.4em]
Global Dynamic Land Cover (GDLC) 
& Global land cover dataset 
& GDLC classes; 100m 
& Data available for 2015-2019. Piecewise-constant assumption: the 2015 map was applied to 2011–2014 and the 2019 map to 2020–2024. \\
\bottomrule
\end{tabular}
\end{table}

\clearpage
\begin{table}[htbp]
\centering
\caption{Reclassification of CORINE Land Cover classes used in the SCARFACE dataset. Original CLC classes were harmonized into nine broader categories to ensure consistency with the GDLC classification.}
\label{tab:clc_reclassification}
\begin{tabular}{p{3.2cm} p{3.9cm} p{2.8cm} p{6.8cm}}
\toprule
\textbf{Reclassified class} & \textbf{Code Reclassified} & \textbf{Original CLC codes} & \textbf{Original CLC classes} \\
\midrule
Urbanized area & \texttt{LC\_CLC\_Urban} & 111, 112, 121, 122, 123, 124, 131, 132, 133, 141, 142 & Continuous urban fabric; Discontinuous urban fabric; Industrial or commercial units; Road and rail networks and associated land; Port areas; Airports; Mineral extraction sites; Dump sites; Construction sites; Green urban areas; Sport and leisure facilities \\[0.3em]
Arable land & \texttt{LC\_CLC\_Arable-Land} & 211, 212, 213 & Non-irrigated arable land; Permanently irrigated land; Rice fields \\[0.3em]
Permanent crops & \texttt{LC\_CLC\_PermCrops} & 221, 222 & Vineyards; Fruit trees and berry plantations \\[0.3em]
Pastures & \texttt{LC\_CLC\_Pastures} & 223, 231 & Olive groves; Pastures \\[0.3em]
Heterogeneous agricultural areas & \texttt{LC\_CLC\_HetAgro} & 241, 242, 243, 244 & Annual crops associated with permanent crops; Complex cultivation patterns; Land principally occupied by agriculture, with significant areas of natural vegetation; Agro-forestry areas \\[0.3em]
Forests & \texttt{LC\_CLC\_Forests} & 311, 312, 313 & Broad-leaved forest; Coniferous forest; Mixed forest \\[0.3em]
Grassland, scrub and open spaces with little or no vegetation & \texttt{LC\_CLC\_GrassScrub OpenSpaceLilVeg} & 321, 322, 323, 324, 331, 332, 333, 334, 335 & Natural grassland; Moors and heathland; Sclerophyllous vegetation; Transitional woodland-scrub; Beaches, dunes, sands; Bare rocks; Sparsely vegetated areas; Burnt areas; Glaciers and perpetual snow \\[0.3em]
Wetlands & \texttt{LC\_CLC\_Wetlands} & 411, 412, 421, 422, 423 & Inland marshes; Peat bogs; Salt marshes; Salines; Intertidal flats \\[0.3em]
Water bodies & \texttt{LC\_CLC\_Water} & 511, 512, 521, 522, 523 & Water courses; Water bodies; Coastal lagoons; Estuaries; Sea and ocean \\
\bottomrule
\end{tabular}
\end{table}

\clearpage
\begin{table}[htbp]
\centering
\caption{Reclassification of Global Dynamic Land Cover (GDLC) classes used in the SCARFACE dataset. Original GDLC land cover classes were harmonized into nine broader categories to ensure consistency with the CORINE Land Cover classification.}
\label{tab:gdlc_reclassification}
\begin{tabular}{p{3.4cm} p{4.0cm} p{8.0cm}}
\toprule
\textbf{Reclassified class} & \textbf{Code Reclassified} & \textbf{Original GDLC classes} \\
\midrule
Urbanized area & \texttt{LC\_GDLC\_Urban} & Built-up areas and artificial surfaces \\[0.3em]
Arable land & \texttt{LC\_GDLC\_ArableLand} & Cropland; irrigated cropland; rainfed cropland \\[0.3em]
Permanent crops & \texttt{LC\_GDLC\_PermCrops} & Tree crops and plantations (e.g., orchards, vineyards) \\[0.3em]
Pastures & \texttt{LC\_GDLC\_Pastures} & Managed grasslands and pasture areas \\[0.3em]
Heterogeneous agricultural areas & \texttt{LC\_GDLC\_HetAgro} & Cropland--natural vegetation mosaics; mixed agricultural areas \\[0.3em]
Forests & \texttt{LC\_GDLC\_Forests} & Evergreen forest; deciduous forest; mixed forest \\[0.3em]
Grassland, scrub and open spaces with little or no vegetation & \texttt{LC\_GDLC\_GrassScrub OpenSpaceLilVeg} & Natural grasslands; shrublands; sparse vegetation; barren land \\[0.3em]
Wetlands & \texttt{LC\_GDLC\_Wetlands} & Inland wetlands; marshes; peatlands \\[0.3em]
Water bodies & \texttt{LC\_GDLC\_Water} & Rivers; lakes; reservoirs; coastal water bodies \\
\bottomrule
\end{tabular}
\end{table}

\clearpage
\section{Time-invariant geographical features}
\begin{table}[htbp]
\centering
\caption{Time-invariant geographical features included in the SCARFACE dataset.}
\label{tab:geographical_features}
\begin{tabular}{p{3.8cm} p{7.5cm} p{3.5cm}}
\toprule
\textbf{Feature} & \textbf{Description} & \textbf{Source} \\
\midrule
Area (m$^2$) & Total land area of the Agrarian Sub Region computed from the ASR polygon geometry (in square meters). & ASR geometries \\
Elevation (m) & Minimum, maximum, mean, and standard deviation of elevation computed within each ASR from a digital elevation model with approximately 30\,m ground resolution. & AWS Terrain Tiles \\
Share \% Area (Plain) & Share of ASR area located at elevation $\leq 200$ m. & DEM derived indicators \\
Share \% Area (Hills) & Share of ASR area located at elevation between 200 m and 600 m. & DEM derived indicators \\
Share \% Area (Mount) & Share of ASR area located at elevation $> 600$ m. & DEM derived indicators \\
Longitude, Latitude & Geographic coordinates of the centroid of each ASR polygon. & ASR geometries \\
NUTS administrative attributes & Hierarchical identifiers and territorial classifications for NUTS0--NUTS3 regions, including urban--rural typology, remoteness, metropolitan status, coastal and mountain indicators, border status, and land surface measures. & Eurostat GISCO \\
\bottomrule
\end{tabular}
\end{table}


\clearpage
\section{Techno-economic specialization of farms and agronomic practices from FADN}
\footnotesize
\setlength{\tabcolsep}{4pt}
\renewcommand{\arraystretch}{1.05}

\begin{longtable}{@{}>{\ttfamily}p{5.5cm}
                    p{3.5cm}
                    p{6.5cm}
                    p{1.5cm}@{}}
\caption{Farm Accountancy Data Network (FADN) variables included in the SCARFACE dataset}
\label{tab:fadn_variables}\\
\toprule
Variable & Domain & Description & Unit \\
\midrule
\endfirsthead
\toprule
Variable & Domain & Description & Unit \\
\midrule
\endhead
\midrule
\multicolumn{4}{r}{Continued on next page} \\
\midrule
\endfoot
\bottomrule
\multicolumn{4}{p{\textwidth}}{
\footnotesize
\textit{Note:} data are obtained from the Italian Farm Accountancy Data Network (FADN) survey (\href{https://www.crea.gov.it/en/web/politiche-e-bioeconomia/-/fadn-farm-accountancy-data-network}{https://www.crea.gov.it/en/web/politiche-e-bioeconomia/-/fadn-farm-accountancy-data-network}), coordinated by the Italian Council for Agricultural Research and Agricultural Economics Analysis (CREA). For every variables, yearly farm-level values are aggregated at the ASR-level using the Horvitz-Thompson estimator of the mean and the total, both augmented by the corresponding estimate of the area-level variance. All the variables cover the period 2011--2023 with time-varying spatial coverages (i.e., the list of areas covered can change across the years). Variables are classified according to thematic domains: agronomic practices, farms structure, economic balanca (revenues, costs, results, and subsidies) and productivity.
}
\endlastfoot
Bio & Agronomic practices & Share of organic farms in the ASR & 0--100\% \\
Fert\_NitrogenTot & Agronomic practices & Fertilizers -- Total quantity of nitrogen employed & Quintals \\
Fert\_PhosphorusTot & Agronomic practices & Fertilizers -- Total quantity of phosphorus employed & Quintals \\
Fert\_PotassiumTot & Agronomic practices & Fertilizers -- Total quantity of potassium employed & Quintals \\
FertFertHectares & Agronomic practices & Fertilizers -- Surface on which fertilizers are spread & Hectares \\
Fert\_NitrogenPerHA & Agronomic practices & Fertilizers -- Nitrogen per hectare & Quintals per ha \\
Fert\_PhosphorusPerHA & Agronomic practices & Fertilizers -- Phosphorus per hectare & Quintals per ha \\
Fert\_PotassiumPerHA & Agronomic practices & Fertilizers -- Potassium per hectare & Quintals per ha \\
Phyto\_QtyPerHA\_ToxicClass0 & Agronomic practices & Phytopharmaceuticals (class 0 -- non-toxic) per hectare & kg per ha \\
Phyto\_QtyPerHA\_ToxicClass1 & Agronomic practices & Phytopharmaceuticals (class 1 -- slightly toxic) per hectare & kg per ha \\
Phyto\_QtyPerHA\_ToxicClass2 & Agronomic practices & Phytopharmaceuticals (class 2 -- moderately toxic) per hectare & kg per ha \\
Phyto\_QtyPerHA\_ToxicClass3 & Agronomic practices & Phytopharmaceuticals (class 3 -- highly toxic) per hectare & kg per ha \\
Phyto\_QtyPerHA\_ToxicClass4 & Agronomic practices & Phytopharmaceuticals (class 4 -- extremely toxic) per hectare & kg per ha \\
Phyto\_QtyTot\_ToxicClass0 & Agronomic practices & Total quantity of phytopharmaceuticals (class 0 -- non-toxic) employed & kg \\
Phyto\_QtyTot\_ToxicClass1 & Agronomic practices & Total quantity of phytopharmaceuticals (class 1 -- slightly toxic) employed & kg \\
Phyto\_QtyTot\_ToxicClass2 & Agronomic practices & Total quantity of phytopharmaceuticals (class 2 -- moderately toxic) employed & kg \\
Phyto\_QtyTot\_ToxicClass3 & Agronomic practices & Total quantity of phytopharmaceuticals (class 3 -- highly toxic) employed & kg \\
Phyto\_QtyTot\_ToxicClass4 & Agronomic practices & Total quantity of phytopharmaceuticals (class 4 -- extremely toxic) employed & kg \\
Phyto\_Hectares\_ToxicClass0 & Agronomic practices & UAA surface where class 0 (non-toxic) phytopharmaceuticals are applied & ha \\
Phyto\_Hectares\_ToxicClass1 & Agronomic practices & UAA surface where class 1 (slightly toxic) phytopharmaceuticals are applied & ha \\
Phyto\_Hectares\_ToxicClass2 & Agronomic practices & UAA surface where class 2 (moderately toxic) phytopharmaceuticals are applied & ha \\
Phyto\_Hectares\_ToxicClass3 & Agronomic practices & UAA surface where class 3 (highly toxic) phytopharmaceuticals are applied & ha \\
Phyto\_Hectares\_ToxicClass4 & Agronomic practices & UAA surface where class 4 (extremely toxic) phytopharmaceuticals are applied & ha \\
ProdSO & Farms structure & Agricultural Standard Output: value of the potential production generated from LSU and UAA & Ths.\ \euro \\
TotalLaborHours & Farms structure & Total hours worked in a year & Hours \\
LSU & Farms structure & Livestock Standard Unit & Units \\
UAA & Farms structure & Utilized Agricultural Area (hectares) & ha \\
AWU & Farms structure & Annual Work Unit: total hours worked divided by 1,800 (employees) and 2,200 (family labor) & Units \\
ManureQL\_QtyInvInit & Agronomic practices & Total quantity of manure (stock) at January 1st & Quintals \\
ManureQL\_ValInvInit & Agronomic practices & Value of manure (stock) at January 1st & \euro \\
ManureQL\_QtyInvFin & Agronomic practices & Total quantity of manure (stock) at December 31st & Quintals \\
ManureQL\_ValInvFin & Agronomic practices & Value of manure (stock) at December 31st & \euro \\
ManureQL\_QtySales & Agronomic practices & Quantity of manure sold during the year & Quintals \\
ManureQL\_ValSales & Agronomic practices & Value of manure (stock) sold during the year & \euro \\
ManureQL\_QtyProd & Agronomic practices & Quantity of manure produced during the year & Quintals \\
ManureQL\_QtyOther & Agronomic practices & Quantity of manure employed in other ways during the year & Quintals \\
ManureQL\_ValOther & Agronomic practices & Value of manure (stock) employed in other ways during the year & \euro \\
Machine\_Number & Agronomic practices & Number of machines & Units \\
Machine\_TotPowerKW & Agronomic practices & Total power of available agricultural machinery & kW \\
Machine\_AvgPowerKW & Agronomic practices & Average power per machine & kW per machine \\
Machine\_Hours & Agronomic practices & Usage hours of agricultural machinery & Hours \\
CarbFoot & Environmental impact & Agricultural carbon footprint & Ths.\ tonnes \\
RevenuesTotal & Econ. bal.: revenues & Total revenue & \euro \\
FGSP\_Total & Econ. bal.: revenues & Total Farm Gross Saleable Production (FGSP) & \euro \\
FGSP\_TransfProducts & Econ. bal.: revenues & FGSP from processed products & \euro \\
FGSP\_DirectSales & Econ. bal.: revenues & FGSP from direct sales of agricultural products & \euro \\
FGSP\_Crops & Econ. bal.: revenues & FGSP from crops & \euro \\
FGSP\_Livestock & Econ. bal.: revenues & FGSP from livestock & \euro \\
FGSP\_RenewEnergy & Econ. bal.: revenues & FGSP from renewable energy produced by the farm & \euro \\
FGSP\_Quality & Econ. bal.: revenues & FGSP from high-quality products & \euro \\
SalesProducts & Econ. bal.: revenues & Revenue from sales of products and services & \euro \\
InventoryChange & Econ. bal.: revenues & Change in inventories & \euro \\
AidsEU & Econ. bal.: subsidies & Operating public grants from EU CAP policies (Pillar I) & \euro \\
Selfconsumes & Econ. bal.: revenues & Self-consumption, gifts, and in-kind wages & \euro \\
IncreasePhysAssets & Econ. bal.: revenues & Capitalized internal work (internally generated assets) & \euro \\
RevenuesAgritourism & Econ. bal.: revenues & Agritourism revenue & \euro \\
RevenuesThirdPartyServices & Econ. bal.: revenues & Contract work revenue (custom farming services) & \euro \\
RevenuesComplementary & Econ. bal.: revenues & Complementary activities revenue & \euro \\
RentActive & Econ. bal.: revenues & Rental income (active leases) & \euro \\
CurrentCosts & Econ. bal.: costs & Total current costs & \euro \\
Seeds & Econ. bal.: costs & Seeds and seedlings costs & \euro \\
Fertilizers & Econ. bal.: costs & Fertilizer costs & \euro \\
PesticidesHerbicides & Econ. bal.: costs & Pesticides and herbicides costs & \euro \\
AnimalFeed & Econ. bal.: costs & Animal feed costs & \euro \\
ForageBedding & Econ. bal.: costs & Forage and bedding costs & \euro \\
Machines & Econ. bal.: costs & Mechanization costs & \euro \\
Utilities & Econ. bal.: costs & Water, electricity, and fuel costs & \euro \\
ConsFactorsAgritourism & Econ. bal.: costs & Agritourism input costs & \euro \\
ExpenditureCommercialTransform & Econ. bal.: costs & Processing, marketing, and storage costs & \euro \\
ExpenditureGeneraliFondiarie & Econ. bal.: costs & General and land-related overheads & \euro \\
ThirdPartyServices & Econ. bal.: costs & Third-party services costs & \euro \\
RentPassive & Econ. bal.: costs & Equipment rental costs (operating leases) & \euro \\
VeterinaryExpenditure & Econ. bal.: costs & Veterinary and health costs & \euro \\
AgritourismCosts & Econ. bal.: costs & Agritourism service costs & \euro \\
Insurance & Econ. bal.: costs & Insurance costs & \euro \\
ValueAdded & Econ. bal.: results & Value added: total revenue -- current costs & \euro \\
MultiYearCosts & Econ. bal.: costs & Multi-year costs (depreciation and provisions) & \euro \\
Depreciation & Econ. bal.: costs & Depreciation & \euro \\
Provisions & Econ. bal.: costs & Provisions & \euro \\
FarmNetIncome & Econ. bal.: results & Net farm income (value added -- multi-year costs) & \euro \\
DistributedIncome & Finance & Distributed income (wages, social security, rent) & \euro \\
SalariesSocialCosts & Econ. bal.: costs & Wages and social security contributions & \euro \\
RentsPassive & Econ. bal.: costs & Rent expenses (passive leases) & \euro \\
OperativeIncome & Econ. bal.: results & Operating income & \euro \\
NonOperatingIncome & Econ. bal.: results & Non-operating income & \euro \\
FinancialRevenues & Econ. bal.: revenues & Financial income & \euro \\
FinancialCosts & Econ. bal.: costs & Financial expenses & \euro \\
AidsPublicCapitalAccount & Econ. bal.: subsidies & Capital grants (public investment subsidies) & \euro \\
OtherAids & Econ. bal.: subsidies & Non-EU operating subsidies & \euro \\
CurrentTaxes & Econ. bal.: costs & Current taxes & \euro \\
NetIncome & Econ. bal.: results & Net income & \euro \\
VariableCosts & Econ. bal.: costs & Variable costs & \euro \\
FixedCosts & Econ. bal.: costs & Fixed costs & \euro \\
GrossOperatingMargin & Econ. bal.: results & Gross operating margin (value added -- labor costs) & \euro \\
FarmNetValueAdded & Econ. bal.: results & Farm net value added & \euro \\
Aids\_EUPillarIprod\_Amount & Econ. bal.: subsidies & EU CAP subsidies (Pillar I -- production) & \euro \\
Aids\_EUPillarIIrurdev\_Amount & Econ. bal.: subsidies & EU CAP subsidies (Pillar II -- rural development) & \euro \\
Aids\_StateRegLocal\_Amount & Econ. bal.: subsidies & National, regional, and local subsidies & \euro \\
AvgFGSPcrops\_UAA & Productivity & Average FGSP from crops per hectare of UAA & \euro per ha \\
AvgFGSPlivestock\_LSU & Productivity & Average FGSP from livestock per LSU & \euro per LSU \\
\end{longtable}

\end{document}